\documentclass[transmag]{IEEEtran}
\usepackage{latexsym}
\usepackage{graphicx}
\usepackage{amsfonts,amssymb,amsmath}
\usepackage{hyperref}

\usepackage{amsmath,amssymb,amsfonts}

\usepackage{textcomp}
\usepackage{comment}
\usepackage{amsthm}

\usepackage{breakurl}
\usepackage{url}
\usepackage{tikz}
\usepackage{multirow}
\usepackage{algpseudocode}
\usepackage{algorithm}

\usepackage{colortbl}
\usepackage[nolist]{acronym}

\usepackage{xcolor}
\usepackage{color}

\usepackage{varwidth}
\usepackage{rotating}

\usepackage{cite}

\usepackage{xurl}

\usepackage[normalem]{ulem}

\usepackage{nameref}
\usepackage{zref-xr,zref-user}
\zxrsetup{toltxlabel=true, tozreflabel=false}
\usepackage{xcite}

\usepackage{tikz-3dplot}

\usepackage{subfigure}

\usepackage[resetlabels,labeled]{multibib}
\newcites{A}{References}

\usetikzlibrary{decorations.pathreplacing,decorations.markings,decorations.text}

\tikzset{arrow data/.style 2 args={%
      decoration={%
         markings,
         mark=at position #1 with \arrow{#2}},
         postaction=decorate}
      }%

\usetikzlibrary{shapes,backgrounds,calc}

\tikzstyle{dummy} = [rectangle, text width=0.1em, draw=white, white,
                      minimum width=0.1em, minimum height=3em, opacity=0.0]

\makeatletter
\newcommand*{\Strut}[1][0.1em]{\vrule\@width\z@\@height#1\@depth\z@\relax}
\makeatother

\definecolor{Linen}{rgb}{0.9803,0.9411,0.9019}
\definecolor{White}{rgb}{1,1,1}
\definecolor{Green}{rgb}{0.5,1,0.5}
\definecolor{Red}{rgb}{1,0.4,0.4}
\definecolor{Coral}{rgb}{1,0.4980,0.3137}
\definecolor{Grayblue}{rgb}{0.9411,0.9411,0.9803}
\definecolor{DarkLinen}{rgb}{0.729,0.7176,0.635}

\usepackage{mdframed}
\begin{document}

\begin{acronym}

\acro{5G-NR}{5G New Radio}
\acro{3GPP}{3rd Generation Partnership Project}
\acro{AC}{address coding}
\acro{ACF}{autocorrelation function}
\acro{ACR}{autocorrelation receiver}
\acro{ADC}{analog-to-digital converter}
\acrodef{aic}[AIC]{Analog-to-Information Converter}     
\acro{AIC}[AIC]{Akaike information criterion}
\acro{aric}[ARIC]{asymmetric restricted isometry constant}
\acro{arip}[ARIP]{asymmetric restricted isometry property}

\acro{ARQ}{automatic repeat request}
\acro{AUB}{asymptotic union bound}
\acrodef{awgn}[AWGN]{Additive White Gaussian Noise}     
\acro{AWGN}{additive white Gaussian noise}

\acro{APSK}[PSK]{asymmetric PSK} 

\acro{waric}[AWRICs]{asymmetric weak restricted isometry constants}
\acro{warip}[AWRIP]{asymmetric weak restricted isometry property}
\acro{BCH}{Bose, Chaudhuri, and Hocquenghem}        
\acro{BCHC}[BCHSC]{BCH based source coding}
\acro{BEP}{bit error probability}
\acro{BFC}{block fading channel}
\acro{BG}[BG]{Bernoulli-Gaussian}
\acro{BGG}{Bernoulli-Generalized Gaussian}
\acro{BPAM}{binary pulse amplitude modulation}
\acro{BPDN}{Basis Pursuit Denoising}
\acro{BPPM}{binary pulse position modulation}
\acro{BPSK}{binary phase shift keying}
\acro{BPZF}{bandpass zonal filter}
\acro{BSC}{binary symmetric channels}              
\acro{BU}[BU]{Bernoulli-uniform}
\acro{BER}{bit error rate}
\acro{BS}{Base Station}

\acro{CP}{Cyclic Prefix}
\acrodef{cdf}[CDF]{cumulative distribution function}   
\acro{CDF}{cumulative distribution function}
\acrodef{c.d.f.}[CDF]{cumulative distribution function}
\acro{CCDF}{complementary cumulative distribution function}
\acrodef{ccdf}[CCDF]{complementary CDF}               
\acrodef{c.c.d.f.}[CCDF]{complementary cumulative distribution function}
\acro{CD}{cooperative diversity}

\acro{CDMA}{Code Division Multiple Access}
\acro{ch.f.}{characteristic function}
\acro{CIR}{channel impulse response}
\acro{cosamp}[CoSaMP]{compressive sampling matching pursuit}
\acro{CR}{cognitive radio}
\acro{cs}[CS]{compressed sensing}                   
\acrodef{cscapital}[CS]{Compressed sensing} 
\acrodef{CS}[CS]{compressed sensing}
\acro{CSI}{channel state information}
\acro{CCSDS}{consultative committee for space data systems}
\acro{CC}{convolutional coding}
\acro{Covid19}[COVID-19]{Coronavirus disease}
\acro{CAPEX}{CAPital EXpenditures}

\acro{DAA}{detect and avoid}
\acro{DAB}{digital audio broadcasting}
\acro{DAS}{Distributed Antenna System}
\acro{DCT}{discrete cosine transform}
\acro{dft}[DFT]{discrete Fourier transform}
\acro{DR}{distortion-rate}
\acro{DS}{direct sequence}
\acro{DS-SS}{direct-sequence spread-spectrum}
\acro{DTR}{differential transmitted-reference}
\acro{DVB-H}{digital video broadcasting\,--\,handheld}
\acro{DVB-T}{digital video broadcasting\,--\,terrestrial}
\acro{DL}{downlink}
\acro{DSSS}{Direct Sequence Spread Spectrum}
\acro{DFT-s-OFDM}{Discrete Fourier Transform-spread-Orthogonal Frequency Division Multiplexing}
\acro{DNA}{Deoxyribonucleic Acid}

\acro{EC}{European Commission}
\acro{ECDF}{Empirical Cumulative Distribution Function}
\acro{EED}[EED]{exact eigenvalues distribution}
\acro{EIRP}{Equivalent Isotropically Radiated Power}
\acro{ELP}{equivalent low-pass}
\acro{eMBB}{Enhanced Mobile Broadband}
\acro{EMF}{Electro-Magnetic Field}
\acro{EU}{European Union}
\acro{ELP}{Exposure Limit-based Power}

\acro{FC}[FC]{fusion center}
\acro{FCC}{Federal Communications Commission}
\acro{FEC}{forward error correction}
\acro{FFT}{fast Fourier transform}
\acro{FH}{frequency-hopping}
\acro{FH-SS}{frequency-hopping spread-spectrum}
\acrodef{FS}{Frame synchronization}
\acro{FSsmall}[FS]{frame synchronization}  
\acro{FDMA}{Frequency Division Multiple Access}  
\acro{FSPL}{Free Space Path Loss}

\acro{GA}{Gaussian approximation}
\acro{GF}{Galois field }
\acro{GG}{Generalized-Gaussian}
\acro{GIC}[GIC]{generalized information criterion}
\acro{GLRT}{generalized likelihood ratio test}
\acro{GPS}{Global Positioning System}
\acro{GMSK}{Gaussian minimum shift keying}
\acro{GSMA}{Global System for Mobile communications Association}

\acro{HAP}{high altitude platform}

\acro{IDR}{information distortion-rate}
\acro{IFFT}{inverse fast Fourier transform}
\acro{iht}[IHT]{iterative hard thresholding}
\acro{i.i.d.}{independent, identically distributed}
\acro{IoT}{Internet of Things}                      
\acro{IR}{impulse radio}
\acro{lric}[LRIC]{lower restricted isometry constant}
\acro{lrict}[LRICt]{lower restricted isometry constant threshold}
\acro{ISI}{intersymbol interference}
\acro{ITU}{International Telecommunication Union}
\acro{ICNIRP}{International Commission on Non-Ionizing Radiation Protection}
\acro{IEEE}{Institute of Electrical and Electronics Engineers}
\acro{ICES}{IEEE international committee on electromagnetic safety}
\acro{IEC}{International Electrotechnical Commission}
\acro{IARC}{International Agency on Research on Cancer}
\acro{IS-95}{Interim Standard 95}

\acro{LEO}{low earth orbit}
\acro{LF}{likelihood function}
\acro{LLF}{log-likelihood function}
\acro{LLR}{log-likelihood ratio}
\acro{LLRT}{log-likelihood ratio test}
\acro{LOS}{Line-of-Sight}
\acro{LRT}{likelihood ratio test}
\acro{wlric}[LWRIC]{lower weak restricted isometry constant}
\acro{wlrict}[LWRICt]{LWRIC threshold}
\acro{LPWAN}{low power wide area network}
\acro{LoRaWAN}{Low power long Range Wide Area Network}
\acro{NLOS}{non-line-of-sight}
\acro{LB}{Lower Bound}

\acro{MB}{multiband}
\acro{MC}{multicarrier}
\acro{MDS}{mixed distributed source}
\acro{MF}{matched filter}
\acro{m.g.f.}{moment generating function}
\acro{MI}{mutual information}
\acro{MIMO}{multiple-input multiple-output}
\acro{MISO}{multiple-input single-output}
\acrodef{maxs}[MJSO]{maximum joint support cardinality}                       
\acro{ML}[ML]{maximum likelihood}
\acro{MMSE}{minimum mean-square error}
\acro{MMV}{multiple measurement vectors}
\acrodef{MOS}{model order selection}
\acro{M-PSK}[${M}$-PSK]{$M$-ary phase shift keying}                       
\acro{M-APSK}[${M}$-PSK]{$M$-ary asymmetric PSK} 
\acro{MSP}{Minimum Sensitivity-based Power}

\acro{M-QAM}[$M$-QAM]{$M$-ary quadrature amplitude modulation}
\acro{MRC}{maximal ratio combiner}                  
\acro{maxs}[MSO]{maximum sparsity order}                                      
\acro{M2M}{machine to machine}                                                
\acro{MUI}{multi-user interference}
\acro{mMTC}{massive Machine Type Communications}      
\acro{mm-Wave}{millimeter-wave}
\acro{MP}{mobile phone}
\acro{MPE}{maximum permissible exposure}
\acro{MAC}{media access control}
\acro{NB}{narrowband}
\acro{NBI}{narrowband interference}
\acro{NLA}{nonlinear sparse approximation}
\acro{NLOS}{Non-Line of Sight}
\acro{NTIA}{National Telecommunications and Information Administration}
\acro{NTP}{National Toxicology Program}
\acro{NHS}{National Health Service}
\acro{NSA}{Non-StandAlone}

\acro{OC}{optimum combining}                             
\acro{OC}{optimum combining}
\acro{ODE}{operational distortion-energy}
\acro{ODR}{operational distortion-rate}
\acro{OFDM}{orthogonal frequency-division multiplexing}
\acro{omp}[OMP]{orthogonal matching pursuit}
\acro{OSMP}[OSMP]{orthogonal subspace matching pursuit}
\acro{OQAM}{offset quadrature amplitude modulation}
\acro{OQPSK}{offset QPSK}
\acro{OFDMA}{Orthogonal Frequency-division Multiple Access}
\acro{OPEX}{OPerating EXpenditures}
\acro{OQPSK/PM}{OQPSK with phase modulation}

\acro{PAM}{pulse amplitude modulation}
\acro{PAR}{peak-to-average ratio}
\acrodef{pdf}[PDF]{probability density function}                      
\acro{PDF}{probability density function}
\acrodef{p.d.f.}[PDF]{probability distribution function}
\acro{PDP}{power dispersion profile}
\acro{PMF}{probability mass function}                             
\acrodef{p.m.f.}[PMF]{probability mass function}
\acro{PN}{pseudo-noise}
\acro{PPM}{pulse position modulation}
\acro{PRake}{Partial Rake}
\acro{PSD}{power spectral density}
\acro{PSEP}{pairwise synchronization error probability}
\acro{PSK}{phase shift keying}
\acro{PD}{Power Density}
\acro{8-PSK}[$8$-PSK]{$8$-phase shift keying}

\acro{FSK}{frequency shift keying}

\acro{QAM}{Quadrature Amplitude Modulation}
\acro{QPSK}{quadrature phase shift keying}
\acro{OQPSK/PM}{OQPSK with phase modulator }

\acro{RD}[RD]{raw data}
\acro{RDL}{"random data limit"}
\acro{ric}[RIC]{restricted isometry constant}
\acro{rict}[RICt]{restricted isometry constant threshold}
\acro{rip}[RIP]{restricted isometry property}
\acro{ROC}{receiver operating characteristic}
\acro{rq}[RQ]{Raleigh quotient}
\acro{RS}[RS]{Reed-Solomon}
\acro{RSC}[RSSC]{RS based source coding}
\acro{RFP}{Radio Frequency ``Pollution''}
\acro{r.v.}{random variable}                               
\acro{R.V.}{random vector}
\acro{RMS}{root mean square}
\acro{RFR}{radiofrequency radiation}
\acro{RIS}{Reconfigurable Intelligent Surface}
\acro{RNA}{RiboNucleic Acid}
\acro{RSSI}{Received Signal Strength Indicator}
\acro{RSRP}{Reference Signal Received Power}
\acro{RSRQ}{Reference Signal Received Quality}

\acro{SCBSES}[SCBSES]{Source Compression Based Syndrome Encoding Scheme}
\acro{SCM}{sample covariance matrix}
\acro{SEP}{symbol error probability}
\acro{SG}[SG]{sparse-land Gaussian model}
\acro{SIMO}{single-input multiple-output}
\acro{SINR}{Signal-to-Interference plus Noise Ratio}
\acro{SIR}{signal-to-interference ratio}
\acro{SISO}{single-input single-output}
\acro{SMV}{single measurement vector}
\acro{SNR}[\textrm{SNR}]{signal-to-noise ratio} 
\acro{sp}[SP]{subspace pursuit}
\acro{SS}{spread spectrum}
\acro{SW}{sync word}
\acro{SAR}{Specific Absorption Rate}
\acro{SSB}{synchronization signal block}
\acro{SA}{StandAlone}

\acro{TH}{time-hopping}
\acro{ToA}{time-of-arrival}
\acro{TR}{transmitted-reference}
\acro{TW}{Tracy-Widom}
\acro{TWDT}{TW Distribution Tail}
\acro{TCM}{trellis coded modulation}
\acro{TDD}{time-division duplexing}
\acro{TDMA}{Time Division Multiple Access}

\acro{UAV}{unmanned aerial vehicle}
\acro{uric}[URIC]{upper restricted isometry constant}
\acro{urict}[URICt]{upper restricted isometry constant threshold}
\acro{UWB}{ultrawide band}
\acro{UWBcap}[UWB]{Ultrawide band}   
\acro{URLLC}{Ultra Reliable Low Latency Communications}
         
\acro{wuric}[UWRIC]{upper weak restricted isometry constant}
\acro{wurict}[UWRICt]{UWRIC threshold}                
\acro{UE}{user equipment}
\acro{UL}{uplink}
\acro{UB}{Upper Bound}
\acro{UTM}{Universal Transverse Mercator}

\acro{WiM}[WiM]{weigh-in-motion}
\acro{WLAN}{wireless local area network}
\acro{wm}[WM]{Wishart matrix}                               
\acroplural{wm}[WM]{Wishart matrices}
\acro{WMAN}{wireless metropolitan area network}
\acro{WPAN}{wireless personal area network}
\acro{wric}[WRIC]{weak restricted isometry constant}
\acro{wrict}[WRICt]{weak restricted isometry constant thresholds}
\acro{wrip}[WRIP]{weak restricted isometry property}
\acro{WSN}{wireless sensor network}                        
\acro{WSS}{wide-sense stationary}
\acro{WHO}{World Health Organization}
\acro{Wi-Fi}{wireless fidelity}

\acro{sss}[SpaSoSEnc]{sparse source syndrome encoding}

\acro{VLC}{visible light communication}
\acro{VPN}{virtual private network} 
\acro{RF}{Radio-Frequency}
\acro{FSO}{free space optics}
\acro{IoST}{Internet of space things}

\acro{GSM}{Global System for Mobile Communications}
\acro{2G}{second-generation cellular network}
\acro{3G}{third-generation cellular network}
\acro{4G}{fourth-generation cellular network}
\acro{5G}{5th-generation cellular network}	
\acro{gNB}{next-generation Node-B}
\acro{NR}{New Radio}
\acro{UMTS}{Universal Mobile Telecommunications Service}
\acro{LTE}{Long Term Evolution}
\acro{QoS}{Quality of Service}
\end{acronym}

\def\BibTeX{{\rm B\kern-.05em{\sc i\kern-.025em b}\kern-.08em
    T\kern-.1667em\lower.7ex\hbox{E}\kern-.125emX}}

\def\BibTeX{{\rm B\kern-.05em{\sc i\kern-.025em b}\kern-.08em
    T\kern-.1667em\lower.7ex\hbox{E}\kern-.125emX}}

\def\BibTeX{{\rm B\kern-.05em{\sc i\kern-.025em b}\kern-.08em T\kern-.1667em\lower.7ex\hbox{E}\kern-.125emX}}
\markboth{OJCOMS-01060-2022}
{OJCOMS-01060-2022}



\title{How Much Exposure from 5G Towers is Radiated over Children, Teenagers, Schools and Hospitals?}

\author{Luca Chiaraviglio$^{(1,2)}$, \IEEEmembership{Senior Member, IEEE}, Chiara Lodovisi,$^{(1,2)}$ Daniele Franci,$^{(3)}$ Enrico Grillo,$^{(3)}$\\ Settimio Pavoncello,$^{(3)}$ Tommaso Aureli,$^{(3)}$ Nicola Blefari-Melazzi,$^{(1,2)}$ Mohamed-Slim Alouini$^{(4)}$, \IEEEmembership{Fellow, IEEE}

\thanks{This work was supported by the PLAN-EMF project (KAUST Award No. OSR-2020-CRG9-4377).}
\thanks{L. Chiaraviglio, C. Lodovisi and N. Blefari-Melazzi are with the Department of Electronic Engineering, University of Rome Tor Vergata, Rome, Italy, email \{luca.chiaraviglio,chiara.lodovisi,blefari\}@uniroma2.it and with Consorzio Nazionale Interuniversitario per le Telecomunicazioni, Italy.}
\thanks{D. Franci, E. Grillo, S. Pavoncello, and T. Aureli are with Agenzia per la Protezione Ambientale del Lazio (ARPA Lazio), Rome, Italy, \{daniele.franci,enrico.grillo,settimio.pavoncello,tommaso.aureli\}@arpalazio.it}
\thanks{M.-S. Alouini is with the Computer, Electrical, and Mathematical Science and Engineering (CEMSE) Division, King Abdullah University of Science and Technology (KAUST), Thuwal, Makkah Province, Saudi Arabia  email slim.alouini@kaust.edu.sa}}

\IEEEtitleabstractindextext{\begin{abstract}
The rolling-out of 5G antennas over the territory is a fundamental step to provide 5G connectivity. However, little efforts have been done so far on the exposure assessment from 5G cellular towers over young people and ``sensitive'' buildings, like schools and medical centers.  {To face such issues, we provide} a sound methodology for the numerical evaluation of 5G (and pre-5G) {downlink} exposure over children, teenagers, schools and medical centers.  {We then apply the proposed methodology over two real scenarios}. {Results} reveal that the exposure from 5G cellular towers will increase in the forthcoming years, in parallel with the {growth of the 5G} adoption levels. However, the exposure levels are well below the maximum ones defined by international regulations. Moreover, the exposure over children and teenagers is similar to the one of the whole population, while the exposure over schools and medical centers can be lower than the one of the whole set of buildings. Finally, the exposure from 5G is strongly lower {than} the pre-5G one {when the building attenuation is introduced and a maturity adoption level for 5G is assumed.}
\end{abstract}

\begin{IEEEkeywords}
5G networks, EMF analysis, population analysis
\end{IEEEkeywords}
}


\maketitle

\section{Introduction}
\label{sec:intro}


The deployment of cellular towers generally triggers a mixture of positive and negative reactions among the population \cite{9518367}. For example, different municipalities across the world authorize newly installed \acp{BS} only when such installations are enough far from schools and hospitals \cite{gsmaexcl}.  {Clearly, innovative 5G services like Industry 4.0 and smart healthcare can be provided only if new 5G \acp{BS} are pervasively installed over the territory. However,} generalized bans were promoted by several Italian municipalities to deny the installation of 5G sites \cite{gerli2021municipal}, which required the central government to emanate new laws against such bans. {In addition,} the minimization of exposure from cellular towers in social, recreation and medical places is pursued by different national regulations (see e.g., Italy \cite{italian2003limits} and Greece \cite{kapetanakis2022assessment} regulations), on the basis of a supposed ``precautionary'' principle. 

Although there is no a scientific evidence supporting such restrictions, the exposure from 5G towers over children, teenagers, schools and hospitals is a matter of debate at both municipality and community levels. As reported by \ac{GSMA} in a comprehensive report \cite{gsmadebate}, ensuring a dialogue with the population and the local authorities is {a} fundamental {task} for the deployment of \acp{BS} close to schools and medical centers. Obviously, the expert in the exposure assessment field may object that the amount of time spent in schools and hospitals is low compared to the other daily activities. However, the topic is often dominated by irrational arguments, which give extra importance to the exposure levels that are observed in such ``sensitive'' environments.    

In this context, the assessment of 5G exposure over children and teenagers is rapidly gaining attention \cite{9518367}, as such categories will (likely) receive a continuous exposure from 5G from the start until the end of 5G adoption. For example, a recent competitive Call of \ac{EU} Commission \cite{eucall} has identified the exposure assessment over children as a key activity to be pursued for the financed projects. In addition, the assessment of \ac{BS} exposure over young people is a key aspect for current (and future) epidemiological studies \cite{karipidis20215g}, aimed at studying possible (but still not proven at present time) correlations between the levels of 5G whole-body exposure and the emergence of long-term health diseases.  

The goal of this paper is to fill the gap between the concerns that are associated with exposure from 5G towers over children, teenagers, schools and medical centers and the technical evidence of such exposure levels. More concretely, we tackle the following questions: What is the exposure that young people and ``sensitive'' areas receive from 5G cellular towers? How such exposure levels will evolve in the future - in parallel with an increase of utilization of the 5G network? To answer such questions, we design an innovative geospatial-based methodology, which integrates: children and teenagers data taken from national census, real positioning of the buildings (including schools and hospitals), real positioning of 5G (and pre-5G) \acp{BS} over the territory, realistic radio configurations (including mm-Wave antennas), and a conservative computation of exposure, tailored to the technical features implemented by 5G antennas.

{Our innovative contributions include: \textit{i}) the definition of an innovative end-to-end framework, able to characterize the exposure over children, teenagers, schools and medical centers, by considering the realistic deployment of base stations (and their configuration) over the territory, as well as real population data and real building positioning, \textit{ii}) an exposure computation tailored to 5G antennas - including mm-Wave base stations, \textit{iii}) the investigation of the impact of the 5G utilization levels on the exposure from base stations over the population and over the buildings.}

Results, obtained over two realistic case studies, reveal that the exposure from 5G networks will increase in the forthcoming years. However, this outcome should not be interpreted as an alarm, but rather as a natural consequence of the growth in the number of users connected through 5G. In fact, the predicted exposure levels are always far below the maximum \ac{EMF} {whole-body} limits defined by international guidelines. More importantly, the exposure over schools and hospitals is always comparable to the one observed in the other building types. Eventually, the \ac{EMF} over children and teenagers is greatly affected by the building attenuation, which has a deeper effect on mm-Wave frequencies compared to lower ones.
 
We believe that our outcomes may be useful for a variety of innovative actions. First, we open a communication channel with the municipalities and community involved in our study, by clearly showing at what extent the deployment of 5G network and its level of utilization will impact the predicted exposure levels. Second, we provide a technical tool to the research community, which could be easily adopted (and eventually refined) for the assessment of 5G exposure in other relevant areas. Third, we stimulate a societal engagement approach, in which the computation of exposure levels is massively evaluated over the territory, and then translated into indicators (exposure over young people, exposure over schools and medical centers) that can be easily interpreted by the population.

The rest of the paper is organized as follows. Sec.~\ref{sec:rel_works} overviews the related works. Sec.~\ref{sec:methodology} presents our methodology for the exposure assessment from 5G towers. The adopted hardware/software tools and the realistic scenarios under investigation are reported in Sec.~\ref{sec:scenarios}. Sec.~\ref{sec:results} presents the results, which are obtained over the selected scenarios. {Sec.~\ref{sec:discussion} includes a discussion of our approach.} Finally, Sec.~\ref{sec:conclusions} concludes our work.

\section{Related Works}
\label{sec:rel_works}

We provide coverage about the related works falling in the following categories: \textit{i}) \ac{BS} exposure assessment over children and teenagers, \textit{ii}) \ac{BS} exposure assessment over schools and medical centers and \textit{iii}) geospatial methodologies for \ac{BS} \ac{EMF} assessments. We intentionally leave apart the works focusing on exposure from other wireless devices (like mobile terminals \cite{morelli2021numerical}, small cells and indoor access points \cite{bonato2020human}), whose \ac{EMF} is additive with respect to the \ac{BS} exposure considered in this work.

\subsection{BS exposure over children and teenagers}

We initially cover the works that evaluate \acp{BS} exposure over children and/or teenagers \cite{calvente2015outdoor,schoeni2016symptoms,birks2018spatial,schmutz2022personal}.
More concretely, Calvente \textit{et al.} \cite{calvente2015outdoor} measure the environmental exposure (including pre-5G \ac{BS} sources) close to 123 dwellings hosting families, finding levels well below the limits defined in international guidelines. However, they recognize the importance of characterizing incident \ac{EMF} levels in children. Schoeni \textit{et al.} \cite{schoeni2016symptoms} perform an \ac{EMF} assessment from fixed transmitters (e.g. pre-5G \acp{BS} and radio/TV towers) over 439 teenagers by asking participants to complete questionnaires, followed by a post-analysis based on a numerical evaluation of \ac{EMF}. No consistent association between the self-reported symptoms in questionnaires and the exposure from the transmitters is found. Personal pre-5G exposure over children is collected through exposimeters and thoroughly analyzed by Birks \textit{et al.} \cite{birks2018spatial}. The authors performed a wide study over 529 children located in different countries, finding that the largest contribution of children exposure is the \ac{BS} downlink. Schumtz \textit{et al.} \cite{schmutz2022personal} record personal \ac{EMF} measurements from 148 teenagers in the United Kingdom, by adopting exposimeters covering up to 3.5~[GHz]. Their goal is to assess the impact of restriction rules on the usage of the smartphone on the collected exposure levels.  Interestingly, the authors conclude that restrictions on the smartphone usage do not imply a lower exposure levels for the teenagers. 

Overall, works  \cite{calvente2015outdoor,schoeni2016symptoms,birks2018spatial,schmutz2022personal} reveal the importance and significance of studies tailored to the investigation of \ac{BS} exposure over children and teenagers. In constrast to them, in this work we go three steps further by: \textit{i}) focusing on the evaluation of 5G mid-band and mm-Wave \ac{BS} exposure, \textit{ii}) designing an innovative methodology for the assessment of children/teenagers exposure, tacking into account the technical features of 5G antennas (like \ac{MIMO} and beamforming), \textit{iii}) applying the proposed methodology to areas inhabitated by thousands of children and teenagers, thus widening the scope of the considered analysis.

\subsection{BS exposure over schools and medical centers}

We then move our attention on the works tailored to the evaluation of \ac{BS} exposure over schools and/or medical centers \cite{gallastegi2018children,bhatt2017radiofrequency,kiouvrekis2020statistical,kurnaz2020exposure,ramirez2020measurements,kapetanakis2022assessment}.
More in depth, Gallastegi \textit{et al.} \cite{gallastegi2018children} perform a thorough assessment of pre-5G exposure over places in which children spend most of their time (including schools), by performing spot measurements. Results show that \ac{BS} downlink is among the main sources of exposure. Bhatt \textit{et al.} \cite{bhatt2017radiofrequency} evaluate pre-5G exposure in kinder-gardens, by considering a set of 20 buildings located in Australia. The authors apply a methodology based on an exposure assessment through exposimeters, finding that the largest amount of exposure is due to pre-5G \acp{BS}. Kiouvrekis \textit{et al.} \cite{kiouvrekis2020statistical} analyze the \ac{EMF} exposure levels in a set of Greek schools located in urban environments, by considering pre-5G sources operating up to 3~[GHz] of frequency. In all cases, the measured exposure levels are always lower than the international limits.

\begin{figure*}[t]
\centering
\includegraphics[width=17cm]{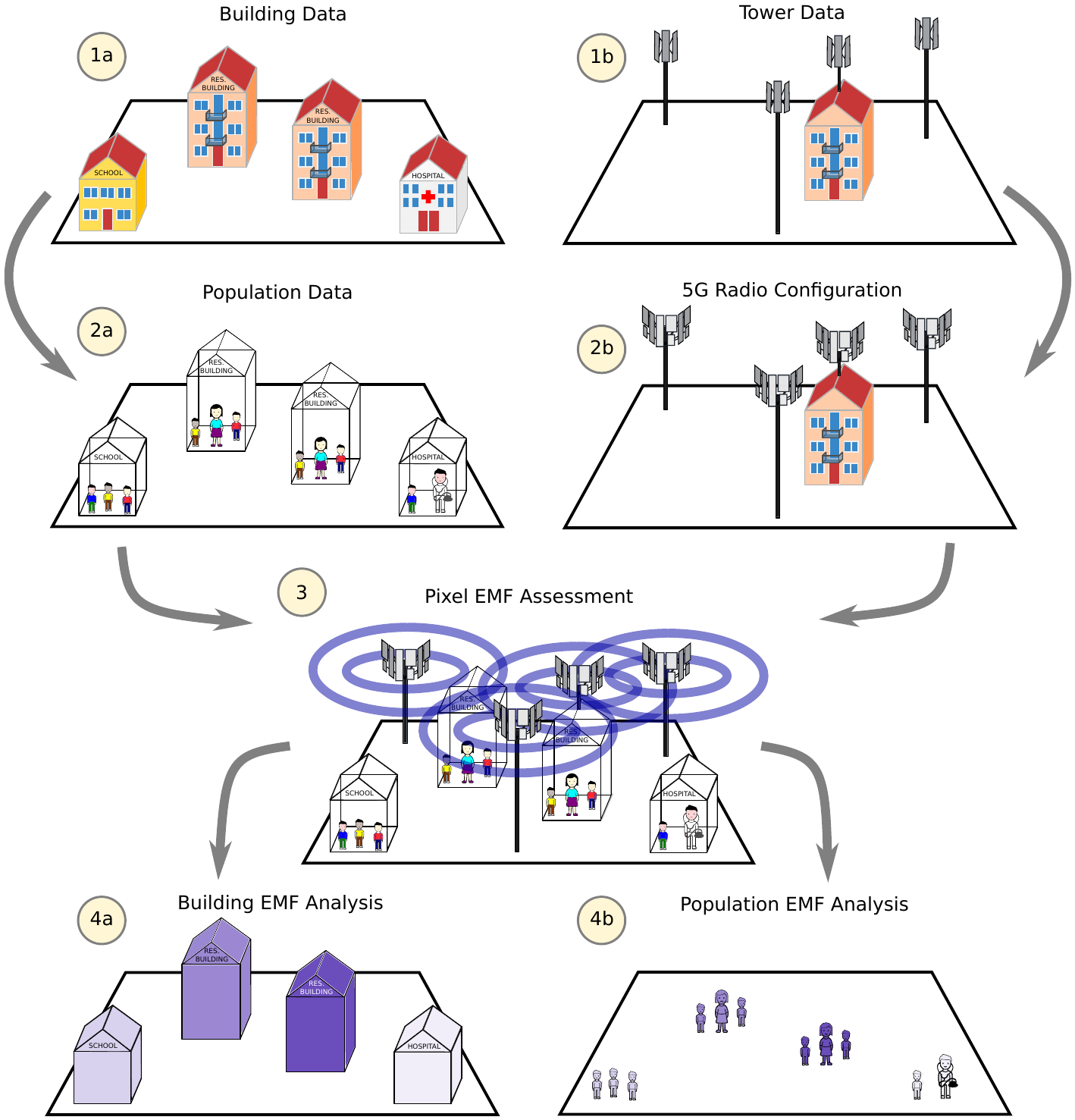}
\caption{Proposed methodology for the building and population exposure assessment.}
\label{fig:general_scheme}
\end{figure*}

Eventually, Kurnaz \textit{et al.} \cite{kurnaz2020exposure} target exposure assessments in hospitals, by performing \ac{EMF} measurements in a set of medical centers located in Turkey. The considered frequencies include all pre-5G sources operating up to 3~[GHz] of frequency. In line with previous works, pre-5G \acp{BS} are the main contribution to the total \ac{EMF}. However, the measured \ac{EMF} is largely below the maximum limits defined in international guidelines. Pre-5G exposure in schools is measured by Ramirez-Vazquez \textit{et al.} \cite{ramirez2020measurements}. \ac{EMF} levels well below the maximum exposure limits are observed. Kapetanakis \textit{et al.} \cite{kapetanakis2022assessment} estimate \ac{EMF} levels in proximity to kinder-gardens and schools, by considering both urban and suburban environments. The authors perform a set of broad-band and narrow-band measurements, by considering pre-5G frequencies up to 3~[GHz]. Results reveal that the highest levels of exposure are concentrated over mobile and broadcasting frequencies. Interestingly, different exposure patterns tend to emerge among urban and suburban locations. 

Summarizing, \cite{gallastegi2018children,bhatt2017radiofrequency,kiouvrekis2020statistical,kurnaz2020exposure,ramirez2020measurements,kapetanakis2022assessment} indicate that the measured pre-5G exposure over schools and hospitals is highly influenced by pre-5G \acp{BS}. Although we recognize the importance of the aforementioned measurement-based approaches, in this work we design a new methodology for the numerical assessment of exposure over schools and medical centers. In this way, we are able to analyze the predicted exposure levels over the years, in parallel with the utilization level of the 5G network. Moreover, most of previous works are focused on pre-5G sources \cite{gallastegi2018children,bhatt2017radiofrequency,kiouvrekis2020statistical,kurnaz2020exposure,ramirez2020measurements,kapetanakis2022assessment}. In contrast to them, in this work we evaluate the exposure from 5G sources, including mm-Wave antennas, which are currently being installed in different countries in the world. However, since \cite{gallastegi2018children,bhatt2017radiofrequency,kiouvrekis2020statistical,kurnaz2020exposure,ramirez2020measurements,kapetanakis2022assessment} indicate that the amount of exposure from pre-5G sources is not neglibile, we also integrate in our evaluations the exposure from legacy 2G/4G technologies, which will operate in parallel to 5G for many years to come.



\subsection{Geospatial Methodologies for BS Exposure Assessments}

Finally, we consider the literature employing geospatial methodologies for the \ac{BS} exposure assessment over the territory \cite{beekhuizen2013geospatial,beekhuizen2014modelling,beekhuizen2015input,martens2015validity,guxens2016outdoor}. The main idea shared by such works is to exploit a tool, called NISMap, able to numerically compute the exposure on the buildings of the considered portions of territory. In more detail, Beekhuizen \textit{et al.} \cite{beekhuizen2013geospatial} evaluate the \ac{EMF} levels from pre-5G \acp{BS} over five outdoor areas in Netherlands, characterized by different building features (e.g., low-rise and high-rise). Beekhuizen \textit{et al.} \cite{beekhuizen2014modelling} extend the outcomes of  \cite{beekhuizen2013geospatial} by also considering indoor locations (such as schools). Beekhuizen \textit{et al.} \cite{beekhuizen2015input} conclude that the geospatial modelling of \ac{EMF} exposure is a fundamental tool for ranking exposure levels in epidemiological studies. However, the authors recognize the importance of precisely settings in their simulations the \ac{BS} configuration parameters, which include antenna height above ground, adopted frequency, antenna location and antenna orientation. Martens \textit{et al.} \cite{martens2015validity} apply geospatial methodologies to estimate the exposure in a set of homes located in the Netherlands. Interestingly, authors conclude that a meaningful ranking of personal \ac{EMF} exposure can be obtained. Eventually, Guxens \textit{et al.} \cite{guxens2016outdoor} apply a geospatial model to predict the exposure from pre-5G \acp{BS} over a set of children located in Amsterdam. 

Overall, \cite{beekhuizen2013geospatial,beekhuizen2014modelling,beekhuizen2015input,martens2015validity,guxens2016outdoor} indicate that exposure from pre-5G \acp{BS} can be estimated through geospatial-based methodologies. In line with them, we design a geospatial-based methodology for the numerical evaluation of exposure over the territory (and hence in cascade over children, teenagers, schools and medical buildings). However, differently from \cite{beekhuizen2013geospatial,beekhuizen2014modelling,beekhuizen2015input,martens2015validity,guxens2016outdoor}, our model is tailored to 5G sources. In particular, one of the key innovations brought by 5G antennas is the high directionality of the 5G signals, which implies that the statistical variation of both antenna radiation diagrams and output power play a great role in determining the exposure from a given \ac{BS}. This aspect is explicitly considered in our work (while obviously neglected by \cite{beekhuizen2013geospatial,beekhuizen2014modelling,beekhuizen2015input,martens2015validity,guxens2016outdoor}, which are focused on pre-5G deployments). In addition, we include in the exposure evaluation key 5G parameters that capture the impact of MIMO and dynamic beamforming, which are largely employed by 5G antennas (particularly those ones operating over over mm-Wave).

\section{Methodology}
\label{sec:methodology}

\subsection{Overview and Rationale}

Fig.~\ref{fig:general_scheme} sketches the methodology pursued in this work. Our final goal is to assess the \ac{EMF} exposure level over the buildings (including schools and medical centers) and the population (including children and teenagers). To achieve such goal, we proceed as follows. We first obtain the 3D model of the buldings in the considered scenario (step \textsc{1a} in the figure). We then retrieve the number of children, teenagers and adults in each building (step \textsc{2a}). In parallel to steps \textsc{1a}-\textsc{2a}, we obtain the real positioning of the cellular towers currently installed in the scenario under investigation (step \textsc{1b}). Obviously, not all the towers are currently fully supporting 5G service. To overcome this issue, we design and apply a suitable 5G radio configuration (including mm-Wave antennas) for the considered towers (step \textsc{2b}). Given steps \textsc{1b} and \textsc{2b}, we then numerically evaluate the \ac{EMF} exposure for the pixels of territory in the considered scenario  (step \textsc{3}). Given this information, we finally analyze building exposure (step \textsc{4a}) and population exposure (step \textsc{4b}).

In the following subsections, we provide more details about steps \textsc{1a}-\textsc{4b}.  {The whole notation used throughout the section is also reported in Appendix~A.}

\subsection{Buildings Modeling}

\begin{figure}[t]
\centering
\includegraphics[width=6cm]{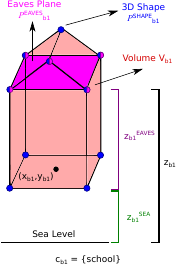}
\caption{Visualization of the required building data for a simple case of one school building.}
\label{fig:building_scheme}
\end{figure}

\begin{table}[t]
    \caption{Building Input Data.}
    \label{tab:buildings_information}
    \scriptsize
    \centering
    \begin{tabular}{|m{2cm}|m{2cm}|c|}
\hline
\rowcolor{Linen} \textbf{Metric Name} & \textbf{Format} & \textbf{Notation}/\textbf{Equation} \\
\hline
Altitude above sea level & Float [m] & $z^{\text{SEA}}_b$\\
\hline
Positioning (center) & Float[m] (global UTM coordinates) & $x_b$, $y_b$\\
\hline
Eaves height & Float [m] (above ground) &  $z^{\text{EAVES}}_b$\\
\hline
Eaves global height & Float [m] (above sea level) & $z_b=h^{\text{SEA}}_b+z^{\text{EAVES}}_b$\\
\hline
3D shape & Array of float (points in local coordinates) & $\mathcal{P}^{\text{SHAPE}}_b$ \\
\hline
Eaves plane & Array of float (subset of points from 3D shape) & $\mathcal{P}^{\text{EAVES}}_b$ \\
\hline
Volume & Float [m]$^3$ & $V_b$\\
\hline
Type & Category $\{$school, med. center, other$\}$ & $C_b$ \\
\hline
\end{tabular}
\end{table}

In the first step, the building set $\mathcal{B}$ is retrieved from relevant geo-spatial databases storing the building information for the considered scenario. The required information is reported in Tab.~\ref{tab:buildings_information} and sketched in Fig.~\ref{fig:building_scheme}. The features that are required for each building $b \in \mathcal{B}$ include: \textit{i}) altitude above sea level $h^{\text{SEA}}_b$ and height of the bulding eaves above ground  $h^{\text{EAVES}}_b$, \textit{ii}) coordinates $x_b$, $y_b$ of the building positioning, \textit{iii}) 3D shape $\mathcal{P}^{\text{SHAPE}}_b$ (in terms of array of points describing the volume of the building), \textit{iv}) eaves plane $\mathcal{P}^{\text{EAVES}}_b$ (subset of points from $\mathcal{P}^{\text{SHAPE}}_b$ denoting the eaves level), \textit{v}) volume $V_b$ (which can be computed out from $\mathcal{P}^{\text{SHAPE}}_b$), and type $C_b$ (i.e., school, medical center, other).

Up to this point, a natural question is then: Why do we consider the level of the eaves {as height reference}? The answer is directly connected to the \ac{EMF} evaluation, which, in our case, will be based on a conservative approach: we assess the exposure from the \acp{BS} in the highest accessible zone of the building by the population, which is assumed to be at the eaves height. Therefore, the eaves information is required for each building.

\subsection{Population Modeling}

\begin{table}[t]
    \caption{Population Input Data.}
    \label{tab:populaton_information}
    \scriptsize
    \centering
    \begin{tabular}{|m{3.1cm}|m{0.8cm}|c|}
\hline
\rowcolor{Linen} \textbf{Metric Name} & \textbf{Format} & \textbf{Notation}/\textbf{Equation} \\
\hline
Number of children/teenagers in building $b$ & Integer & $N^{\text{CHD-TN}}_b$\\
\hline
Number of adults in building $b$ & Integer & $N^{\text{AD}}_b$\\
\hline
\end{tabular}
\end{table}

The following step of our approach requires the collection of population information, as reported in Tab.~\ref{tab:populaton_information}. Normally, the number of adults and the number of children/teenagers are available in government databases with a data granularity of a census zone, i.e., a small portion of territory including multiple buildings. More formally, the set of census zones in the scenario under consideration is denoted as $\mathcal{N}$. Focusing on a generic zone $n \in \mathcal{N}$, let us express the number of adults  and the number of children/teenagers in $n$ as $N^{\text{AD}}_n$ and $N^{\text{CHD-TN}}_n$, respectively. Although census-based metrics are relevant for our analysis, we actually need to work on a higher level of detail, by assessing the number of adults and children/teenagers in each building $b \in \mathcal{B}_n$, where $\mathcal{B}_n$ is the subset of buildings falling inside census zone $n$. This information is in fact instrumental to properly compute the estimated exposure for all the inhabitants (adults, children and teenagers) inside the building, as the exposure strongly depends on the building features (e.g., size, height, etc.). Let us denote with $N^{\text{AD}}_b$ and $N^{\text{CHD-TN}}_b$ the number of adults and the number of children/teenagers in building $b$, respectively.  In order to compute $N^{\text{AD}}_b$ and $N^{\text{CHD-TN}}_b$ out from $N^{\text{AD}}_n$ and $N^{\text{CHD-TN}}_n$, we proceed as follows. First,
we compute the total volume of the buildings belonging to census zone $n$ as: 
\begin{equation}
V^{\text{TOT}}_n = \sum_{b \in \mathcal{B}_n} V_b
\end{equation}
Second, we assume that the number of adults and children/teenagers in the building is proportional to the building volume, i.e., larger buildings generally host more people compared to smaller ones. Third, we compute $N^{\text{AD}}_b$ and $N^{\text{CHD-TN}}_b$ as:
\begin{equation}
\label{eq:n_ad}
N^{\text{AD}}_b = N^{\text{AD}}_n \times \frac{V_b}{V^{\text{TOT}}_n}; \quad \forall b \in \mathcal{B}_n, \forall n \in \mathcal{N}
\end{equation}
\begin{equation}
\label{eq:n_chd_tn}
N^{\text{CHD-TN}}_b = N^{\text{CHD-TN}}_n \times \frac{V_b}{V^{\text{TOT}}_n}; \quad \forall b \in \mathcal{B}_n, \forall n \in \mathcal{N}
\end{equation}


\subsection{Tower Characterization}

\begin{figure}[t]
\centering
\includegraphics[width=5cm]{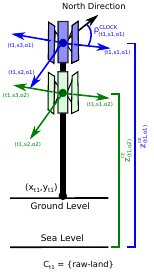}
\caption{Visualization of the required tower data for a simple case with two operators on a raw-land.}
\label{fig:tower_scheme}
\end{figure}

\begin{table}[t]
    \caption{BS Tower Input Data.}
    \label{tab:tower_information}
    \scriptsize
    \centering
    \begin{tabular}{|m{2.5cm}|m{3cm}|c|} 
\hline
\rowcolor{Linen} \textbf{Metric Name} & \textbf{Format} & \textbf{Notation}/\textbf{Equation} \\
\hline
Positioning & Float [m] (global UTM coordinates) & $x_t$, $y_t$\\
\hline
Height (electrical center) of operator $o$ & Float[m] & $z^{\text{CE}}_{(t,o)}$\\
\hline
Orientation of sector $s$ of operator $o$ & Array of integers (degrees) & $\rho^\text{CLOCK}_{(t,s,o)}$ \\
\hline
Type & Category $\{$raw-land, roof-top, fake chimney$\}$ & $C_t$\\
\hline
\end{tabular}
\end{table}

\begin{figure*}[t]
	\centering
 	\subfigure[Raw-land]
	{
		\includegraphics[width=3.65cm]{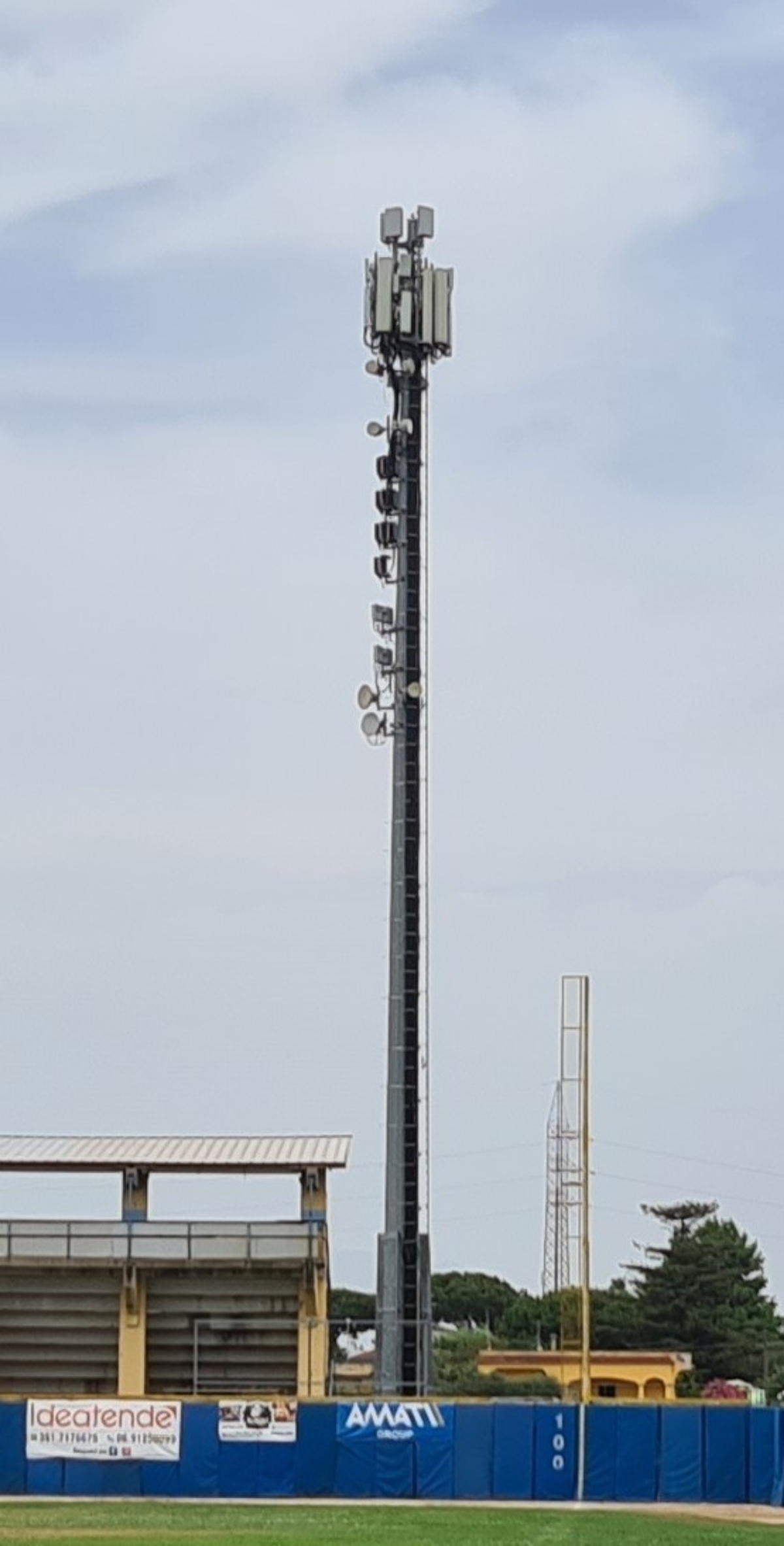}
		\label{fig:raw-land-installation}
	}
 	\subfigure[Roof-top]
	{
		\includegraphics[width=5.5cm]{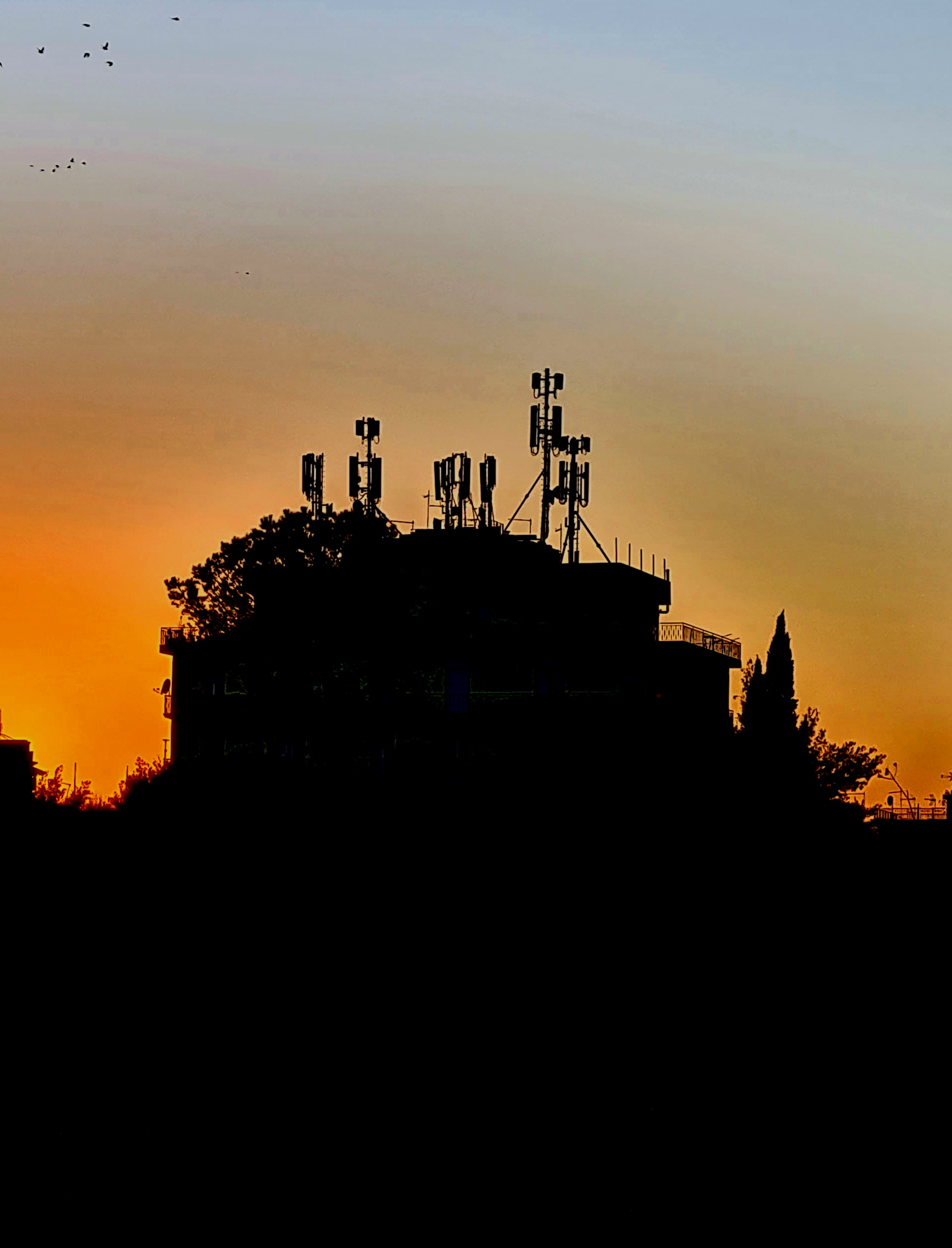}
		\label{fig:roof-top-installation}

	}
 	\subfigure[Fake Chimney]
	{
		\includegraphics[width=6.1cm]{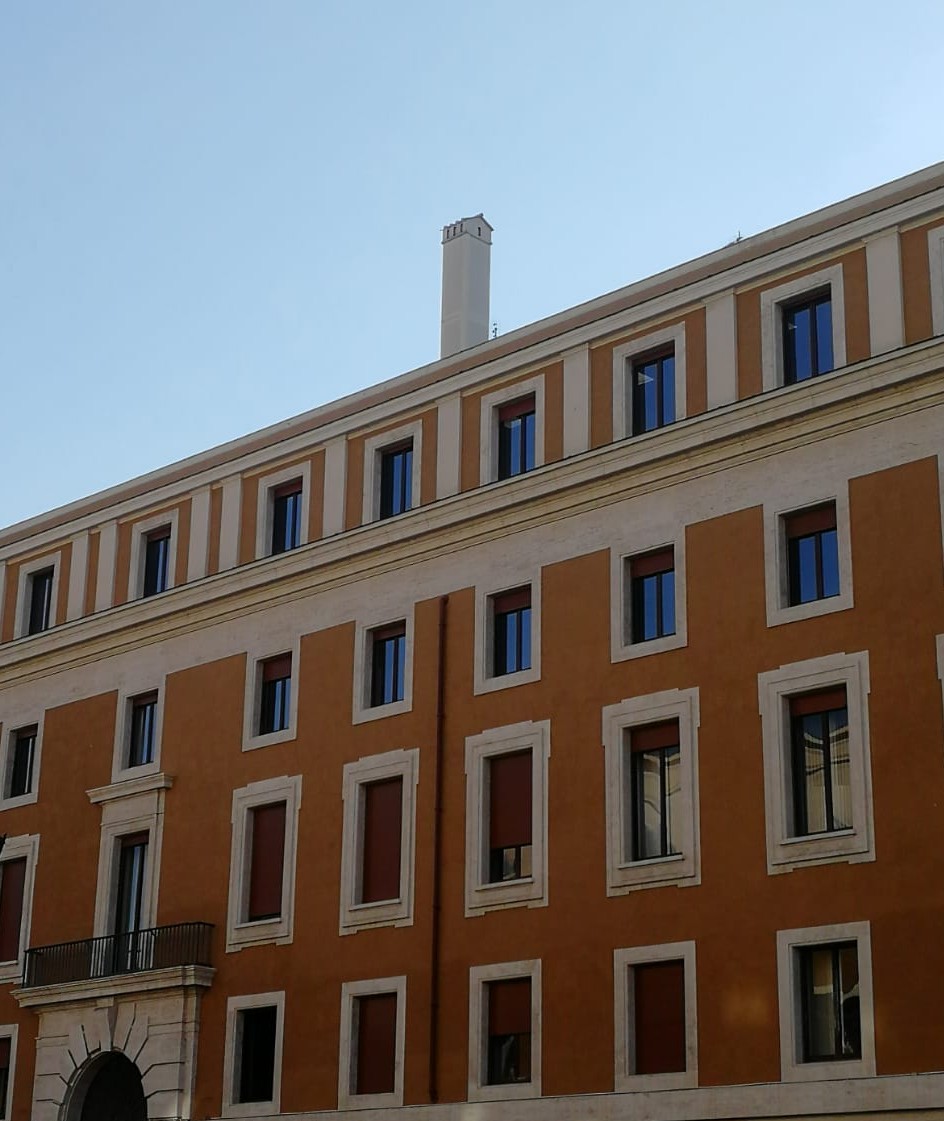}
		\label{fig:fake-chimney-installation}

	}
	\caption{Main types of cellular towers.}
	\label{fig:types-installation}
\end{figure*}

In parallel to the building and population information, we collect the \ac{BS} tower data, whose metrics are sketched in Fig.~\ref{fig:tower_scheme} and detailed in Tab.~\ref{tab:tower_information}. The set of towers in the considered scenario is denoted as $\mathcal{T}$. For each tower $t \in \mathcal{T}$, we require the positioning in terms of global \ac{UTM} coordinates, denoted as $x_t$ and $y_t$. In addition, we need to characterize the sectors for each operator hosted in the tower $t$. Let us denote with $\mathcal{O}$ the set of operators. The matrix $z^{\text{CE}}_{(t,o)}$ stores the altitude above sea level of the electrical center of the antenna panels owned by operator $o$ on tower $t$.\footnote{The electrical center is a reference point normally made available by operators in the authorization requests for installing the panel(s).} The set of sectors in the scenario is then denoted with $\mathcal{S}$. In addition, the orientation $\rho^\text{CLOCK}_{(t,s,o)}$ for each sector $s \in \mathcal{S}$ of each operator $o \in \mathcal{O}$ hosted on tower $t \in \mathcal{T}$ has to be provided, in terms of angle w.r.t. the North in clockwise direction.

Eventually, the last information to be provided is the tower type $C_t$. As shown in Fig.~\ref{fig:types-installation}, the main tower types currently adopted by operators in Italy include: \textit{i}) \acp{BS} placed on self pole, a.k.a. raw-land (Fig.~\ref{fig:raw-land-installation}), \textit{ii}) \acp{BS} installed on poles above buildings, a.k.a. roof-top (Fig.~\ref{fig:roof-top-installation}) and \textit{iii}) \acp{BS} installed on poles above buildings but inside fake chimneys (Fig.~\ref{fig:fake-chimney-installation}). In this work, fake chimney are assimilated to roof-top installations. However, since it is almost impossible to obtain the sector orientation $\rho^\text{CLOCK}_{(t,s,o)}$ of fake chimneys from a visual check (as the antenna panels are hidden at sight), we provide a practical approach to characterize sectorization. We refer the interested reader to Sec.~\ref{sec:tower_data_analysis}, while here we report the main intuition. In brief, we exploit a driving-based approach with real measurements, which allows characterizing the sectorization $\rho^\text{CLOCK}_{(t,s,o)}$ also for those panels installed in fake chimneys.


\subsection{5G Radio Configuration Modeling}

Given the characterization of real towers retrieved in the previous step, in this part we design a suitable 5G radio configuration. The rational of our approach is the following one: we apply the same radio configuration for each sector and each operator, in order to make a uniform analysis of the impact of 5G antennas over the considered scenarios. However, we consider the deployment of panels including also pre-5G signals, as legacy technologies will continue to be active in the forthcoming years, in parallel to the deployment of the 5G networks.

\begin{figure}[t]
	\centering
 	\subfigure[Roof-top sector with two operators.]
	{
		\includegraphics[width=4cm]{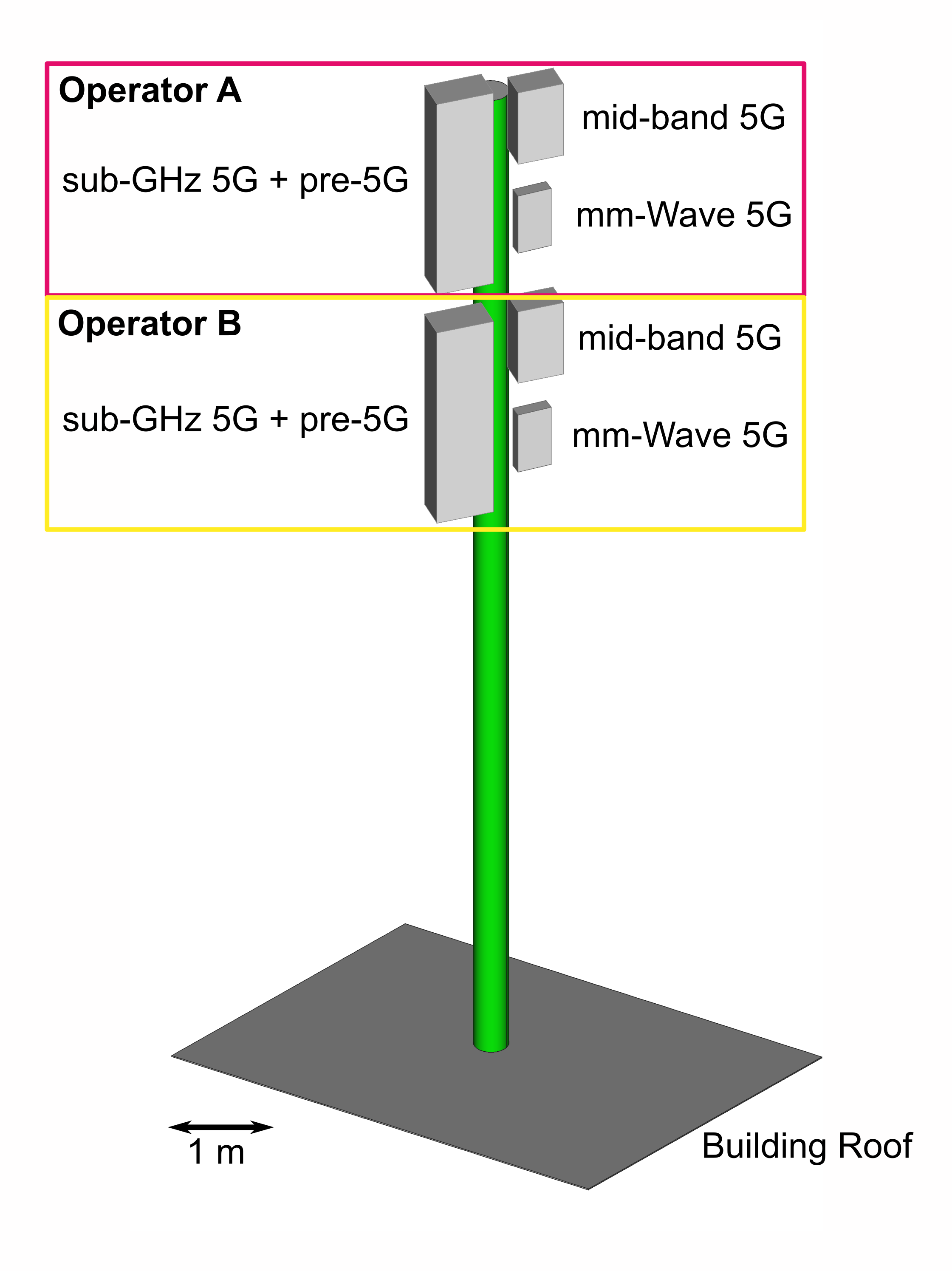}
		\label{fig:roof-top}

	}
 	\subfigure[Raw-land sector with three operators.]
	{
		\includegraphics[width=4cm]{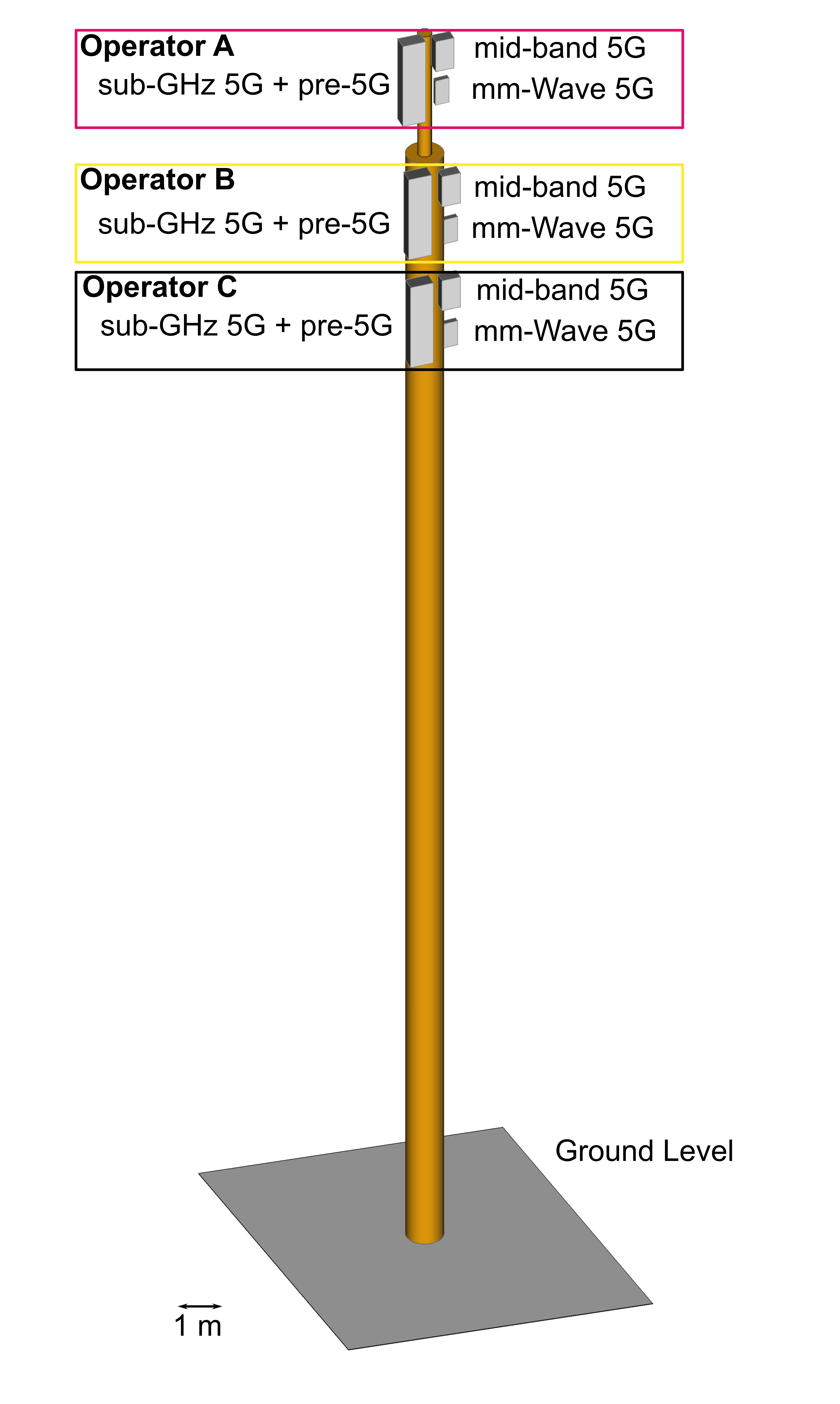}
		\label{fig:raw-land}
	}
	\caption{Antenna panel installations for each operator (example with one sector per operator): roof-top tower (left) vs. raw-land tower (right).}
	\label{fig:raw_land_roof_top_examples}
	\vspace{-4mm}
\end{figure}

Fig.~\ref{fig:raw_land_roof_top_examples} reports an example (in scale) of a realization of a sector for a roof-top with two operators (left) and for a raw-land with three operators (right). More in depth, we adopt the same set of panels across the two deployment types, while, obviously, different heights of the electrical centers are used. For each sector and for each operator, we deploy three distinct panels: a quadri-band panel - able to radiate 2G/4G and sub-GHz 5G signals, a mid-band panel - providing 5G coverage over mid-band frequencies, and a mm-Wave panel - radiating 5G signals over mm-Wave frequencies. The deployed panels match the expected configuration fully realizing the 5G coverage in the forthcoming years.

\begin{table}[t]
    \caption{Panel Configuration Parameters.}
    \label{tab:radio_conf}
    \scriptsize
    \centering
    \begin{tabular}{|m{2.5cm}|m{3cm}|c|} 
\hline
\rowcolor{Linen} \textbf{Metric Name} & \textbf{Format} & \textbf{Notation}/\textbf{Equation} \\
\hline
Antenna frequency & Integer (MHz) & $F_a$ \\
\hline
Antenna technology & Category $\{$ 2G, 4G, 5G $\}$ & $T_a$ \\
\hline
Panel type & Category $\{$ Quadri-band, 5G Mid-band, 5G mm-Wave $\}$ & $C_p$ \\
\hline
Mechanical tilt of panel $p$ from sector $s$ & Integer (degrees) & $\sigma^\text{M}_{(t,s,o,p)}$ \\
\hline
Electrical tilt of antenna $a$ on panel $p$ from sector $s$ & Integer (degrees) & $\sigma^\text{E}_{(t,s,o,p,a)}$ \\
\hline
Maximum output power of antenna $a$ from sector $s$ & Integer (W) &  $P^{\text{MAX}}_{(t,s,o,p,a)}$ \\
\hline
Maximum gain of antenna $a$ from sector $s$ & Float (dBi) & $G^{\text{MAX}}_{(t,s,o,p,a)}$ \\
\hline
Horizontal diagram of antenna $a$ from sector $s$ & MSI file  & $D^{\text{H}}_{(t,s,o,p,a)}$ \\
\hline
Vertical diagram of antenna $a$ from sector $s$ & MSI file  & $D^{\text{V}}_{(t,s,o,,p,a)}$ \\
\hline
& & \\[-0.8em]
Global panel height & $z_{(t,s,o,p)}$ & $z^{\text{CE}}_{(t,o)} + z^{\text{LOCAL}}_{(s,p)}$  \\ 
\hline
Power reduction factor for antenna $a$ on panel $p$ & Float & $R_{(t,s,o,p,a)}$\\
\hline
\end{tabular}
\end{table}

\begin{table*}[t]
    \caption{Received Gain Computation.}
    \label{tab:norm_gain_assessment}
    \scriptsize
    \centering
    \begin{tabular}{|m{0.5cm}|m{3cm}|m{1.5cm}|c|} 
\hline
\rowcolor{Linen} \textbf{Step} & \textbf{Metric} & \textbf{Notation} & \textbf{Equation} \\
\hline
3.1 & Normalized sector orientation & $\rho_{(t,s,o)}$ & $\begin{cases} \frac{\pi}{2} - \rho^{\text{CLOCK}}_{(t,s,o)} \quad \text{if} \quad 0 \leq \rho^{\text{CLOCK}}_{(t,s,o)} \leq \pi/2\\ 
\frac{5}{2}\pi - \rho^{\text{CLOCK}}_{(t,s,o)} \quad \text{if} \quad \pi/2 < \rho^{\text{CLOCK}}_{(t,s,o)} \leq 2 \pi \end{cases}$ \\
\hline
3.2 & Rotation matrix & $\Gamma_{(t,s,o,p)}$ & $\begin{bmatrix} \cos{\rho_{(t,s,o)}}\cos{\sigma^\text{M}_{(t,s,o,p)}} & \sin{\rho_{(t,s,o)}}\cos{\sigma^\text{M}_{(t,s,o,p)}} &  -\sin{\sigma^\text{M}_{(t,s,o,p)}} \\  -\sin{\rho_{(t,s,o)}} &  \cos{\rho_{(t,s,o)}} &  0 \\ \cos{\rho_{(t,s,o)}}\sin{\sigma^\text{M}_{(t,s,o,p)}} &  \sin{\rho_{(t,s,o)}}\sin{\sigma^\text{M}_{(t,s,o,p)}} & \cos{\sigma^\text{M}_{(t,s,o,p)}} \end{bmatrix}$\\
\hline
& & &\\[-0.8em]
3.3 &  Coordinates transformation & $\underbrace{\begin{bmatrix}
\tilde{x}_{(c,t,s,o,p)}\\
\tilde{y}_{(c,t,s,o,p)}\\
\tilde{z}_{(c,t,s,o,p)}
\end{bmatrix}}_{\text{Local coordinates}}$ & $\underbrace{\Gamma_{(t,s,o,p)}}_{\text{Rotation}}
\times
\underbrace{\begin{bmatrix}
x_{c}-x_{t}\\
y_{c}-y_{t}\\
z_{c}-z_{(t,s,o,p)}
\end{bmatrix}}_{\text{Translation}}$ \\ 
\hline
& & & \\[-0.8em]
3.4 & 3D Distance & $r_{(c,t,s,o,p)}$ & $\sqrt{\left(x_{c}-x_{t}\right)^2+\left(y_{c}-y_{t}\right)^2+\left(z_{c}-z_{(t,s,o,p)}\right)^2}$ \\
& & &\\[-0.8em]
\hline
& & & \\[-0.8em]
3.5 & Vertical Plane Angle & $\theta_{(c,t,s,o,p)}$ & $\text{asin}\left(\frac{\tilde{z}_{(c,t,s,o,p)}}{r_{(c,t,s,o,p)}}\right)$ \\
& & & \\[-0.8em]
\hline
& & & \\[-0.8em]
3.6 & Horizontal Plane Angle & $\phi_{(c,t,s,o,p)}$ &  $\text{asin}\left(\frac{\tilde{y}_{(c,t,s,o,p)}}{\sqrt{r^2_{(c,t,s,o,p)}-\tilde{z}^2_{(c,t,s,o,p)}}}\right)\text{sgn}(\tilde{x}_{(c,t,s,o,p)})+ \frac{\pi}{2}\text{sgn}(\tilde{y}_{(c,t,s,o,p)})(1-\text{sgn}(\tilde{x}_{(c,t,s,o,p)}))$\\
& & & \\[-0.8em]
\hline
& & &\\[-0.8em]
3.7 & Pixel Received Gain & $G_{(c,t,s,o,p,a)}$ & $G^{\text{MAX}}_{(t,s,o,p,a)} \cdot D^\text{V}_{(t,s,o,p,a)}(\theta_{(c,t,s,o,p)}) \cdot D^\text{H}_{(t,s,o,p,a)}(\phi_{(c,t,s,o,p)})$\\[0.2em]
\hline
\end{tabular}
\end{table*}

Tab.~\ref{tab:radio_conf} reports the configuration parameters for each panel. Each antenna $a \in \mathcal{A}$ installed on the panel is characterized by a given frequency $F_a$ and a given technology $T_a$ (either 2G, 4G or 5G in our case). Let us denote with $C_p$ the type of panel $p \in \mathcal{P}$, which can be either quadri-band, 5G mid-band, or 5G mm-Wave. The installed panels are then characterized by a mechanical tilting value, which is denoted as $\sigma^\text{M}_{(t,s,o,p)}$. The antennas of the panels are configured in terms of electrical tilting $\sigma^\text{E}_{(t,s,o,p,a)}$, maximum output power $P^{MAX}_{(t,s,o,p,a)}$ and maximum gain $G^{\text{MAX}}_{(t,s,o,p,a)}$, respectively. Moreover, the antenna radiation diagrams are denoted as  $D^{\text{H}}_{(t,s,o,p,a)}$ and $D^{\text{V}}_{(t,s,o,,p,a)}$, respectively in the horizontal and vertical planes. In addition, the panel global height $z_{(t,s,o,p)}$ results from the summation of the height of the electrical center $z^{\text{CE}}_{(t,o)}$ and the relative positioning of the panel w.r.t. the electrical centers, which is denoted as $z^{\text{LOCAL}}_{(s,p)}$. Finally, the last line of Tab.~\ref{tab:radio_conf} reports an antenna reduction factor, denoted as $R_{(t,s,o,p,a)}$ - a key metric capturing the impact of traffic and the statistical variation of radiation for 5G antennas, e.g., as consequence of \ac{MIMO} and beamforming functionalities.

\subsection{Pixel Exposure Computation}
\label{sec:pec_comp}

Given tower, population and building data, the next step is the computation of the exposure over the territory. In this work, we focus on the exposure received at the eaves level of the building, as this height normally represents the maximum one at which people live and/or stay for a long amount of time. Let us assume that the eaves plane of each building is divided into squared pixels. Each pixel is a small portion of the plane (i.e., few meters), characterized by similar exposure conditions w.r.t. the radiating sources. The entire set of pixels (on all buildings), is denoted with $\mathcal{C}$. 

Let us focus on a generic pixel $c \in \mathcal{C}$ that is radiated by antenna $a$ installed on panel $p$, sector $s$, tower $t$ and owned by operator $o$. Let us denote with $(x_c,y_c,z_b)$ the (global) coordinates of the pixel, where we remind that $z_b$ is the eaves global height of the building $b$ hosting the pixel $c$. In addition, the panel coordinates are expressed as $(x_t,y_t,z_{(t,s,o,p)})$, where we remind that $z_{(t,s,o,p)}$ is the global height of panel $p$, installed in sector $s$ and tower $t$, owned by operator $o$.

In line with both national \cite{ceiv1,ceiv2,ceiv3} and international exposure assessment procedures \cite{iecexclusion,iec62669}, the pixel power density $S_{(c,t,s,o,p,a)}$ is formally computed as:\footnote{}
\begin{equation}
\label{eq:power_density}
S_{(c,t,s,o,p,a)}=\frac{P^{\text{MAX}}_{(t,s,o,p,a)} \times G_{(c,t,s,o,p,a)} \times R_{(t,s,o,p,a)}}{4 \pi \times r^2_{(c,t,s,o,p)} \times A_{(c,t,s,o,p,a)}} \ \left[\frac{{W}}{{m}^2}\right]
\end{equation}
where $G_{(c,t,s,o,p,a)}$ is the gain that is received over pixel $c$, $r_{(c,t,s,o,p)}$ is the 3D distance between the radiating source and the pixel, $A_{(c,t,s,o,p,a)}$ is the attenuation of the pixel from the radiating source, while the remaining parameters have been already introduced in Tab.~\ref{tab:radio_conf}.

By observing in more detail Eq.~(\ref{eq:power_density}), we note that the realized gain $G_{(c,t,s,o,p,a)}$, and not the maximum one $G^{\text{MAX}}_{(t,s,o,p,a)}$, is taken under consideration for the exposure assessment. This is a central point of our work, as the gain over the pixel strongly depends on the antenna diagrams (obviously) and the relative positioning of the pixel w.r.t. the radiating source. Therefore, a natural question is then: How to compute $G_{(c,t,s,o,p,a)}$ from $G^{\text{MAX}}_{(t,s,o,p,a)}$? To answer this question, we derive a simple methodology, formally introduced in Tab.~\ref{tab:norm_gain_assessment} through steps 3.1-3.7.

\begin{figure}[t]
	\centering
 	\subfigure[Translation]
	{
		\includegraphics[width=4.7cm]{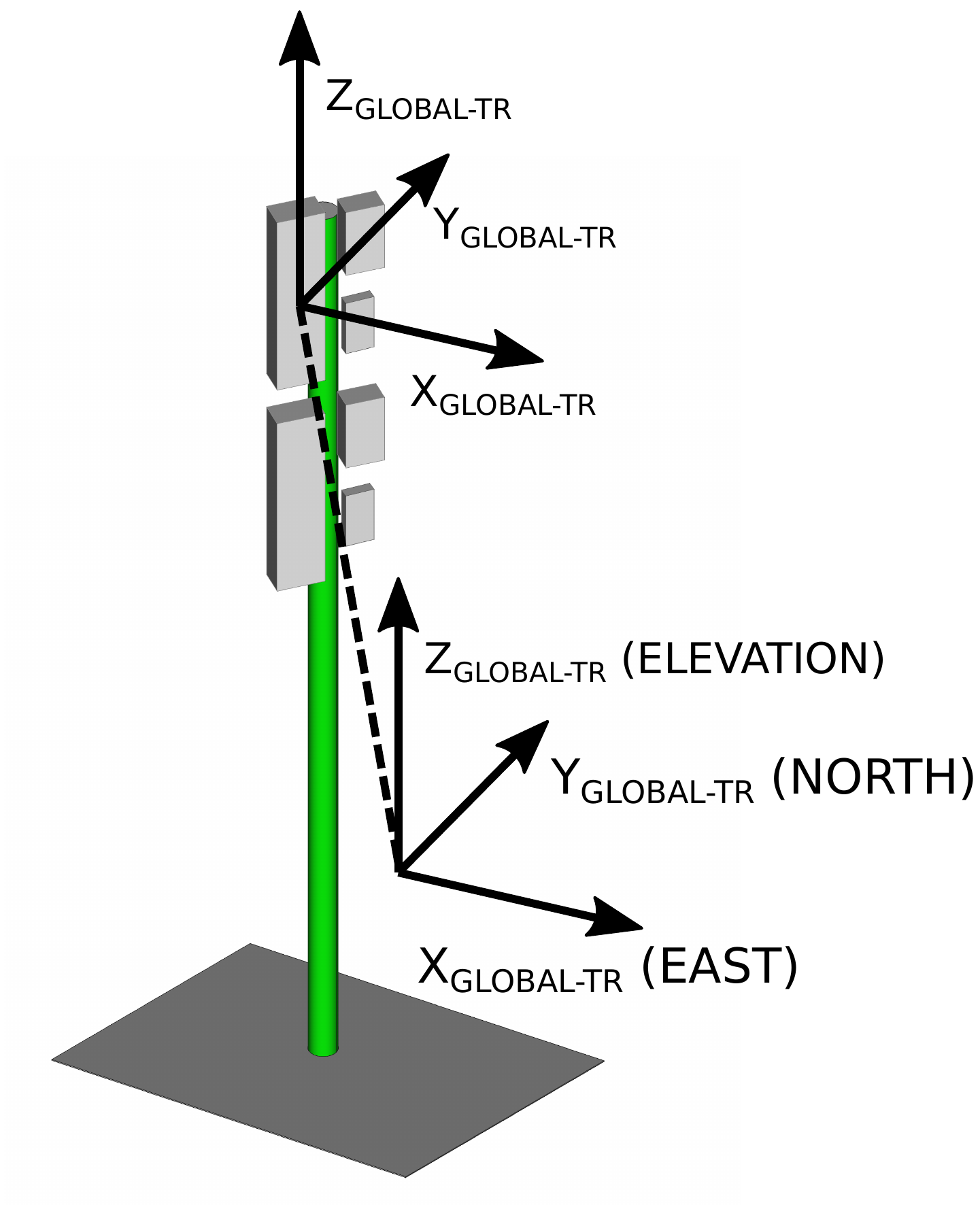}
		\label{fig:trc}

	}
 	\subfigure[Rotation]
	{
		\includegraphics[width=3cm]{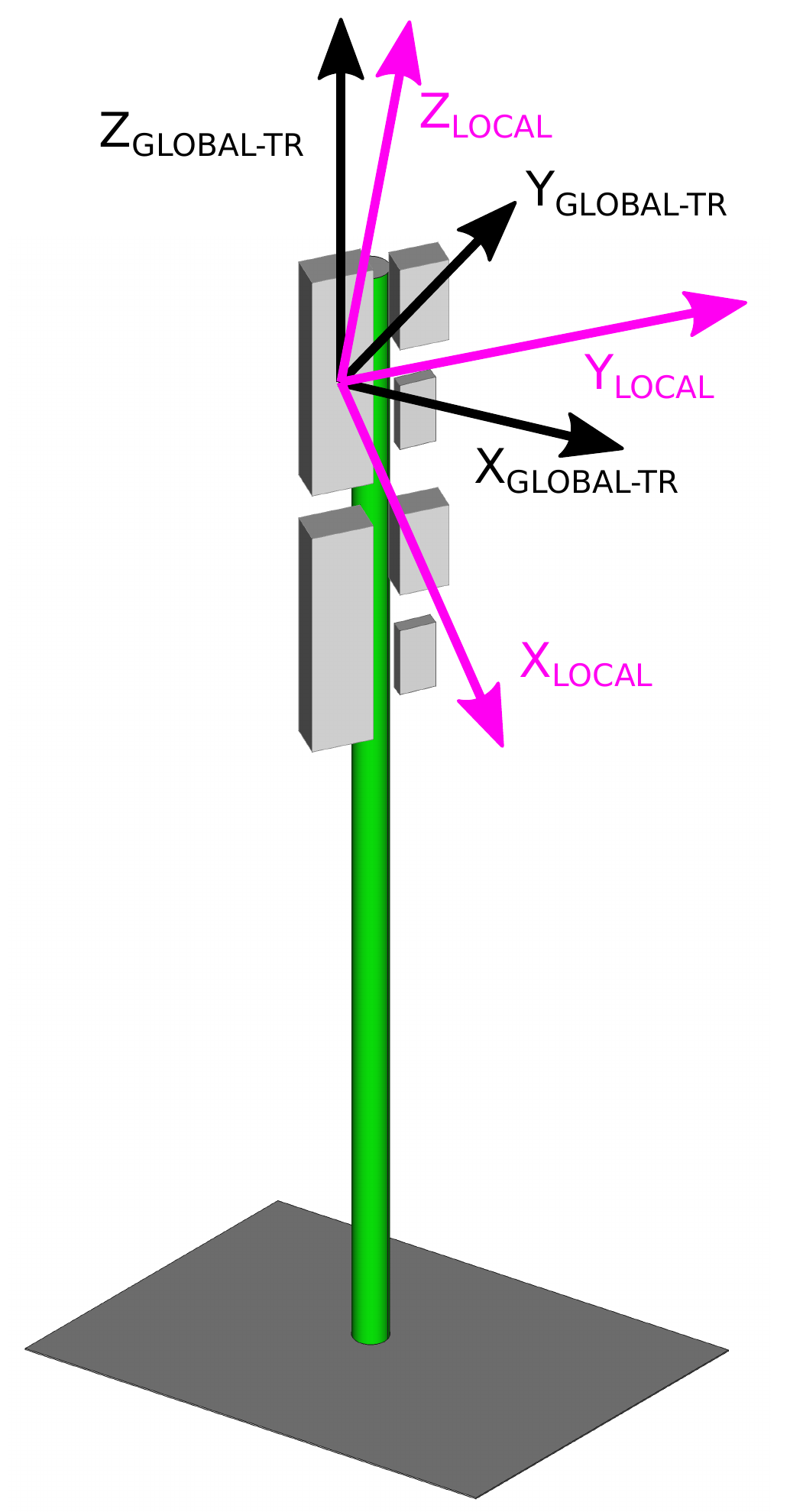}
		\label{fig:rtc}
	}

	\caption{Change from global coordinate system to the local coordinate system of a radiating source.}
	\label{fig:cs_change}
	\vspace{-4mm}
\end{figure}

\begin{figure}[t]
	\centering
 	\subfigure[First Rotation]
	{
		\includegraphics[width=4.1cm]{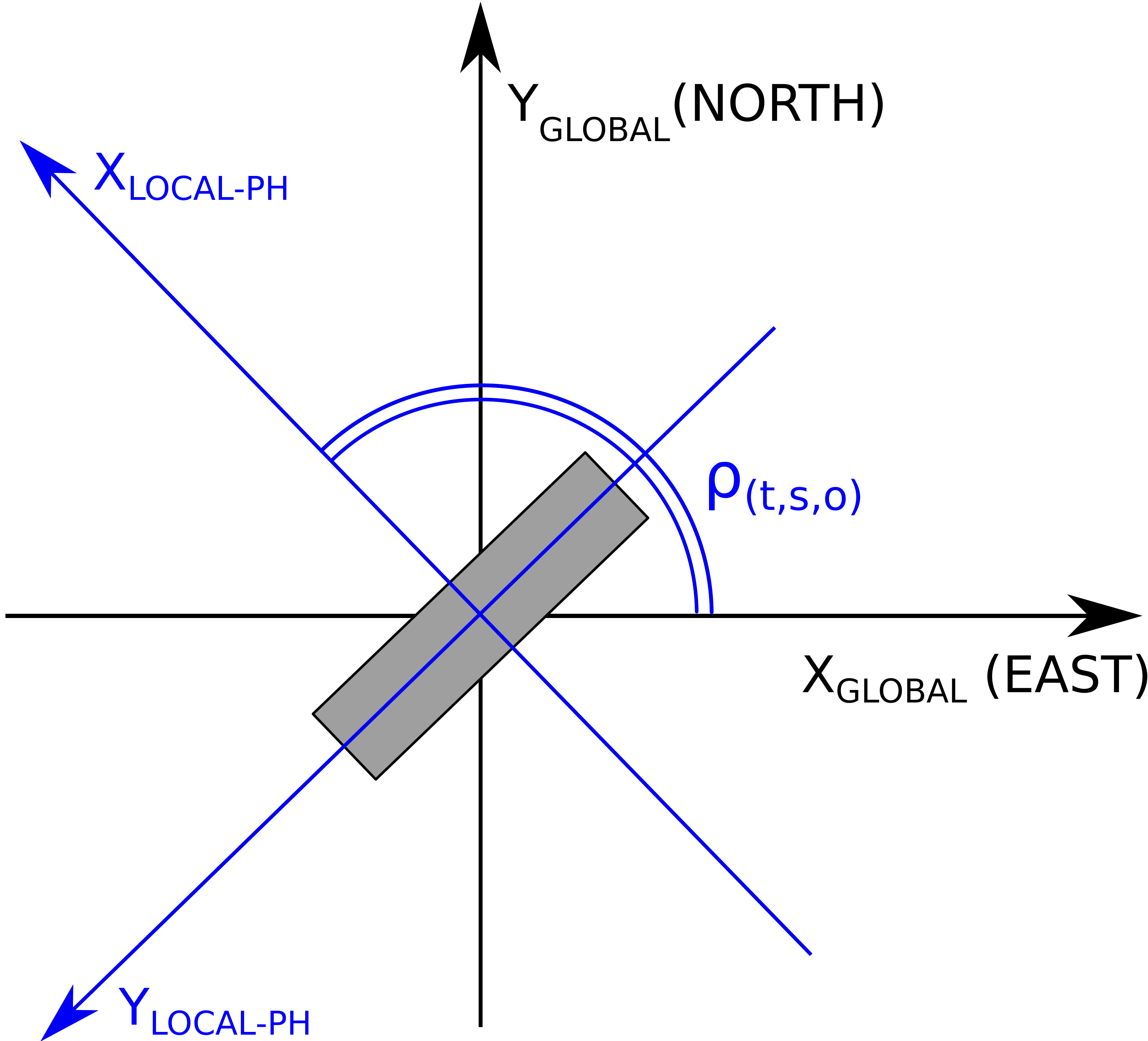}
		\label{fig:fr}

	}
 	\subfigure[Second Rotation]
	{
		\includegraphics[width=4.1cm]{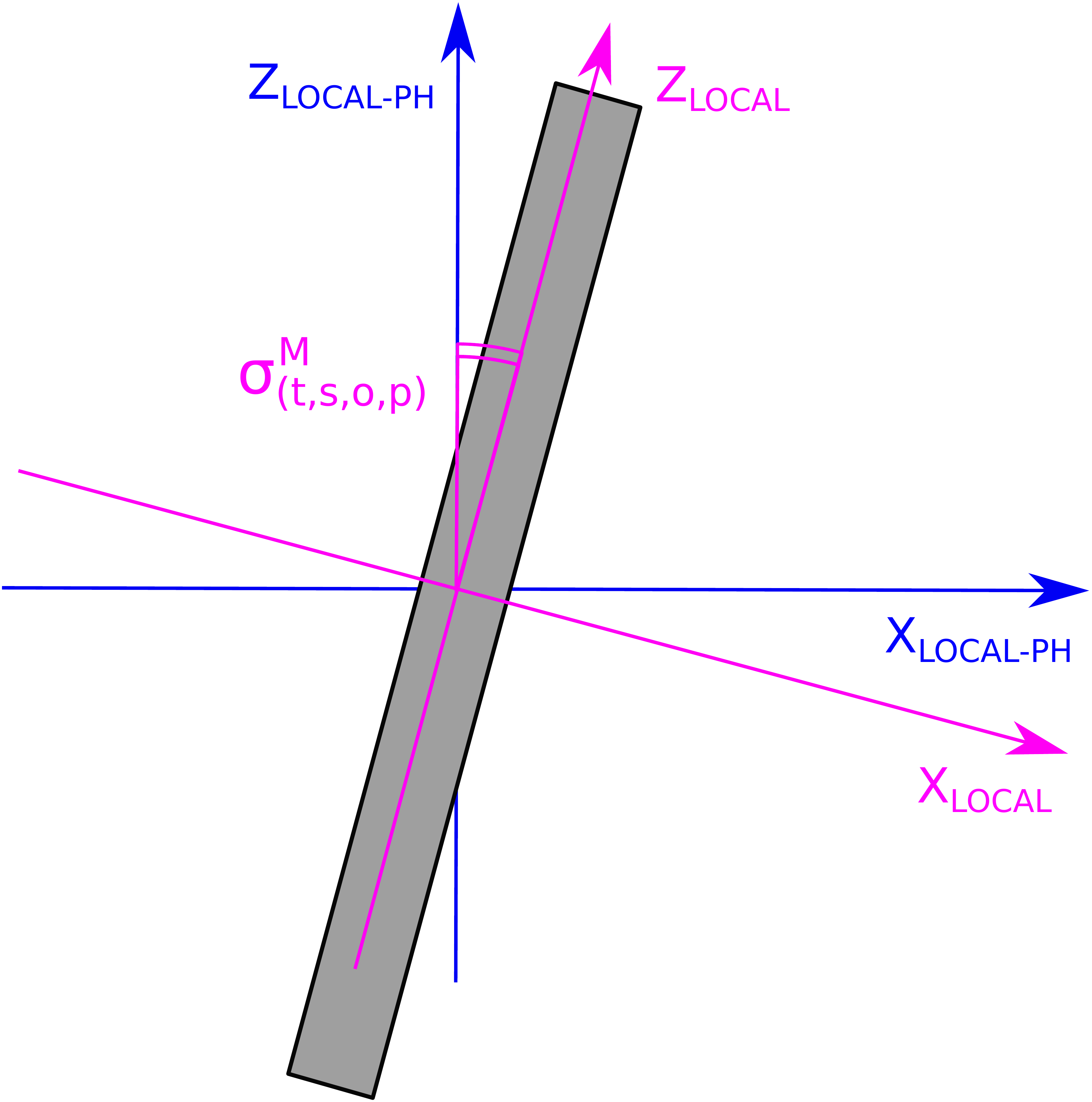}
		\label{fig:sr}
	}

	\caption{Graphical sketch of the rotations applied in $\Gamma_{(t,s,o,p)}$.}
	\label{fig:rotations}
\end{figure}

\begin{figure}[t]
	\centering
 	\subfigure[$\theta_{(c,t,s,o,p)}$ Computation]
	{
		\hspace{-0.9cm}
		\includegraphics[width=4.1cm]{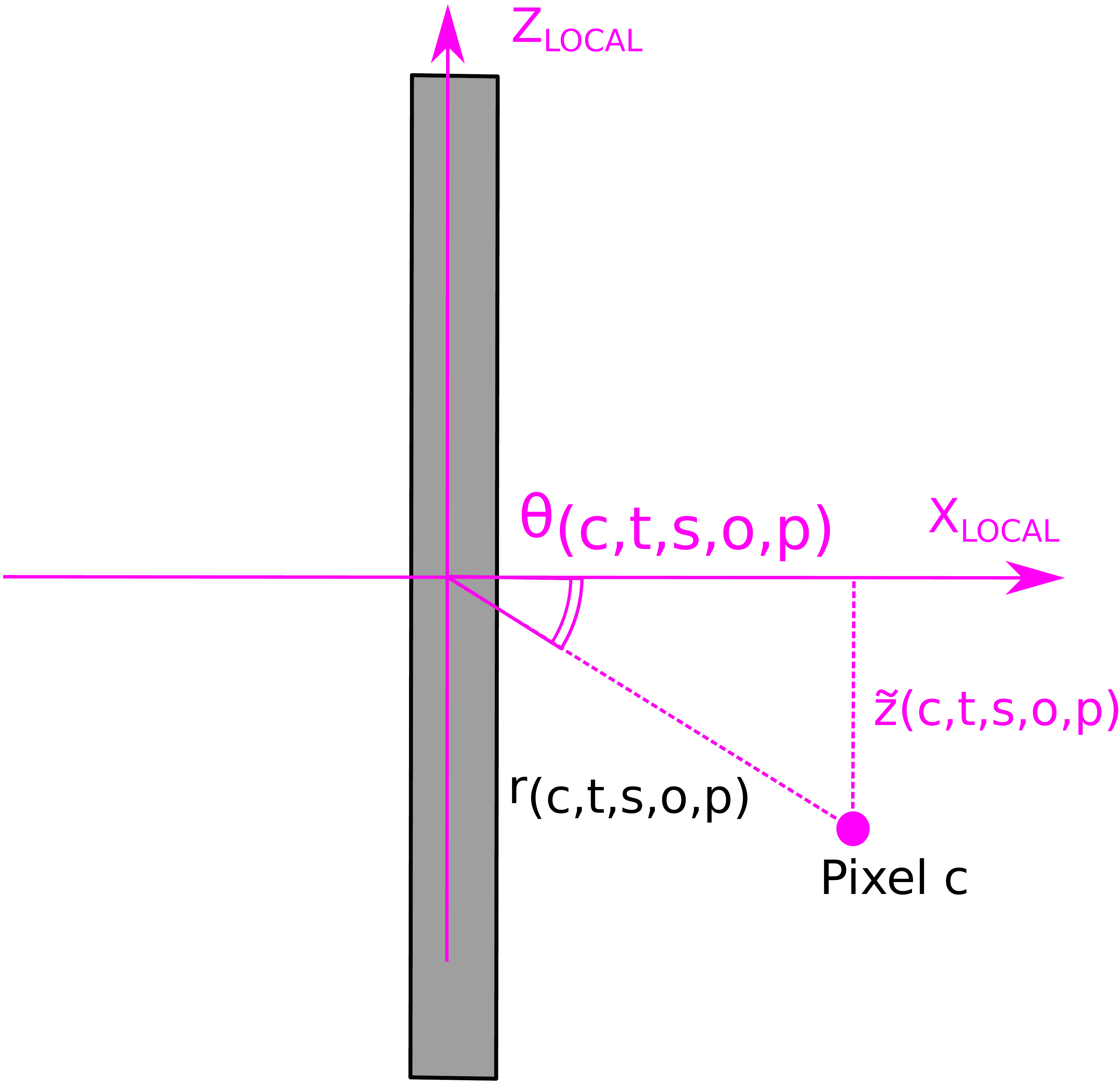}
		\label{fig:theta_comp}

	}
 	\subfigure[$\phi_{(c,t,s,o,p)}$ Computation]
	{
		\includegraphics[width=4.1cm]{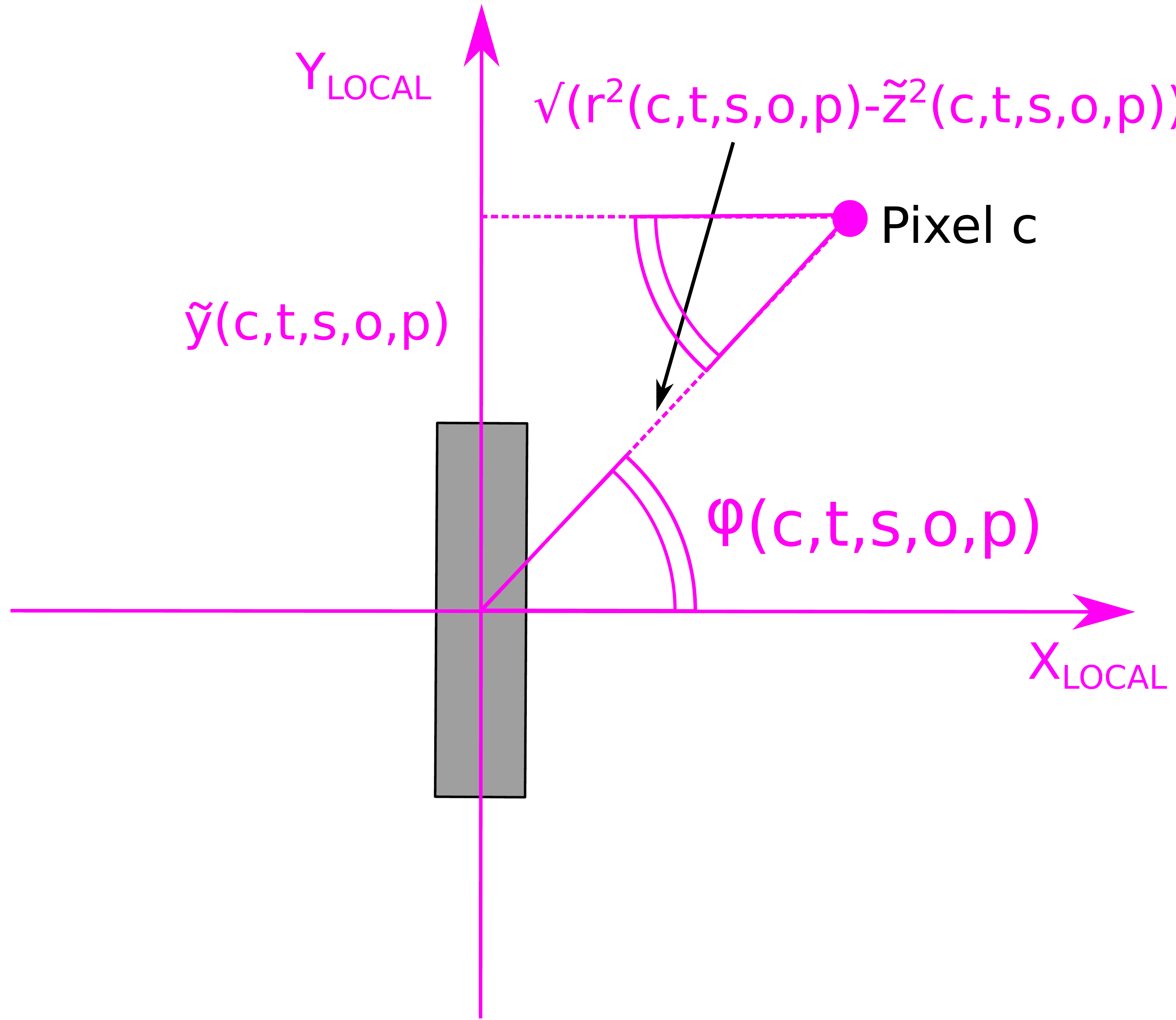}
		\label{fig:phi_comp}
	}

	\caption{Graphical sketch for the computations of the angles $\theta_{(c,t,s,o,p)}$ and $\phi_{(c,t,s,o,p)}$.}
	\label{fig:angle_comp}
	\vspace{-4mm}
\end{figure}

More in depth, we first compute the normalized sector orientation $\rho_{(t,s,o)}$ in counter-wise degrees from East axis (step 3.1). In the following, we apply a transformation of the coordinates from the global system to the local one of the panel (steps 3.2-3-3). As sketched in Fig.~\ref{fig:cs_change}, such operation is composed by a translation (Fig.~\ref{fig:trc}), followed then by a rotation  (Fig.~\ref{fig:rtc}). Focusing on the rotation, Fig.~\ref{fig:rotations} highlights the involved angles. In particular, the first rotation aligns the coordinate system with the sector orientation (Fig.~\ref{fig:fr}), while the second one takes into account the mechanical tilting of the panel (Fig.~\ref{fig:sr}). We then compute the 3D distance $r_{(c,t,s,o,p)}$ between the pixel and the considered source (step 3.4). In the next part, the relative orientation of the pixel with respect to the source is computed. This metric is expressed in terms of horizontal and vertical angles, denoted as $\theta_{(c,t,s,o,p)}$ and $\phi_{(c,t,s,o,p)}$, respectively. The equations to compute $\theta_{(c,t,s,o,p)}$ and $\phi_{(c,t,s,o,p)}$ are reported in steps 3.5-3-6 of Tab.~\ref{tab:norm_gain_assessment}, while Fig.~\ref{fig:angle_comp} shows a visual sketch of the computations. Intuitively,  $\theta_{(c,t,s,o,p)}$ and $\phi_{(c,t,s,o,p)}$ are derived from basic angle properties of right triangles. At last (step 3.7),  $\theta_{(c,t,s,o,p)}$ and $\phi_{(c,t,s,o,p)}$ are used to index the antenna numeric gain diagrams  $D^\text{H}_{(t,s,o,p,a)}$ and $D^\text{V}_{(t,s,o,p,a)}$, which in turn scale the maximum gain by considering the one that is realized over the pixel $c$ under consideration (step 3.7).

Up to this point, we have computed the power density $S_{(c,t,s,o,p,a)}$ that is received over $c$ from a single source (see Eq.~\ref{eq:power_density}). The total power density over the pixel from all the sources is then expressed as:
\begin{equation}
\label{eq:tot_pd_comp}
S^{\text{TOT}}_{c}=\sum_t \sum_s \sum_o \sum_p \sum_a S_{(c,t,s,o,p,a)}
\end{equation}

On the other hand, the total power density only from 5G sources is expressed as:
\begin{equation}
\label{eq:tot_pd_comp_5G}
S^{\text{5G}}_{c}=\sum_t \sum_s \sum_o \sum_p \sum_{a : T_a = \{ \text{5G} \}} S_{(c,t,s,o,p,a)}
\end{equation}

Similarly, the total power density from pre-5G sources is computed as:
\begin{equation}
\label{eq:tot_pd_comp_pre_5G}
S^{\text{PRE-5G}}_{c}=\sum_t \sum_s \sum_o \sum_p \sum_{a : T_a = \{ \text{2G},\text{4G} \}} S_{(c,t,s,o,p,a)}
\end{equation}

Given the power density of Eq.~(\ref{eq:tot_pd_comp})-(\ref{eq:tot_pd_comp_pre_5G}), the electric field computed in far-field conditions is expressed as:
\begin{equation}
\label{eq:electric_field}
E^{(\cdot)}_c=\sqrt{S^{(\cdot)}_c \times 377} \ \left[\frac{{V}}{{m}}\right]
\end{equation}
where $(.)=\{ \text{TOT}, \text{5G}, \text{PRE-5G}\}$





\subsection{Buildings and Population Metrics}
\label{sec:bp_metrics}

In the final part of our methodology, we derive the relevant exposure metrics applied to buildings and population. Focusing on the building assessment, we first compute the average building exposure $E^{(\cdot)}_b$ as the root mean square of the exposure for all the pixels belonging to the eaves plane $\mathcal{P}^{\text{EAVES}}_b$ of building $b$:
\begin{equation}
E^{(\cdot)}_b = \sqrt{\frac{1}{|\mathcal{C}|_b^{\text{EAVES}}} \sum_{c \in \mathcal{C}_b^{\text{EAVES}}} E^{(\cdot)}_c \times E^{(\cdot)}_c}
\end{equation}
where $\mathcal{C}_b^{\text{EAVES}}$ is the subset of pixels belonging to the eaves plane $\mathcal{P}^{\text{EAVES}}_b$ and $(.)=\{ \text{TOT}, \text{5G}, \text{PRE-5G}\}$. 

Given the statistics of exposure for all the buildings $\mathcal{B}$, we compute the linear average building exposure as:
\begin{equation}
E^{(\cdot)}_\text{AVG} = \frac{1}{|\mathcal{B}|}\sum_b E^{(\cdot)}_b 
\end{equation}

Similarly, we compute the average building exposure for schools and medical centers as:
\begin{equation}
E^{(\cdot)}_\text{AVG-SCHOOL} = \frac{1}{N^{\text{SCHOOL}}}\sum_{b : C_b = \{ \text{SCHOOL} \}}  E^{(\cdot)}_b
\end{equation}
\begin{equation}
E^{(\cdot)}_\text{AVG-MED} = \frac{1}{N^{\text{MED-CENTER}}}\sum_{b : C_b = \{ \text{MED-CENTER} \}} E^{(\cdot)}_b
\end{equation}
where $N^{\text{SCHOOL}}$ and $N^{\text{MED-CENTER}}$ are the number of school buildings and the number of medical center buildings in the scenario under consideration.

Focusing on the population analysis, we start from the number of children/teenagers $N_b^{\text{CHD-TN}}$ and the number of adults $N^{\text{AD}}_b$ in each building. We then assume that the building exposure $E^{(\cdot)}_\text{AVG}$ is applied to all the children/teenagers and adults of the building (i.e., a conservative assumption). Let us denote with $\mathcal{I}_b^{\text{CHD-TN}}$ and $\mathcal{I}_b^{\text{AD}}$ the set of children/teenagers and adults in building $b$. The exposure of each individual $i \in \mathcal{I}_b^{\text{CHD-TN}}, \mathcal{I}_b^{\text{AD}}$ is then expressed as:
\begin{equation}
E^{(\cdot)}_i = E^{(\cdot)}_b, \quad \forall i \in \mathcal{I}_b^{\text{CHD-TN}}, \mathcal{I}_b^{\text{AD}} 
\end{equation}

Given such information, we then compute the linear average exposure over the entire population:
\begin{equation}
E^{(\cdot)}_\text{AVG-POP} = \frac{1}{\sum_b (|\mathcal{I}_b^{\text{AD}}|+|\mathcal{I}_b^{\text{CHD-TN}}|)} \sum_b \sum_{i \in \mathcal{I}_b^{\text{AD}},\mathcal{I}_b^{\text{CHD-TN}}} E^{(\cdot)}_i
\end{equation}

In a similar way, the average exposure over children/teenagers and the average exposure over adults are computed as:
\begin{equation}
E^{(\cdot)}_\text{AVG-AD} = \frac{1}{\sum_b |\mathcal{I}_b^{\text{AD}}|} \sum_b \sum_{i \in \mathcal{I}_b^{\text{AD}}} E^{(\cdot)}_i
\end{equation}
\begin{equation}
E^{(\cdot)}_\text{AVG-CHD-TN} = \frac{1}{\sum_b |\mathcal{I}_b^{\text{CHD-TN}}|} \sum_b \sum_{i \in \mathcal{I}_b^{\text{CHD-TN}}} E^{(\cdot)}_i
\end{equation}


\begin{figure}[t]
	\centering
 	\subfigure[Spinaceto Scenario.]
	{
		\includegraphics[width=8cm]{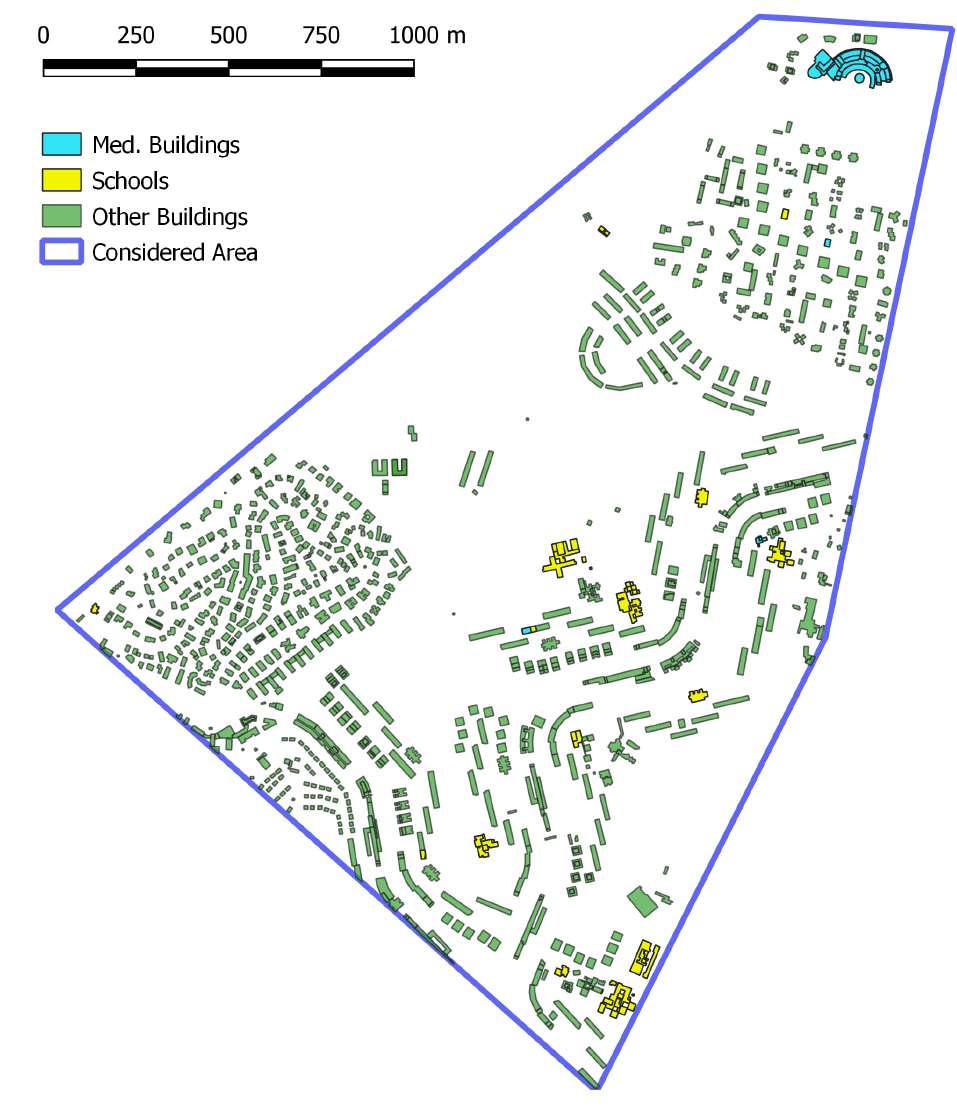}
		\label{fig:buildings_w_considered_area_sp}
	}
 	\subfigure[Ponte-Parione.]
	{
		\includegraphics[width=8cm]{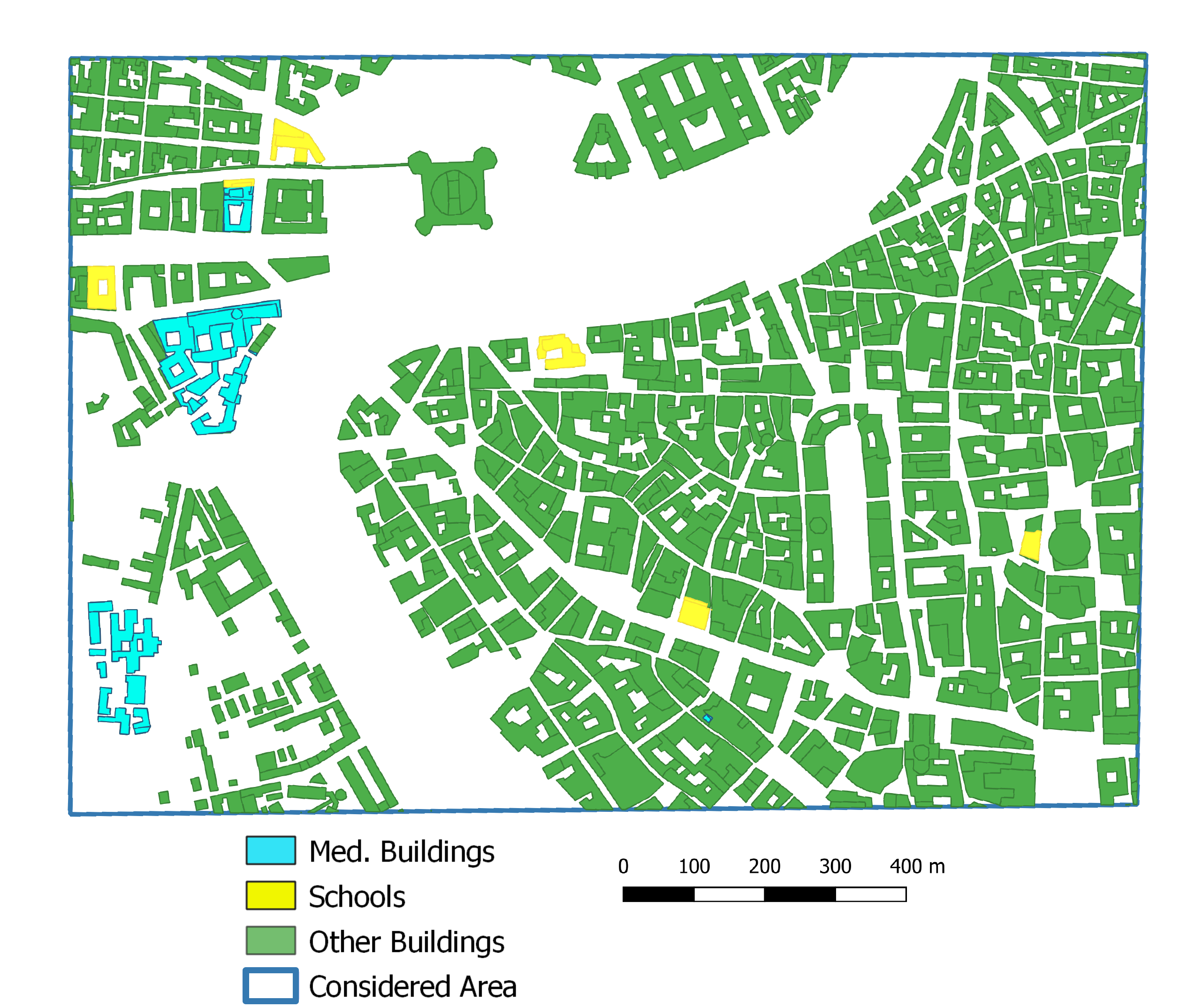}
		\label{fig:buildings_w_considered_area_pp}

	}
	\caption{Building map of Spinaceto and Ponte-Parione scenarios: the eaves planes and the building type are highlighted}.
	\label{fig:areas_buildings}
\end{figure}

\section{Tools and Scenarios}
\label{sec:scenarios}

\subsection{Buildings and Population Data}

\begin{figure}
\centering
\subfigure[Spinaceto]
{
\includegraphics[width=6cm]{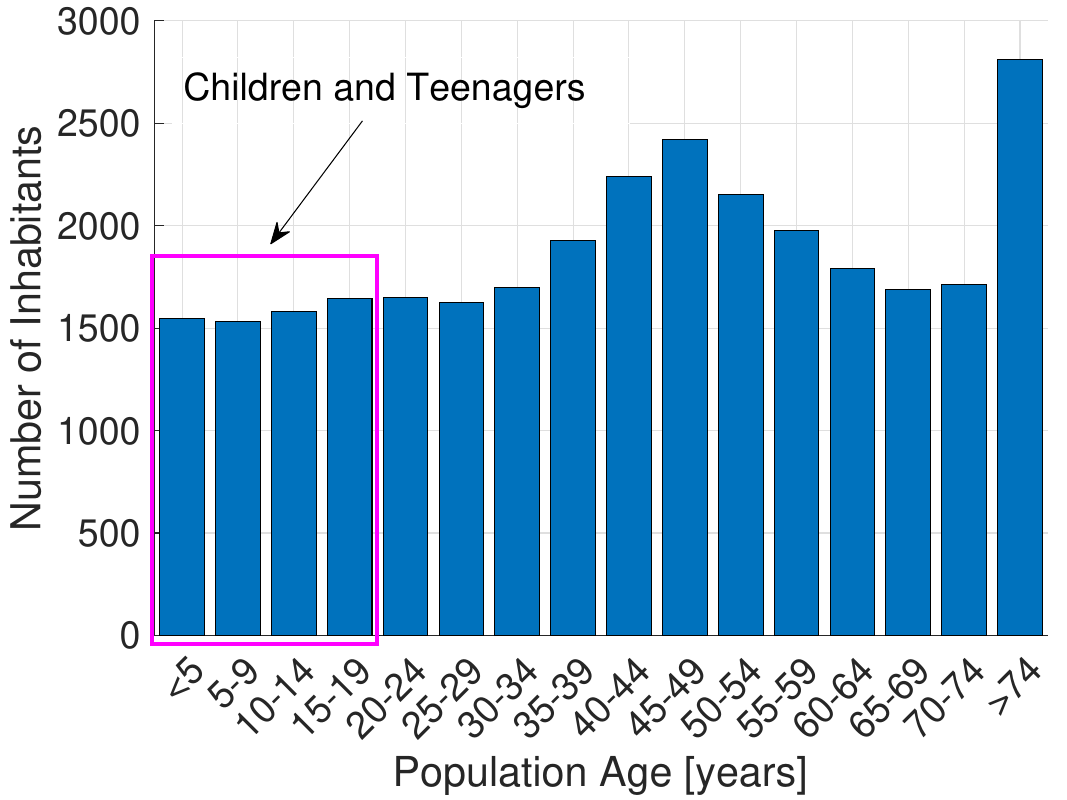}
\label{fig:pop_breakdown_bars_sp}
}
\subfigure[Ponte-Parione]
{
\includegraphics[width=6.5cm]{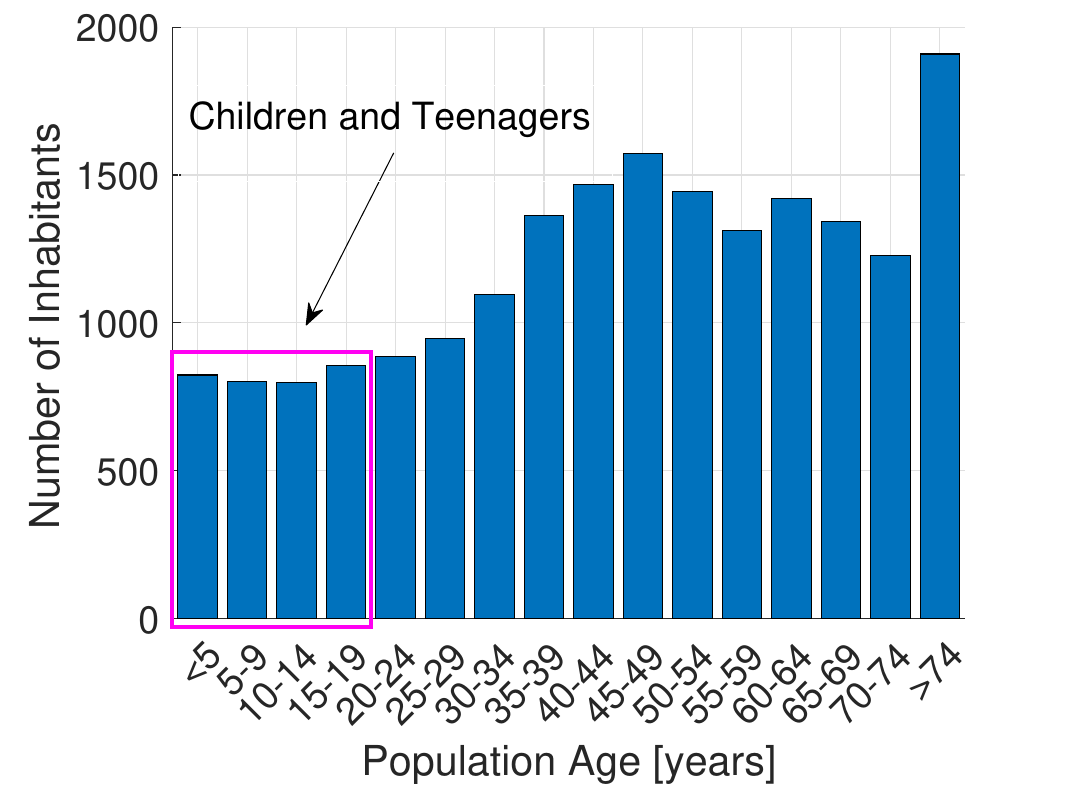}
\label{fig:pop_breakdown_bars_pp}
}
\caption{Breakdown of the population age. The population bins belonging to children and teenagers category are highlighted.}
\label{fig:pop_breakdown_bars}
\end{figure}

We consider two relevant scenarios located in the city of Rome, called Spinaceto and Ponte-Parione. Fig.~\ref{fig:areas_buildings} shows the building map (at the eaves level) and the building type. The rationale for choosing such scenarios is to evaluate the exposure of 5G under different types of building deployment. More in depth, Ponte-Parione (Fig.~\ref{fig:buildings_w_considered_area_pp}) is a dense neighborhood located in the city center, while Spinaceto (Fig.~\ref{fig:buildings_w_considered_area_sp}) is a sparse district located in the outskirts, characterized by a lower and less uniform building density.

In both cases, the building information is derived from the national geoportal database \cite{geoportal}, which provides the building input data of Tab.~\ref{tab:buildings_information}, except from the building type $C_b$ and the building volume $V_b$, which are instead not available. To solve this issue, we proceed in the following way. Focusing on the building type $C_b$, we derive such information from the list of schools and medical buildings available at the involved municipalities. Obviously, when a given school or medical center is composed of different co-located buildings, we mark each of them with the same building type. Focusing instead on the building volume $V_b$, we extract it by loading the 3D shapefile of the buldings in QGis software, and then by running the volume calculation tool.

Focusing on population data, we have exploited the 2011 census done by ISTAT - the Italian national statistics institute - and freely available from \cite{istatvariables}, in order to obtain the number of children/teenagers and adults in the census zones of the scenarios. Although such information is not apparently up-to-date (as more than 10 years passed from the last census), we believe that it is in any case reliable and able to capture the current population conditions for the areas under investigation, due to the following reasons: \textit{i}) national population has not substantially increased in the last 10 years, \textit{ii}) both Spinaceto and Ponte-Parione have been pervasively inhabited much earlier than 10 years ago, \textit{iii}) the urbanization level (in terms of variations in the building set and/or and building category) has not substantially changed over the last 10 years in the considered neighborhoods. 

\begin{table}[t]
    \caption{Buildings and population comparison over the considered scenarios.}
    \label{tab:buildings_pop_metrics}
    \scriptsize
    \centering
    \begin{tabular}{|c|c|c|}
\hline
\rowcolor{Linen}  & \multicolumn{2}{c|}{\textbf{Metric Value}}  \\
\rowcolor{Linen} \multirow{-2}{*}{\textbf{Metric Name}}& \textbf{Spinaceto} & \textbf{Ponte-Parione}  \\
\hline
Size of the area & 3.51~km$^2$ & 1.62~km$^2$  \\
Number of buildings $|\mathcal{B}|$ & 1068 & 755  \\
Building density & 304/km$^2$ & 466/km$^2$\\
Number of schools & 16 & 6   \\
Number of school buildings $N^{\text{SCHOOL}}$ & 58 & 14  \\
Number of medical centers & 4 & 4 \\
Number of medical buildings $N^{\text{MED-CENTER}}$ & 30 & 34 \\ 
Number of inhabitants & 29987 & 19265  \\
Inhabitants density & 8543/km$^2$ & 11892/km$^2$\\
Number of children and teenagers $N_b^{\text{CHD-TN}}$ & 6298 & 3278 \\
Children and teenagers density & 1974/km$^2$ & 2023/km$^2$\\
Number of adults $N_b^{\text{AD}}$ & 23689 & 15987 \\
Adults density & 6749/km$^2$ & 9878/km$^2$ \\
\hline
\end{tabular}
\end{table}

To give more insights, Fig.~\ref{fig:pop_breakdown_bars} reports the number of inhabitants retrieved from the national census over the two scenarios, for different age bins. Two considerations hold by analyzing the figure. First, the distribution of inhabitants over the bins is rather similar over the two scenarios. Second, the number of inhabitants belonging to children and teenagers bins is not marginal, i.e., always more than 1500 for Spinaceto and more than 700 in Ponte-Parione for each bin. 

Given the number of children/teenagers $N_n^{\text{CHD-TN}}$ and the number of adults $N_n^{\text{AD}}$ in each census zone $n$, we then compute the number of children/teenagers $N_b^{\text{CHD-TN}}$ and the number of adults $N_b^{\text{AD}}$ in each census zone $b$ in this way: \textit{i}) we load both buildings and census layers in QGis software, \textit{ii}) we impose in QGis the equations Eq.~(\ref{eq:n_ad})-(\ref{eq:n_chd_tn}), which are automatically evaluated for each building and each census zone.

Finally, Tab.~\ref{tab:buildings_pop_metrics} summarizes the main building and population information collected in the two scenarios. More in depth, the area in Spinaceto is larger than Ponte-Parione, being also characterized by an higher number of buildings. However, the building density in Spinaceto is lower than the one in Ponte-Parione (as expected). Focusing then on the ``sensitive'' buildings, Spinaceto hosts a larger number of schools compared to Ponte-Parione. However, the number of medical centers is comparable over the two scenarios. Clearly, each school and medical center is composed of multiple buildings. Eventually, the number of children/teenagers and inhabitants is higher in Spinaceto compared to Ponte-Parione. However, the opposite holds when considering the population density. Finally, the total number of children and teenagers is rather large, i.e., several thousands in each scenario.

\subsection{Tower Data}
\label{sec:tower_data_analysis}

In this part, we describe the collection of tower data reported in Tab.~\ref{tab:tower_information}. In more detail, we initially extract the tower data from the documents sent by operators to ARPA Lazio in order to get authorization approval for the tower.\footnote{No sensitive information reported in the authorization documents is disclosed in our work.} The collected information for each tower $t$ includes tower type $C_t$, the \ac{UTM} positioning $x_t,y_t$, the installed operators, the sector orientation $\rho^{\text{CLOCK}}_{(t,s,o)}$ and the height of the electrical center $z^{\text{CE}}_{(t,o)}$. In the following step, we perform a cross-check of the obtained data through a driving-based approach. A tower check over the territory is in fact mandatory, due to the following reasons: \textit{i}) the information reported in the tower authorization documents may be outdated (as subsequent tower updates performed after the initial authorization may not generate new authorization requests - this include e.g., tower disposal), \textit{ii}) authorization documents are not available for all the towers (e.g., some towers in Ponte-Parione are placed on buildings belonging to the Vatican state - and those authorization requests are not processed by ARPA Lazio).

\begin{figure}[t]
	\centering
 	\subfigure[Placement of the network scanner in the car]
	{
		\includegraphics[width=3.85cm]{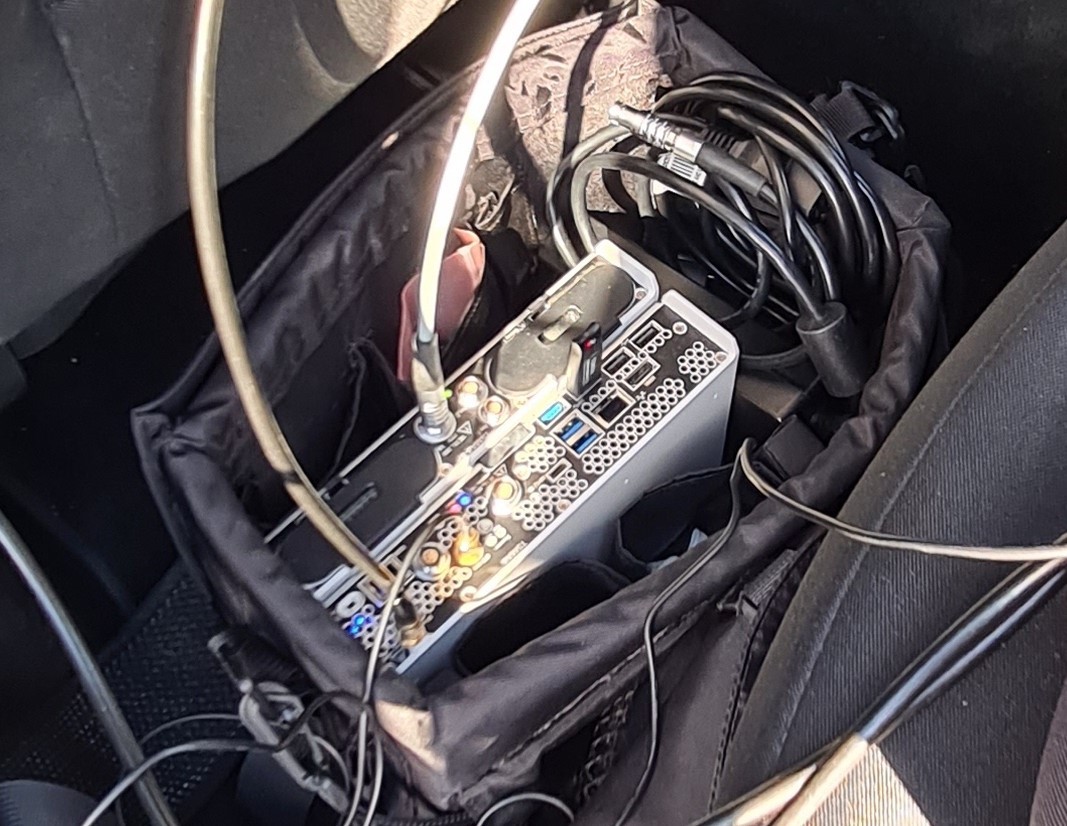}
		\label{fig:placement_network_scanner}

	}
 	\subfigure[Tower view during driving-based measurements]
	{
		\includegraphics[width=3.95cm]{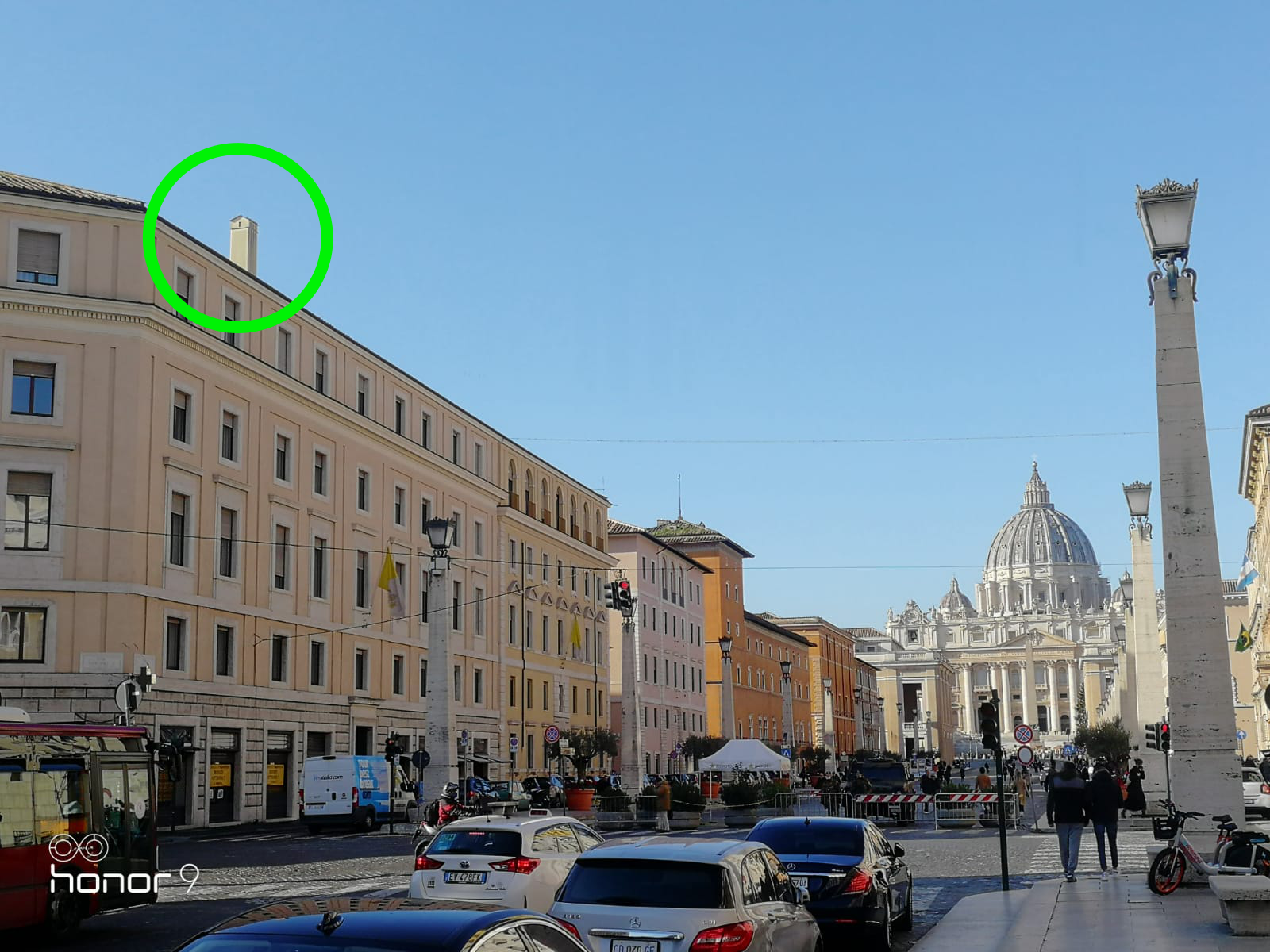}
		\label{fig:war_driving_view}
	}

	\subfigure[Raw sector orientation retrieved by the network scanner (map source: Google Maps)]
	{
		\includegraphics[width=4.1cm]{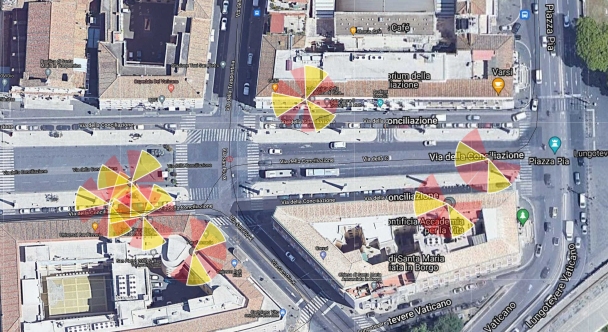}
		\label{fig:network_scanner_raw_info}
	}
 	\subfigure[Aerial view of the towers identified by the network scanner (image source: Google Maps)]
	{
		\includegraphics[width=3.9cm]{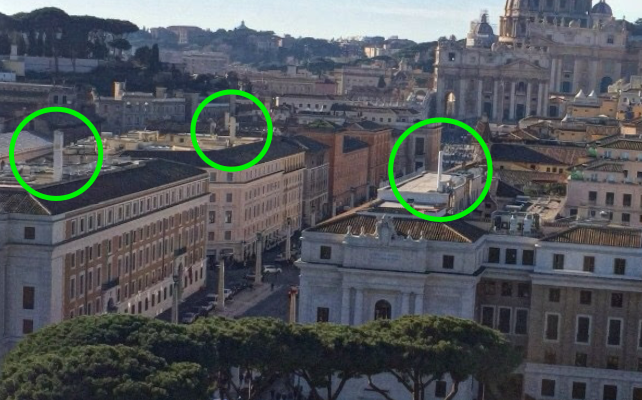}
		\label{fig:aerial_view}

	}
	\caption{Driving-based approach to characterize the towers.}
	\label{fig:new_tower_identification}
	\vspace{-4mm}
\end{figure}

\begin{figure}[t]
	\centering
 	\subfigure[Spinaceto.]
	{
		\includegraphics[width=8cm]{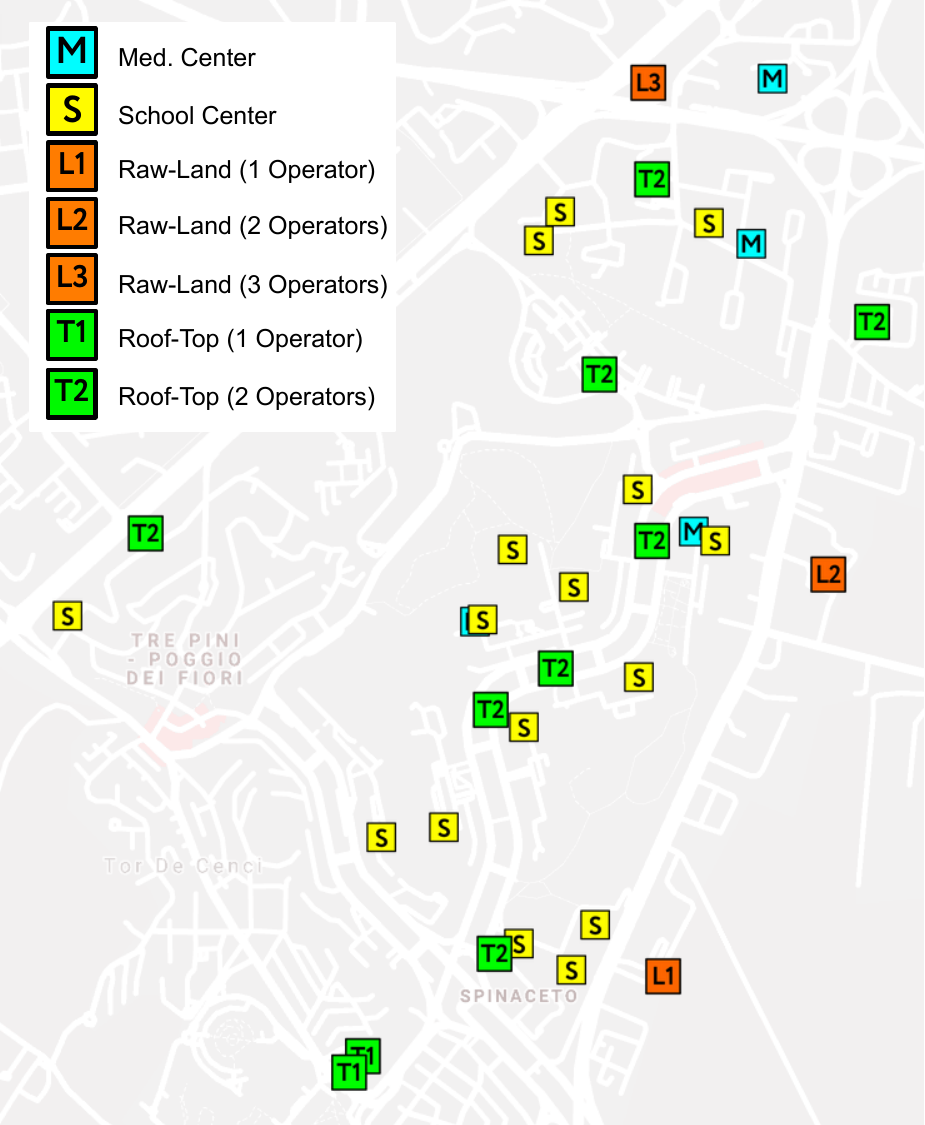}
		\label{fig:towers_w_sensitive_places_sp}

	}
	\subfigure[Ponte-Parione.]
	{
		\includegraphics[width=8cm]{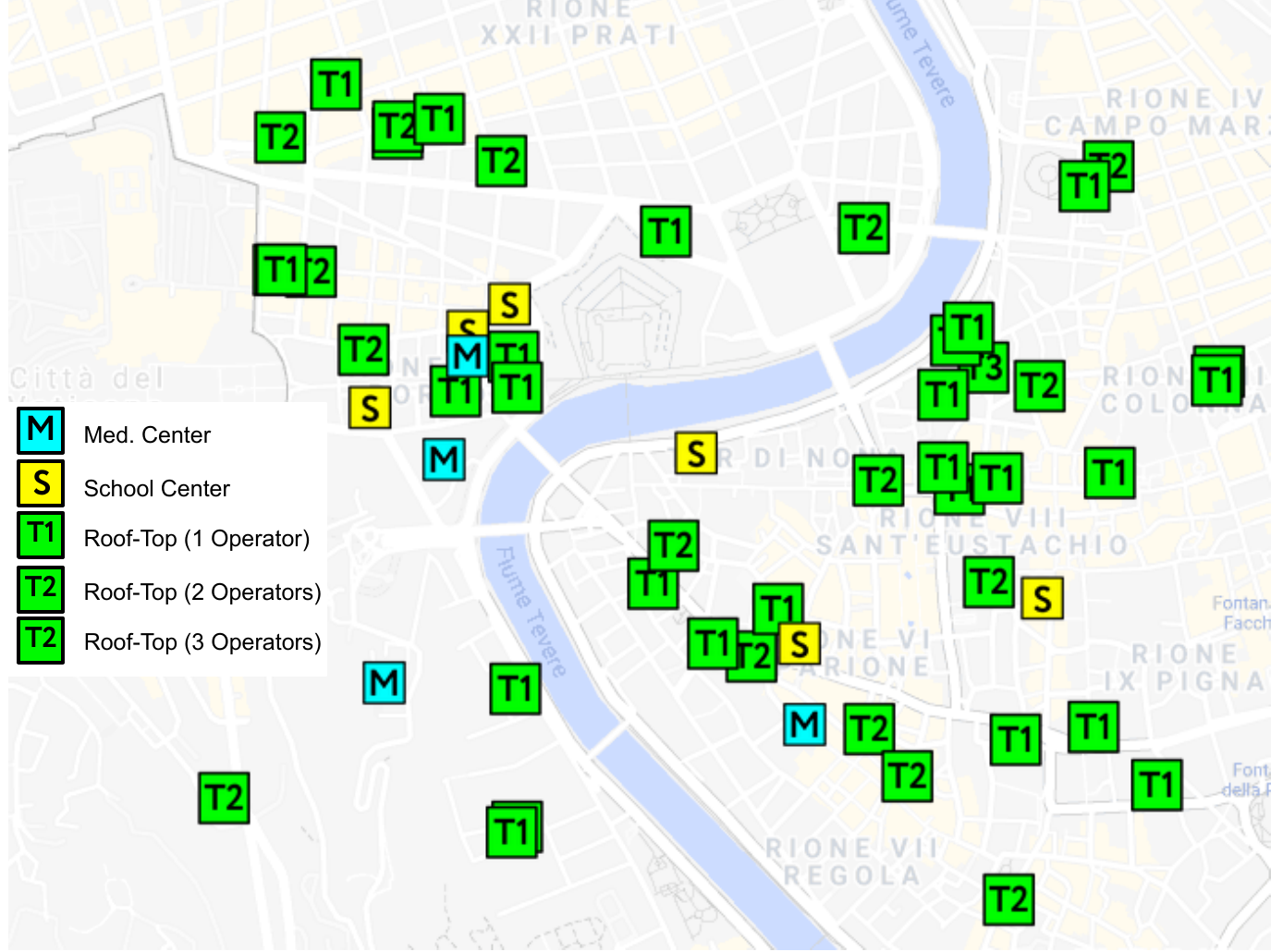}
		\label{fig:towers_w_sensitive_places_pp}

	}
	\caption{Towers and ``sensitive'' buildings in the considered scenarios (sub-figures best viewed in colors).}
	\label{fig:buildings_towers}
	\vspace{-4mm}
\end{figure}

More concretely, the tower check is done in the following way. We install a Rohde\&Schwarz TSMA6 network scanner on a car, as shown Fig.~\ref{fig:placement_network_scanner}. The network scanner is then connected to a PCTEL P286H external omnidirectional antenna (mounted on the roof on the car). In addition, electricity to the network scanner is provided through a plug on the car. The scanner is then complemented by a tablet, wirelessy connected to the scanner, and running the Rohde\&Schwarz ROMES4 commercial software for data acquisition and analysis. We then extensively cover the streets in the considered scenarios, focusing in particular on the zones where we identified the towers (Fig.~\ref{fig:war_driving_view}). Based on the triangularization of the acquired signals, the software is able to position the identified towers on a map, showing also the sector orientation $\rho^{\text{CLOCK}}_{(t,s,o)}$ (Fig.~\ref{fig:network_scanner_raw_info}). This last information is instrumental for towers without available authorization data, particularly those ones for which sector orientation is hidden to sight (like the fake chimney in Fig.~\ref{fig:war_driving_view}). Clearly, the raw information presented by the software (Fig.~\ref{fig:network_scanner_raw_info}) has to be manually filtered, in order to remove duplicates and/or not precise tower positioning. Eventually, Fig.~\ref{fig:aerial_view} reports the towers that were completely characterized through the driving-based approach (i.e., those ones with missing authorization data). For all the other towers (i.e., the ones with available authorization), the driving-based approach allowed identifying the active ones and discarding the disposed ones. In addition, sector information was compared against the one reported in the authorization documents, finding in general a good matching between both sources. Thanks to the tower check, the fake chimneys are completely characterized, and therefore they can be classified as roof-top installation from here onwards.

The real positioning of the retrieved towers and ``sensitive'' buildings is reported in Fig.~\ref{fig:buildings_towers}. More in depth, we have included in our analysis all the towers that are located inside the considered areas, as well as all the outer ones that are installed within 300~m from the area border - in order to avoid a potential exposure underestimation at the borders of the scenario. Several considerations hold by analyzing Fig.~\ref{fig:buildings_towers}. First, the towers are rather spread over Ponte-Parione territory (Fig.~\ref{fig:towers_w_sensitive_places_pp}), while in Spinaceto the towers tend to be mainly placed in proximity to the buildings (Fig.~\ref{fig:towers_w_sensitive_places_sp}). Second, the roof-top installation is the preferred option for Spinaceto - which however hosts some raw-land towers. On the contrary, Ponte-Parione does not include any raw-land, mainly due to the fact that such tower type generally requires to find ground locations without any buildings in the close surroundings - a hard goal to pursue in Ponte-Parione. Third, many towers are located close to the ``sensitive'' buildings (Fig.~\ref{fig:towers_w_sensitive_places_sp},\ref{fig:towers_w_sensitive_places_pp}). Fourth, the majority of installations in Spinaceto host at least two distinct operators on the same tower, while the option with a single operator per tower is the most used in Ponte-Parione.

\subsection{5G Radio Configurations}

\begin{table*}[t]
    \caption{Radio configuration for a sector $s$, owned by operator $o$ and installed on a roof-top tower $t$.}
    \label{tab:example_sector_roof_top}
    \scriptsize
    \centering
    \begin{tabular}{|>{\columncolor{Linen}}c|c|c|c|c|c|c|c|c|}
\hline
 \rowcolor{Linen} \textbf{Antenna ID $a$} & 1 & 2 & 3 & 4 & 5 & 6 & 7 & 8 \\
\hline
\textbf{Frequency Band $F_a$} & {700~MHz} & {800~MHz} & {900~MHz} & {1800~MHz} & {2100~MHz} & {2600~MHz} & {3700~MHz} & {26000~MHz} \\
\hline  
\textbf{Technology $T_a$} & 5G & 4G & 2G & 4G & 4G & 4G & 5G & 5G \\
\hline
\textbf{Panel Type $C_p$} & \multicolumn{6}{c|}{\cellcolor{Green}Quadri-band} & \cellcolor{Grayblue}5G Mid-Band & \cellcolor{DarkLinen}5G mm-Wave \\
\hline
\textbf{Mechanical Tilt $\sigma^{\text{M}}_{(t,s,o,p)}$} & \multicolumn{7}{c|}{0°} & 2°\\
\hline
\textbf{Electrical Tilt $\sigma^{\text{E}}_{(t,s,o,p,a)}$} & 2° & 4° & 2° & \multicolumn{3}{c|}{4°} & 2° & 0°\\
\hline
\textbf{Output Power $P^{\text{MAX}}_{(t,s,o,p,a)}$} & \multicolumn{3}{c|}{40~W} & \multicolumn{3}{c|}{60~W} & 80~W & 5~W \\
\hline
\textbf{Maximum Gain $G^{\text{MAX}}_{(t,s,o,p,a)}$} & 13.8~dBi & 14.4~dBi & 14.9~dBi & 16.4~dBi & 17.5~dBi &  17.7~dBi & 23.9~dBi & 33.5~dBi\\
\hline
 & \cellcolor{Green}MSI file & \cellcolor{Green}MSI file & \cellcolor{Green}MSI file & \cellcolor{Green}MSI file & \cellcolor{Green}MSI file & \cellcolor{Green}MSI file & \cellcolor{Grayblue}MSI file & \cellcolor{DarkLinen} MSI file \\
\multirow{-2}{*}{\textbf{Horiz. Diagram $D^{\text{H}}_{(t,s,o,p,a)}$}} & \cellcolor{Green}$(F_{a_1},t_1)$ & \cellcolor{Green}$(F_{a_2},t_2)$ & \cellcolor{Green}$(F_{a_3},t_3)$ & \cellcolor{Green}$(F_{a_4},t_4)$ & \cellcolor{Green}$(F_{a_5},t_5)$ & \cellcolor{Green}$(F_{a_6},t_6)$ & \cellcolor{Grayblue} $(F_{a_7},t_7)$ & \cellcolor{DarkLinen} $(F_{a_8},t_8)$\\
\hline
 & \cellcolor{Green}MSI file & \cellcolor{Green}MSI file & \cellcolor{Green}MSI file & \cellcolor{Green}MSI file & \cellcolor{Green}MSI file & \cellcolor{Green}MSI file & \cellcolor{Grayblue}MSI file & \cellcolor{DarkLinen} MSI file \\
\multirow{-2}{*}{\textbf{Vert. Diagram $D^{\text{V}}_{(t,s,o,p,a})$}}  & \cellcolor{Green}$(F_{a_1},t_1)$ & \cellcolor{Green}$(F_{a_2},t_2)$ & \cellcolor{Green}$(F_{a_3},t_3)$ & \cellcolor{Green}$(F_{a_4},t_4)$ & \cellcolor{Green}$(F_{a_5},t_5)$ & \cellcolor{Green}$(F_{a_6},t_6)$ & \cellcolor{Grayblue}$(F_{a_7},t_7)$ & \cellcolor{DarkLinen}$(F_{a_8},t_8)$\\
\hline
\textbf{Reduction Factor $R_{(t,s,o,p,a)}$} &\multicolumn{6}{c|}{1} & \multicolumn{2}{c|}{$\alpha^{\text{5G}}_p$} \\
\hline
\end{tabular}
\end{table*}

\begin{table*}[t]
    \caption{Radio configuration for a sector $s$, owned by operator $o$ and installed on a raw-land tower $t$.}
    \label{tab:example_sector_raw_land}
    \scriptsize
    \centering
    \begin{tabular}{|>{\columncolor{Linen}}c|c|c|c|c|c|c|c|c|}
\hline
\rowcolor{Linen}  \textbf{Antenna ID $a$} & 1 & 2 & 3 & 4 & 5 & 6 & 7 & 8 \\
\hline
\textbf{Frequency Band $F_a$} & {700~MHz} & {800~MHz} & {900~MHz} & {1800~MHz} & {2100~MHz} & {2600~MHz} & {3700~MHz} & {26000~MHz} \\
\hline  
\textbf{Technology $T_a$} & 5G & 4G & 2G & 4G & 4G & 4G & 5G & 5G \\
\hline
\textbf{Panel Type $C_p$} & \multicolumn{6}{c|}{\cellcolor{Green}Quadri-band} & \cellcolor{Grayblue}5G mid-Band & \cellcolor{DarkLinen}5G mm-Wave \\
\hline
\textbf{Mechanical Tilt $\sigma^{\text{M}}_{(t,s,o,p)}$} & \multicolumn{7}{c|}{0°} & 2°\\
\hline
\textbf{Electrical Tilt $\sigma^{\text{E}}_{(t,s,o,p,a)}$} & 4° & 6° & 4° & \multicolumn{3}{c|}{6°} & 2° & 0°\\
\hline
\textbf{Output Power $P^{\text{MAX}}_{(t,s,o,p,a)}$} & \multicolumn{3}{c|}{60~W} & \multicolumn{3}{c|}{80~W} & 100~W & 5~W \\
\hline
\textbf{Maximum Gain $G^{\text{MAX}}_{(t,s,o,p,a)}$} & 13.8~dBi & 14.4~dBi & 14.9~dBi & 16.4~dBi & 17.5~dBi &  17.7~dBi & 23.9~dBi & 33.5~dBi\\
\hline
 & \cellcolor{Green}MSI file & \cellcolor{Green}MSI file & \cellcolor{Green}MSI file & \cellcolor{Green}MSI file & \cellcolor{Green}MSI file & \cellcolor{Green}MSI file & \cellcolor{Grayblue}MSI file & \cellcolor{DarkLinen} MSI file \\
\multirow{-2}{*}{\textbf{Horiz. Diagram $D^{\text{H}}_{(t,s,o,p,a)}$}} & \cellcolor{Green}$(F_{a_1},t_1)$ & \cellcolor{Green}$(F_{a_2},t_2)$ & \cellcolor{Green}$(F_{a_3},t_3)$ & \cellcolor{Green}$(F_{a_4},t_4)$ & \cellcolor{Green}$(F_{a_5},t_5)$ & \cellcolor{Green}$(F_{a_6},t_6)$ & \cellcolor{Grayblue} $(F_{a_7},t_7)$ & \cellcolor{DarkLinen} $(F_{a_8},t_8)$\\
\hline
 & \cellcolor{Green}MSI file & \cellcolor{Green}MSI file & \cellcolor{Green}MSI file & \cellcolor{Green}MSI file & \cellcolor{Green}MSI file & \cellcolor{Green}MSI file & \cellcolor{Grayblue}MSI file & \cellcolor{DarkLinen} MSI file \\
\multirow{-2}{*}{\textbf{Vert. Diagram $D^{\text{V}}_{(t,s,o,p,a)}$}}  & \cellcolor{Green}$(F_{a_1},t_1)$ & \cellcolor{Green}$(F_{a_2},t_2)$ & \cellcolor{Green}$(F_{a_3},t_3)$ & \cellcolor{Green}$(F_{a_4},t_4)$ & \cellcolor{Green}$(F_{a_5},t_5)$ & \cellcolor{Green}$(F_{a_6},t_6)$ & \cellcolor{Grayblue}$(F_{a_7},t_7)$ & \cellcolor{DarkLinen}$(F_{a_8},t_8)$\\
\hline
\textbf{Reduction Factor $R_{(t,s,o,p,a)}$} & \multicolumn{6}{c|}{1} & \multicolumn{2}{c|}{$\alpha^{\text{5G}}_p$} \\
\hline
\end{tabular}
\end{table*}

In the following step, we set the 5G radio configuration parameters that have been introduced in Tab.~\ref{tab:radio_conf}. In this work, we apply a uniform setting, in which the panel configuration solely depends on the type of tower (raw-land and roof-top).\footnote{The optimization of the panel configuration for each sector, operator and tower is left for future work}. The resulting radio configurations for the two tower types are reported in  Tab.~\ref{tab:example_sector_roof_top}-Tab.~\ref{tab:example_sector_raw_land}.

Several consideration holds by observing in more detail the tables. Naturally, the quadri-band panel includes sub-GHz band of 5G, as well as the bands for 2G and 4G. In addition, the values for the mechanical/electrical tilting and the output power are taken from typical settings adopted by operators. Eventually, the output power of the mm-Wave panel is set to the maximum one allowed by the equipment under consideration. The antenna gains reported in the tables are retrieved from the datasheets of the panels. Moreover, the radiation diagrams for each source are taken from the real ones made available by the antenna vendors. Such information is included in files with MSI format (i.e., a list of gain values for each angle in horizontal and vertical planes). Each radiation diagram depends on the following features: \textit{i}) the panel model (highlighted with different colors in the tables), \textit{ii}) the operating frequency, and \textit{iii}) the electrical tilting value. Finally, the last row of Tab.~\ref{tab:example_sector_roof_top}-\ref{tab:example_sector_raw_land} reports the reduction factor that is applied to the maximum power. In this work, we assume that the quadri-band panel always radiates at the maximum power (corresponding to the values of output power reported in the tables). On the other hand, the output power of mid-band and mm-Wave 5G panels is scaled by the factor $\alpha^{\text{5G}}_p$, which is set in accordance to the realistic values provided by relevant standards \cite{iec62669appendix}.

\begin{table}[t]
    \caption{Statistical reduction factors for the different levels of 5G adoption.}
    \label{tab:stat_reduction}
    \scriptsize
    \centering
    \begin{tabular}{|c|c|c|c|}
\hline
\rowcolor{Linen} \multicolumn{2}{|c|}{\textbf{5G Adoption Level}} & \multicolumn{2}{c|}{\textbf{$\alpha^{\text{5G}}_p$ values}}  \\
\rowcolor{Linen} \textbf{Name} & \textbf{Index $l$} & \textbf{Ponte-Parione} & \textbf{Spinaceto} \\
\hline
Early & 1 & 0.03 & 0.05\\
Medium & 2 & 0.13 & 0.21\\
Maturity & 3 & 0.18 & 0.31\\
\hline
\end{tabular}
\end{table}

More concretely, $\alpha^{\text{5G}}_p$ captures the statistical power reduction factors that are introduced by the smart antenna management features of 5G over mid-band and mm-Wave. To this aim, Tab.~\ref{tab:stat_reduction} reports the reduction factors over the two scenarios and the different level of 5G adoptions. The numerical values are taken from \cite{iec62669appendix}, by assuming that Spinaceto is representative for a sparse area while Ponte-Parione for a dense one. In addition, the early/medium/maturity levels of 5G adoption correspond to 5\%/50\%/95\% utilization level of \cite{iec62669appendix}, respectively. By observing the values in Tab.~\ref{tab:stat_reduction}, two considerations emerge. First, strong power reduction factors are applied even in the maturity case. Second, the power reduction factors are lower (i.e., stronger) in Ponte-Parione than in Spinaceto, due to the higher and more uniform building density that characterizes the former with respect to the latter.

\begin{table}[t]
    \caption{Tower metrics over the considered scenarios.}
    \label{tab:mobile_net_metrics}
    \scriptsize
    \centering
    \begin{tabular}{|c|c|c|}
\hline
\rowcolor{Linen} & \multicolumn{2}{c|}{\textbf{Metric Value}}  \\
\rowcolor{Linen} \multirow{-2}{*}{\textbf{Metric Name}}& \textbf{Ponte-Parione} & \textbf{Spinaceto} \\
\hline
Number of operators & \multicolumn{2}{c|}{4} \\
Number of towers $|\mathcal{T}|$ & 45 & 13 \\
Number of roof-top towers & 45 & 10 \\
Number of raw-land towers & 0 & 3 \\
Number of pre-5G sources & 934 & 354 \\
Number of 5G sources & 560 & 212 \\
\hline
\end{tabular}
\end{table}

Finally, Tab.~\ref{tab:mobile_net_metrics} reports the relevant tower metrics over Spinaceto and Ponte-Parione. More in depth, all four national operators manage towers in both scenarios.
However, the number of towers in Ponte-Parione is larger than in Spinaceto. Focusing on the tower type, Ponte-Parione includes only roof-top installations, while Spinaceto hosts also some raw-lands.
In addition the table details the number of radiating sources in each scenario, which is computed by counting all antennas for each panel, each sector, each operator and each tower.
Obviously, the total number sources is clearly higher in Ponte-Parione than in Spinaceto. However, we point out that the number of pre-5G sources is always higher than the 5G ones, due to the larger number of frequency bands managed by pre-5G with respect to 5G. 



\subsection{Pixel Exposure Assessment, Population and Building Analysis}

We implement the pixel exposure assessment procedure of Sec.~\ref{sec:pec_comp} in a custom program, written in C++, and taking the required input information (in terms of buildings, population, tower and 5G configuration data) as files in comma-separated values format. The program is compiled and run on a Windows 10 Pro machine, equipped with 16~[GB] of Random Access Memory and 6 Central Processing Unit cores running at 2.1~[GHz]. The pixel granularity is set to 2 $\times$ 2 m$^2$. The total time to run each simulation in Ponte-Parione (i.e., the most complex scenario) is less than 2 minutes. 

Given the values of exposure that are computed in each scenario, the building and population analysis of Sec.~\ref{sec:bp_metrics} is run as a Matlab script on a MacBook Air laptop, equipped with 4~[GB] of Random Access Memory and an Intel Core i5 Central Processing Unit running at 1.3~[GHz]. The total time for running such analysis is lower than one minute on average.


\begin{figure}[t]

	\centering
 	\subfigure[Spinaceto]
	{
		\includegraphics[width=7cm]{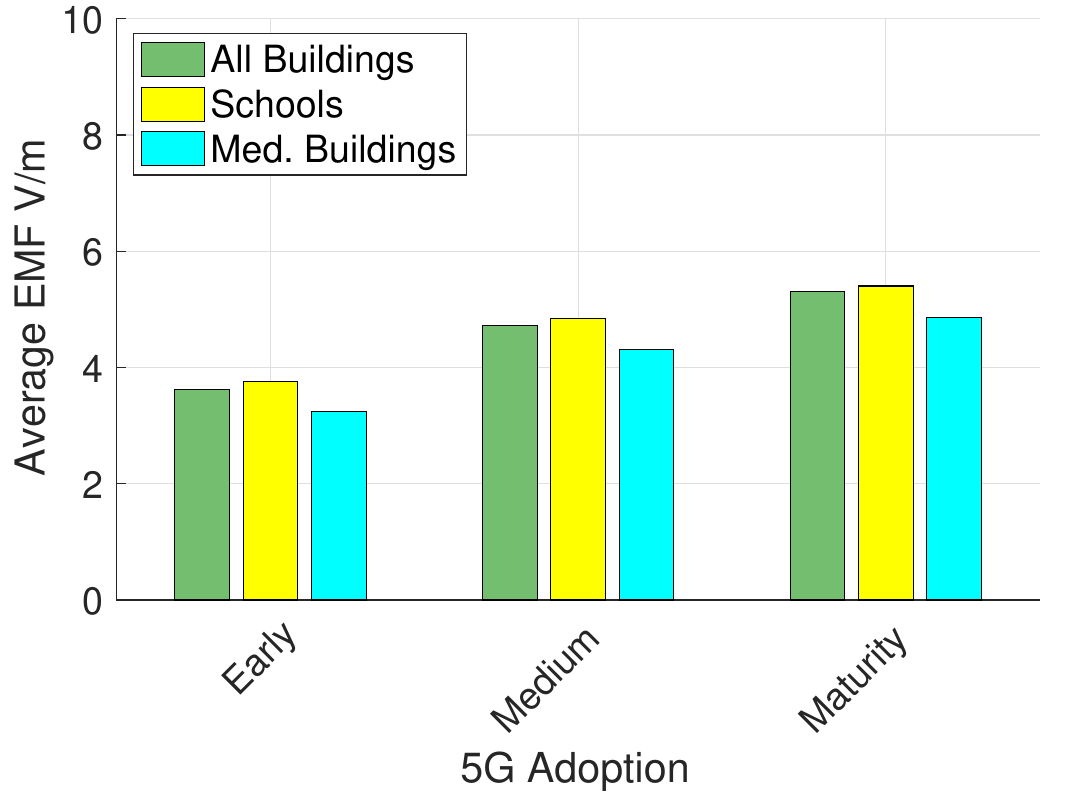}
		\label{fig:avg_emf_building_exposure_sp}
	}
	\subfigure[Ponte-Parione]
	{
		\includegraphics[width=7cm]{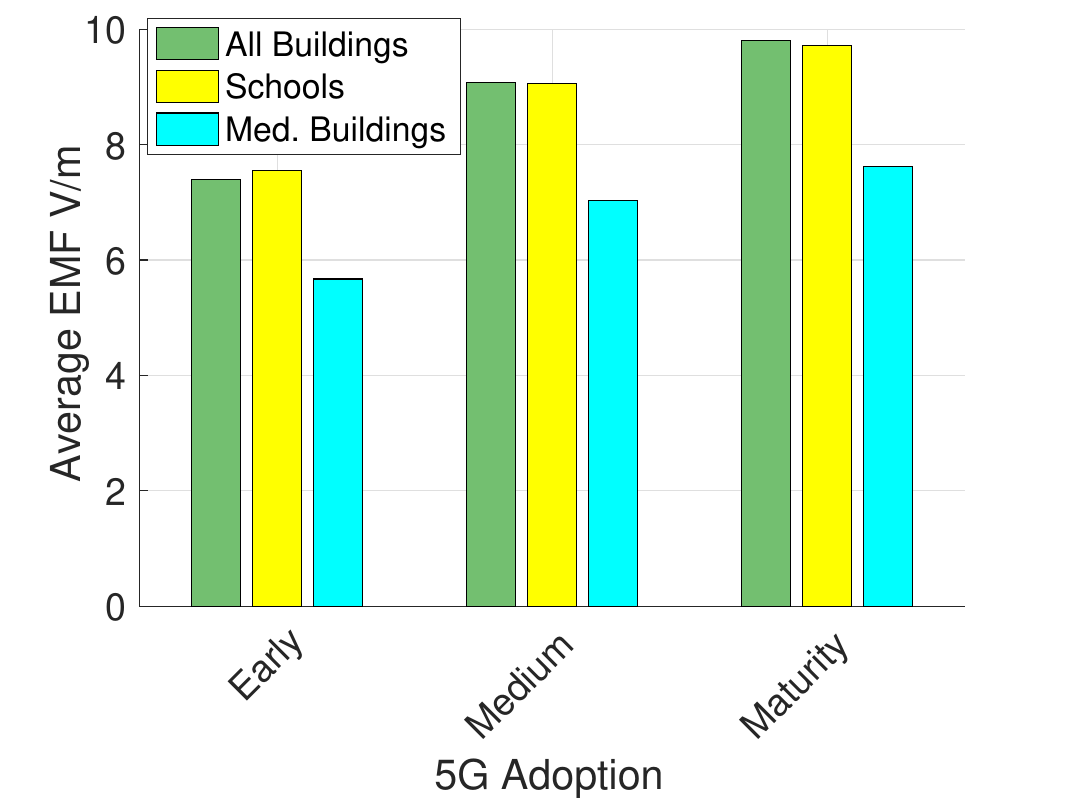}
		\label{fig:avg_emf_building_exposure_pp}
	}
	\caption{Average building \ac{EMF} $E^{\text{TOT}}_{\text{AVG}}$, $E^{\text{TOT}}_{\text{AVG-SCHOOL}}$, $E^{\text{TOT}}_{\text{AVG-MED}}$ vs. level of 5G adoption (sub-figures best viewed in colors).}
	\label{fig:avg_emf_building_exposure}
\end{figure}

\section{Results}
\label{sec:results}

We divide the outcomes of our analysis in the following steps: \textit{i}) evaluation of average exposure levels, \textit{ii}) assessment of the exposure distribution over the whole set of samples, \textit{iii}) visualization of the spatial exposure levels, \textit{iv}) investigation of the impact of building {and weather} attenuation. 




\begin{figure}[t]
	\centering
 	\subfigure[Spinaceto]
	{
		\includegraphics[width=7cm]{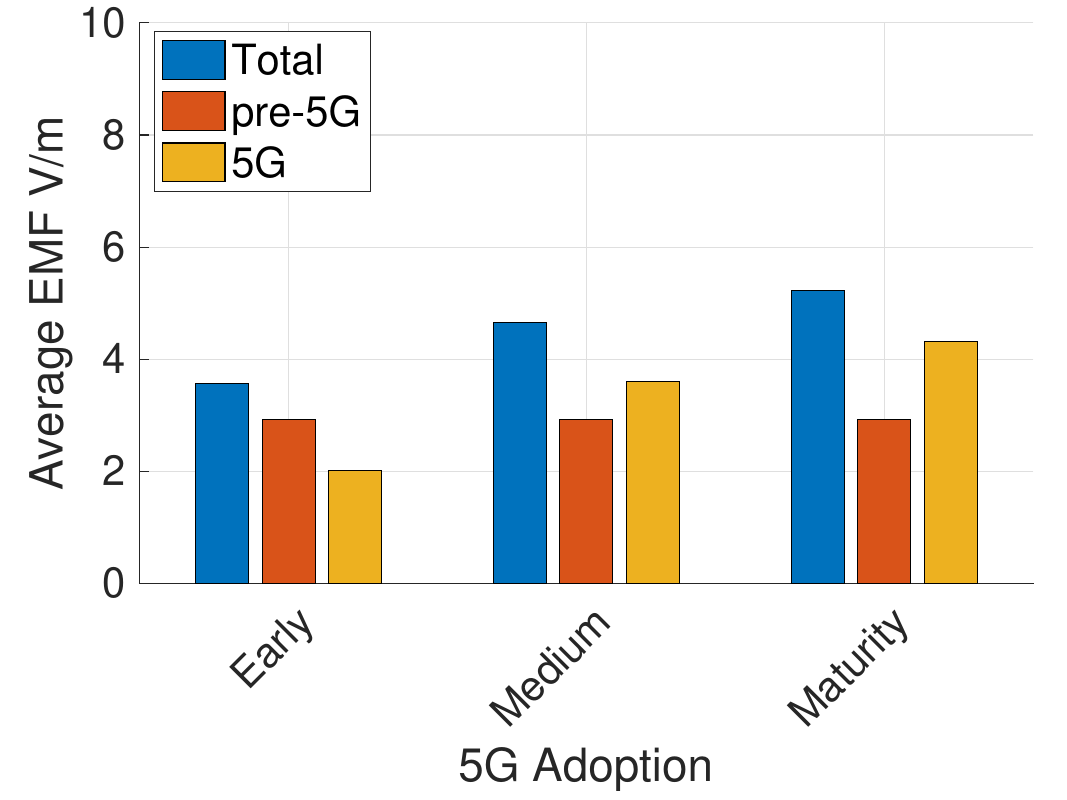}
		\label{fig:avg_emf_chd_adl_vs_5G_adoption_sp}
	}
	\subfigure[Ponte-Parione]
	{
		\includegraphics[width=7cm]{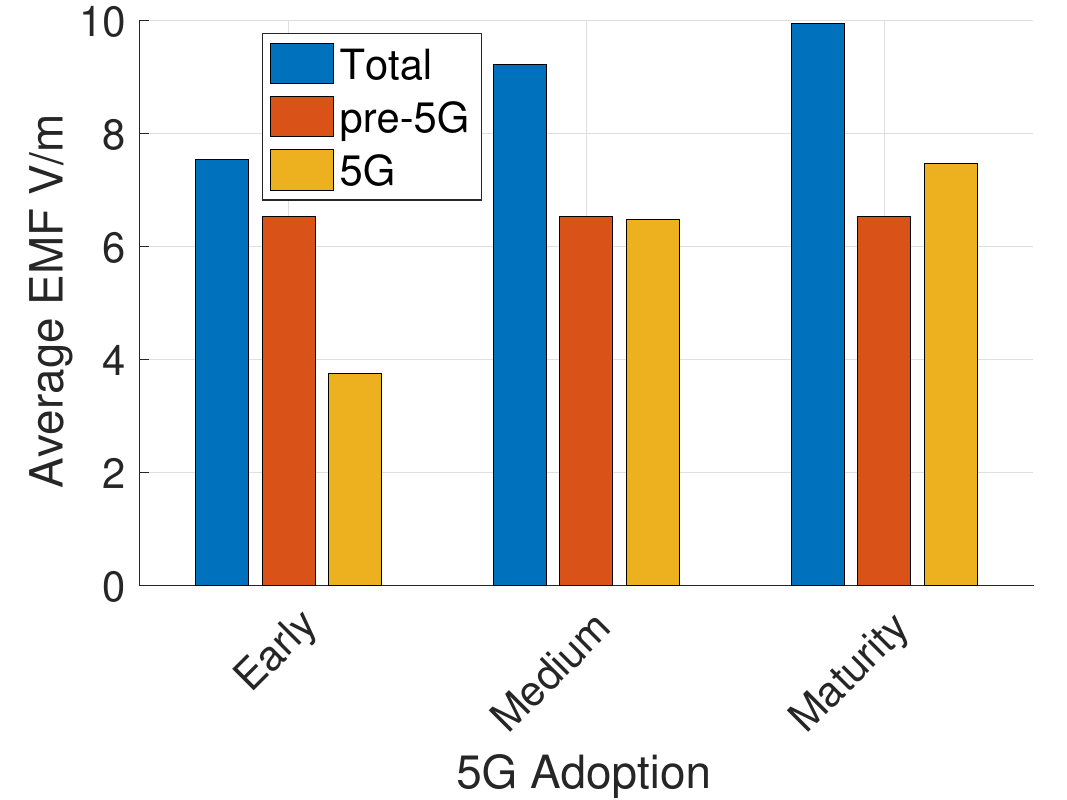}
		\label{fig:avg_emf_chd_adl_vs_5G_adoption_pp}
	}
	
	\caption{Average children/teenagers \ac{EMF} $E^{\text{TOT}}_{\text{AVG-CHD-TN}}$, $E^{\text{5G}}_{\text{AVG-CHD-TN}}$, $E^{\text{PRE-5G}}_{\text{AVG-CHD-TN}}$  vs. level of 5G adoption  (sub-figures best viewed in colors).}
	\label{fig:avg_emf_chd_adl_vs_5G_adoption}
\end{figure}

\subsection{Average Exposure Levels}
\label{sec:avg_exp}

In the first part, we analyse the average exposure metrics over the buildings and the population. Unless otherwise specified, we run our methodology over Spinaceto and Ponte-Parione scenarios, by initially neglecting the building attenuation (i.e., $A_{(c,t,s,o,p,a)}=1$). In this way, we evaluate the exposure under a very conservative assumption. Fig.~\ref{fig:avg_emf_building_exposure} reports the average \ac{EMF} metrics $E^{\text{TOT}}_{\text{AVG}}$, $E^{\text{TOT}}_{\text{AVG-SCHOOL}}$, $E^{\text{TOT}}_{\text{AVG-MED}}$ over the buildings, for different levels of 5G adoption (i.e., early, medium, maturity). Actually, the statistical power reduction factors $\alpha^{\text{5G}}_p$ that are applied to mid-band and mm-Wave 5G panels depend on the adoption level (as reported in Tab.~\ref{tab:stat_reduction}), with a stronger scaling in the early compared to the maturity 5G adoption level. Consequently, the average exposure tends to increase when passing from an early level to the maturity one. 

\begin{figure*}[t]
	\centering
 	\subfigure[Children and teenagers - Spinaceto.]
	{
		\includegraphics[width=5cm]{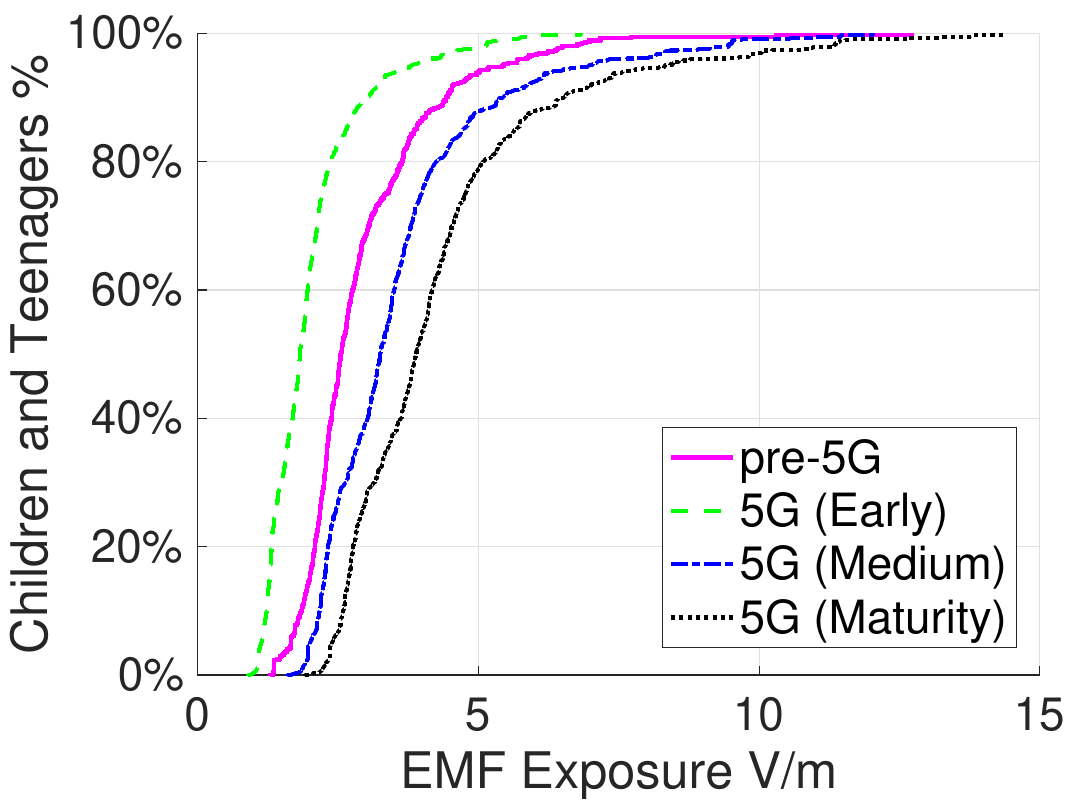}
		\label{fig:ecdf_chd_adl_vs_5G_adoption_sp}

	}
 	\subfigure[Schools - Spinaceto]
	{
		\includegraphics[width=5cm]{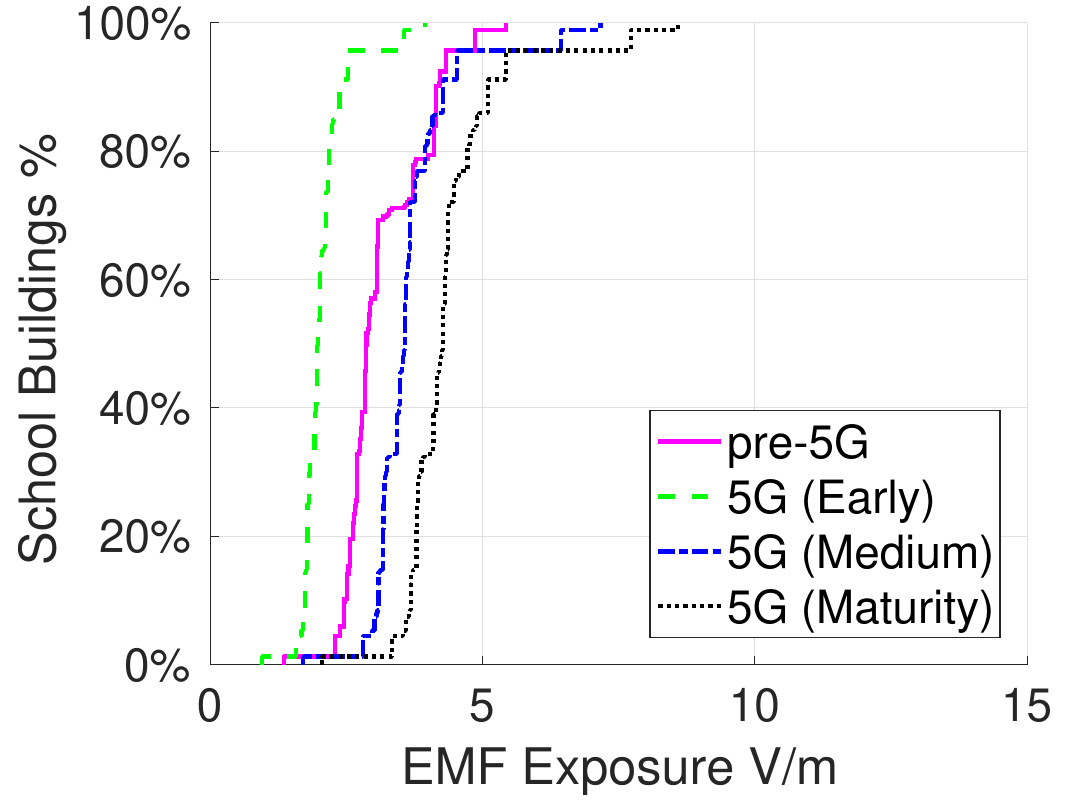}
		\label{fig:ecdf_schools_vs_5G_adoption_sp}
	}
	\subfigure[Med. Buildings - Spinaceto]
	{
		\includegraphics[width=5cm]{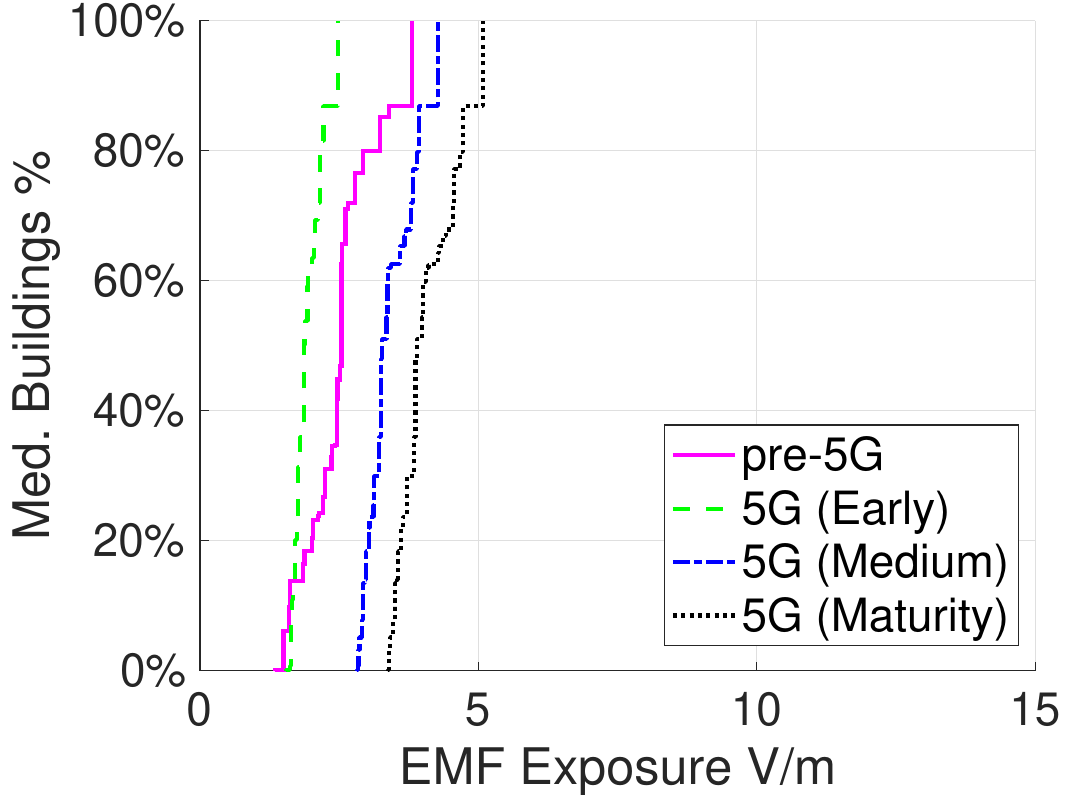}
		\label{fig:ecdf_med_buildings_vs_5G_adoption_sp}
	}

 	\subfigure[Children and teenagers - Ponte-Parione.]
	{
		\includegraphics[width=5cm]{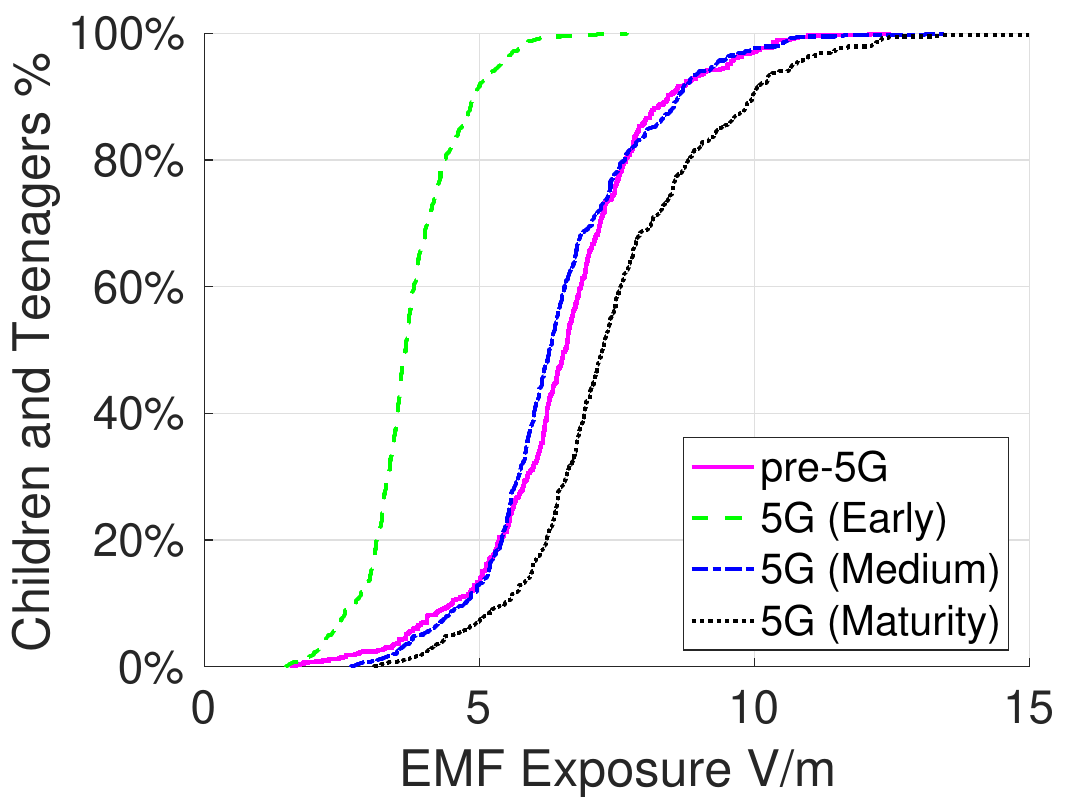}
		\label{fig:ecdf_chd_adl_vs_5G_adoption_pp}

	}
 	\subfigure[Schools - Ponte-Parione]
	{
		\includegraphics[width=5cm]{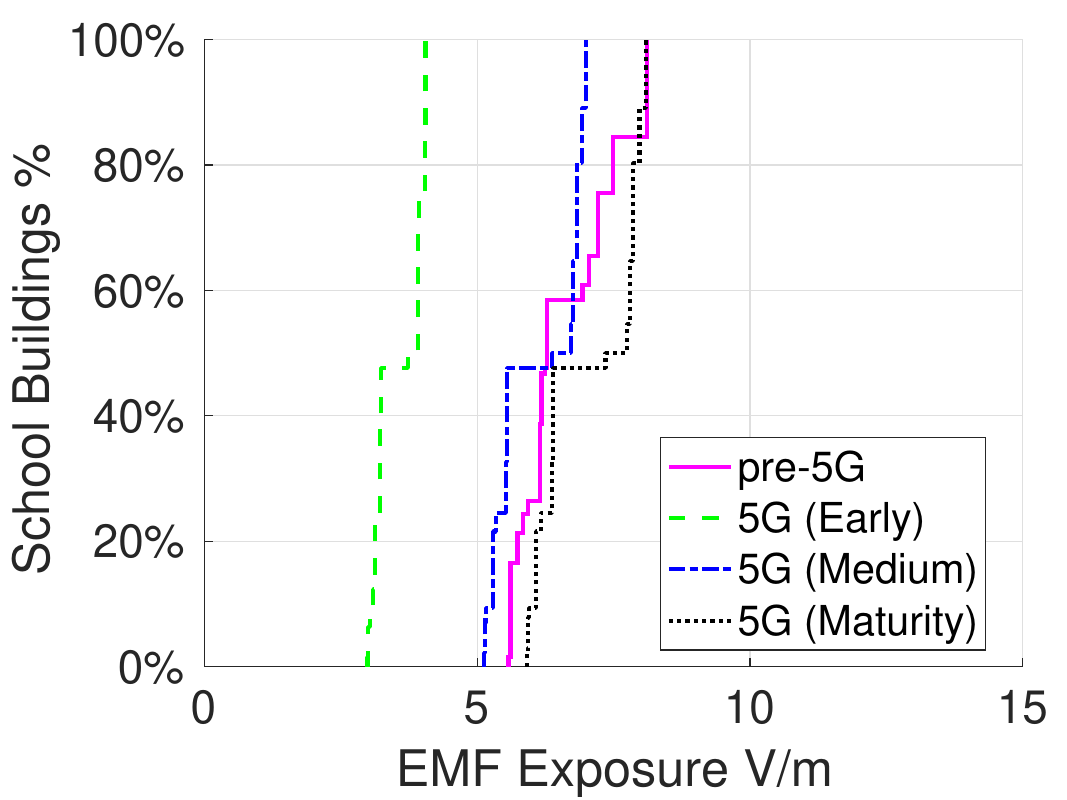}
		\label{fig:ecdf_schools_vs_5G_adoption_pp}
	}
	\subfigure[Med. Buildings - Ponte-Parione]
	{
		\includegraphics[width=5cm]{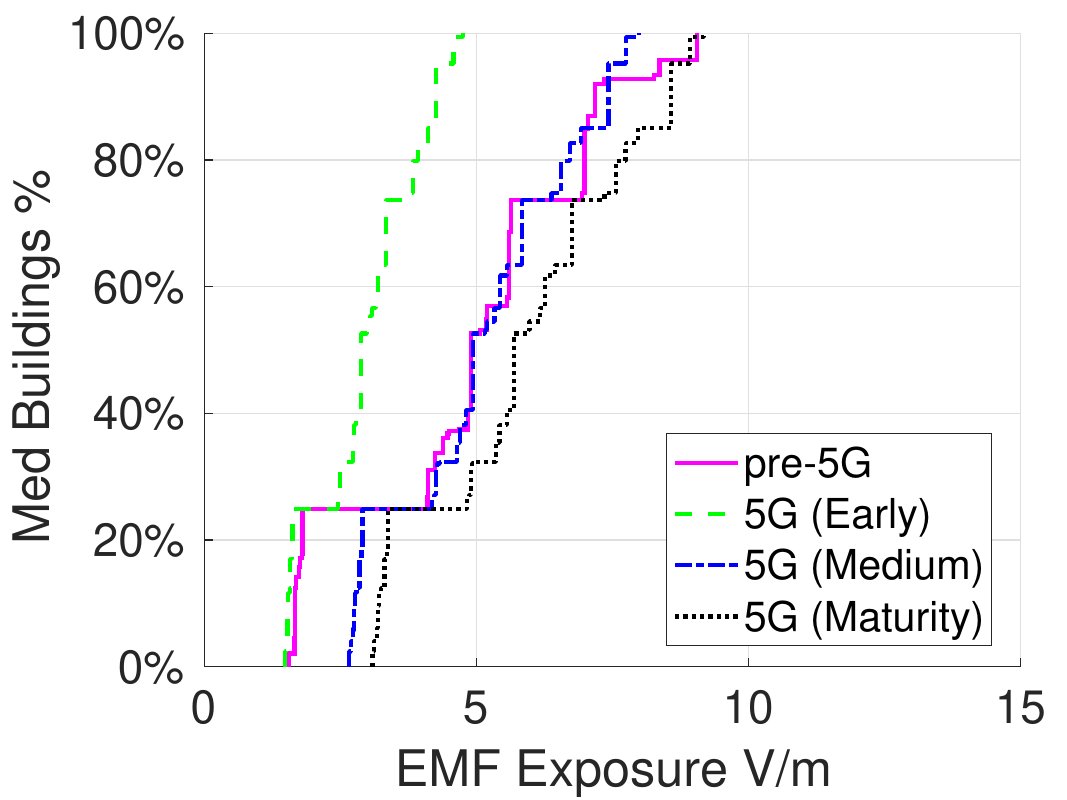}
		\label{fig:ecdf_med_buildings_vs_5G_adoption_pp}
	}
	\caption{Percentage of children and teenagers (left), schools (center) and med. buildings (right) vs. \ac{EMF} exposure.}
	\label{fig:comp_ecdf_exposure_vs_5G_adoption}
\end{figure*}

The figure then further differentiates between Spinaceto (Fig.~\ref{fig:avg_emf_building_exposure_sp}) and Ponte-Parione (Fig.~\ref{fig:avg_emf_building_exposure_pp}). Several considerations hold by comparing Fig.~\ref{fig:avg_emf_building_exposure_sp}-\ref{fig:avg_emf_building_exposure_pp}. First, the exposure over schools is rather similar to the one received by the whole set of buildings in both scenarios. On the contrary, medical buildings receive a lower exposure compared to the other building categories. Second, the maximum exposure increase when passing from an early to a maturity level is typically lower than 2~V/m and 3~V/m for Spinaceto and Ponte-Parione, respectively. Third, the buildings in Ponte-Parione receive an higher average exposure than the ones in Spinaceto, for all 5G adoption levels. This trend may be explained by the higher tower density in Ponte-Parione compared to Spinaceto (see Fig.~\ref{fig:buildings_towers}). Therefore, despite the 5G radiated power per tower is lower in Ponte-Parione than in Spinaceto (due to stronger statistical reduction factors), the composite exposure from all towers is higher in the former compared to the latter.

In the following step, we focus on the different exposure components, by considering the average exposure $E^{\text{TOT}}_{\text{AVG-CHD-TN}}$, $E^{\text{5G}}_{\text{AVG-CHD-TN}}$, $E^{\text{PRE-5G}}_{\text{AVG-CHD-TN}}$ that are received over children and teenagers living in the two areas. To this aim, Fig.~\ref{fig:avg_emf_chd_adl_vs_5G_adoption} differentiates between: \textit{i}) pre-5G exposure $E^{\text{PRE-5G}}_{\text{AVG-CHD-TN}}$, \textit{ii}) 5G exposure $E^{\text{5G}}_{\text{AVG-CHD-TN}}$, \textit{iii}) total exposure $E^{\text{TOT}}_{\text{AVG-CHD-TN}}$. The analysis is then repeated for the different adoption levels of 5G (from early level to maturity one). Interestingly, the contribution of 5G is lower than the pre-5G one for the early 5G adoption case. This outcome corresponds to the current situation, in which 5G represents a fraction of the exposure generated by pre-5G antennas (mainly 4G and 2G) \cite{chiaraviglio2021massive}. Then, when passing from early to medium adoption level, the contribution of 5G exposure tends to increase - as a consequence of the radiated power growth by mid-band and mm-Wave antennas. At last, the exposure from 5G becomes higher that the pre-5G one, i.e., $E^{\text{5G}}_{\text{AVG-CHD-TN}} > E^{\text{PRE-5G}}_{\text{AVG-CHD-TN}}$. This event occurs when a medium adoption level is achieved in Spinaceto (Fig.~\ref{fig:avg_emf_chd_adl_vs_5G_adoption_sp}) and only when a maturity level is reached by Ponte-Parione (Fig.~\ref{fig:avg_emf_chd_adl_vs_5G_adoption_pp}). Intuitively, in fact, the different statistical power reduction factors $\alpha^{\text{5G}}_p$ that are applied in the two scenarios (resulting from the different spatial distribution of the beams) determine the share of 5G exposure with respect to the pre-5G one.

\subsection{Exposure Distribution}
\label{sec:exp_distr}

\begin{figure}[t]
	\centering
 	\subfigure[Percentage of people vs. \ac{EMF} exposure - Spinaceto.]
	{
		\includegraphics[width=4cm]{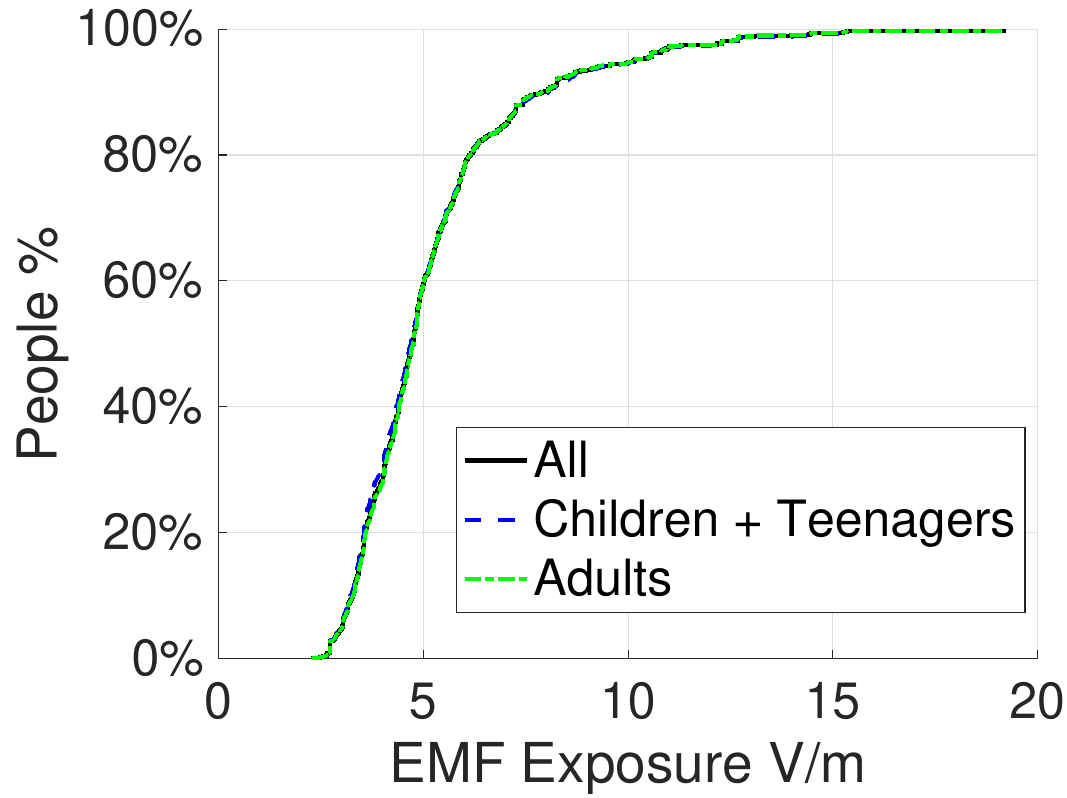}
		\label{fig:ecdf_pop_max_emf_sp}

	}
 	\subfigure[Percentage of buildings vs. \ac{EMF} exposure - Spinaceto.]
	{
		\includegraphics[width=4cm]{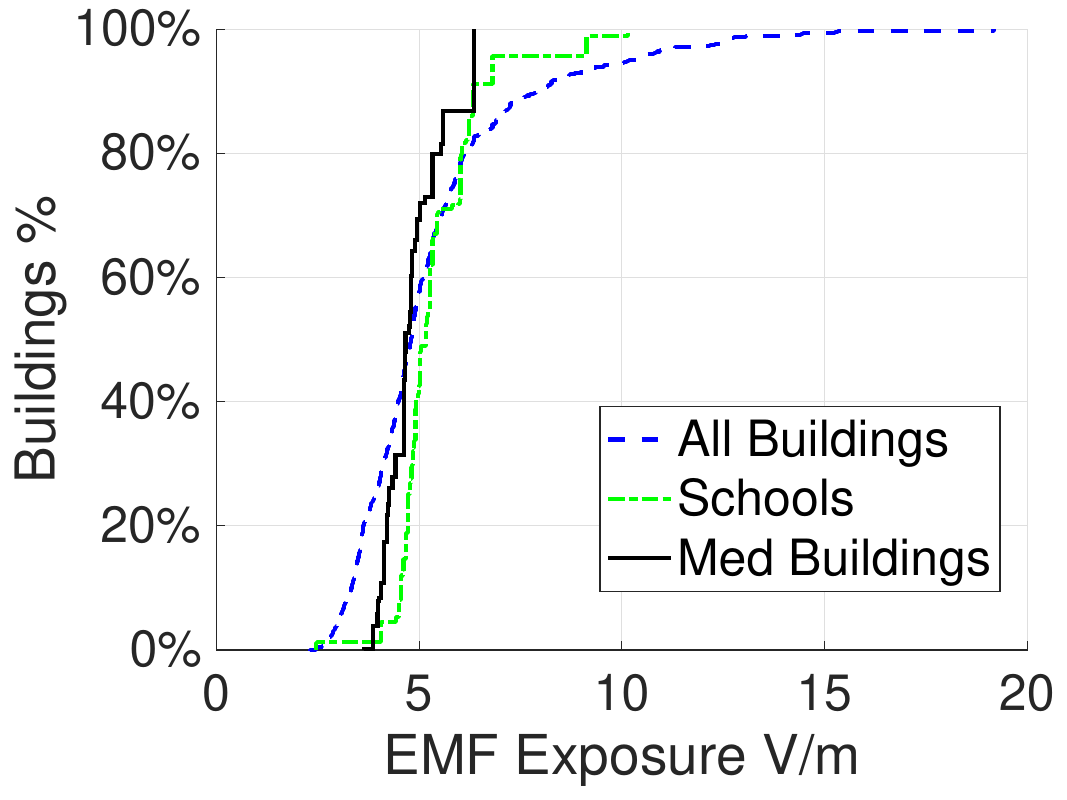}
		\label{fig:ecdf_building_max_sp}
	}

 	\subfigure[Percentage of people vs. \ac{EMF} exposure - Ponte-Parione.]
	{
		\includegraphics[width=4cm]{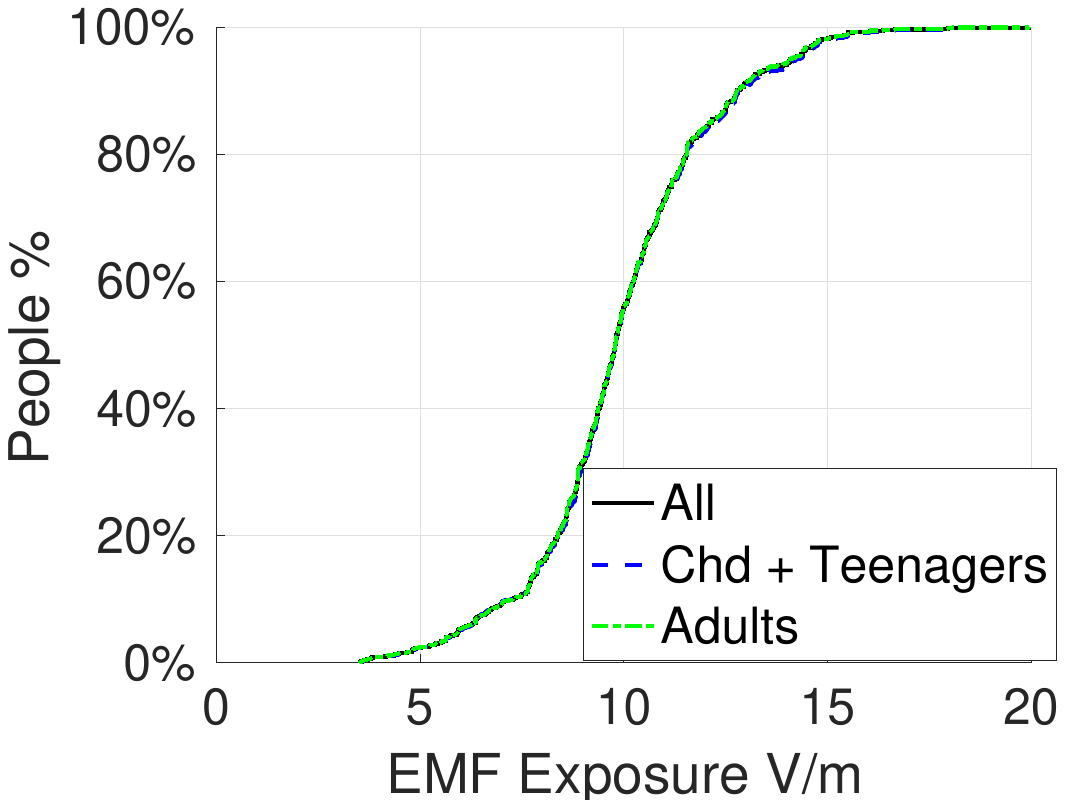}
		\label{fig:ecdf_pop_max_emf_pp}

	}
 	\subfigure[Percentage of buildings vs. \ac{EMF} exposure - Ponte-Parione.]
	{
		\includegraphics[width=4cm]{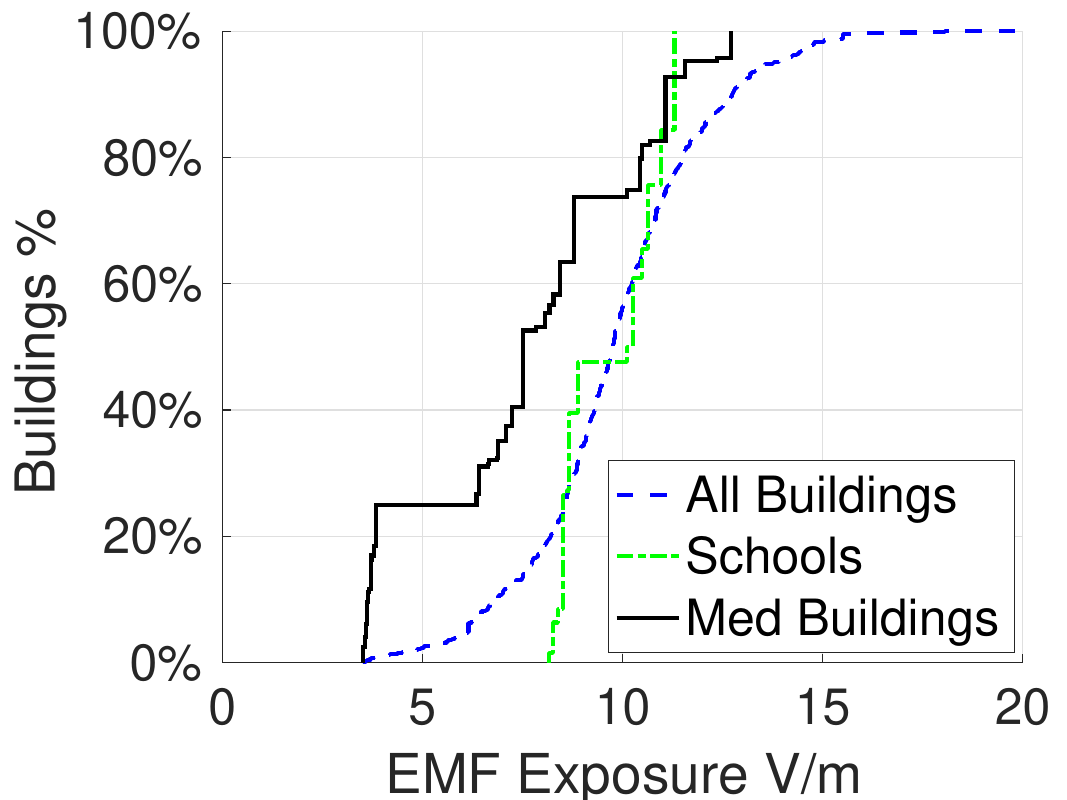}
		\label{fig:ecdf_building_max_pp}
	}
	\caption{Exposure levels for children/teenagers against adults and entire population (left). Comparison of exposure levels vs. the type of the building (right). Figures retrieved with a 5G maturity level.}
	\label{fig:ecdf_avg_emf_comparison}
	\vspace{-4mm}
\end{figure} 

Up to this point, we have considered the average exposure levels, without analyzing the exposure variations of the single samples $E^{(.)}_b$ and $E^{(.)}_i$ in each group (children/teenagers, schools and medical buildings). To shed light on this issue, Fig.~\ref{fig:comp_ecdf_exposure_vs_5G_adoption} reports the \ac{EMF} exposure vs. the percentage of children/teenagers, school buildings and medical buildings over the two scenarios. The figure is retrieved from the collection of all the exposure samples in each group. Intuitively, the curves in Fig.~\ref{fig:comp_ecdf_exposure_vs_5G_adoption} can be interpreted in this way: given a percentage value, the corresponding \ac{EMF} value read on the curve is the maximum exposure that is achieved with that percentage of samples. Clearly, when the percentage equals 100\%, the \ac{EMF} value read on the curve is the maximum \ac{EMF} exposure over all the samples. The slope of the curve captures the \ac{EMF} exposure variation over the samples: when the curve is almost vertical, the variation is low (i.e., many samples have similar exposure values); when the curve is more horizontal, the exposure variation over the samples is increased. Each curve in the subfigures then highlights the following \ac{EMF} metrics: \textit{i}) pre-5G exposure (continuous line), \textit{ii}) 5G exposure - early adoption (dashed line), \textit{iii}) 5G exposure - medium adoption (dashed-dotted line), \textit{iv}) 5G exposure - maturity adoption (dotted line).

Several considerations hold by observing Fig.~\ref{fig:ecdf_chd_adl_vs_5G_adoption_sp}-Fig.~\ref{fig:ecdf_med_buildings_vs_5G_adoption_pp}. First of all, the curves capturing the exposure over children and teenagers are smoother compared to the ones of schools and medical buildings. This is due to the fact that children and teenagers are rather spread over the buildings in the considered scenarios, while on the contrary schools and medical centers represent a fraction of the total buildings. As a result, the number of samples is larger for the formers and smaller for the latters, and thus resulting in smooth lines. Second, \ac{EMF} exposure due to 5G (early adoption) is lower than pre-5G one for all percentages of children/teenagers (Fig.~\ref{fig:ecdf_chd_adl_vs_5G_adoption_sp},\ref{fig:ecdf_chd_adl_vs_5G_adoption_pp}) and school buildings (Fig.~\ref{fig:ecdf_schools_vs_5G_adoption_sp},\ref{fig:ecdf_schools_vs_5G_adoption_pp}). In practical words, a child/teenager or a school building always receives an higher amount of exposure from pre-5G antennas than 5G ones during the early 5G adoption level. Third, the evolution of 5G adoption towards maturity results in larger \ac{EMF} exposure levels by 5G compared to pre-5G ones for children/teenagers and ``sensitive'' places, in both scenarios. Fifth, the maximum 5G exposure is lower than 15~V/m for children and teenagers, while always lower than 10~V/m for schools and medical buildings - well below the {28~[V/m] - 61~[V/m]} \ac{ICNIRP} {whole-body} limits \cite{ICNIRPGuidelines:20} for the general public over the considered frequencies .\footnote{We have also verified that the composite exposure of 5G and pre-5G adheres to the \ac{EMF} limits.} Sixth, the exposure in Ponte-Parione tends to be higher than Spinaceto for almost all the samples when considering the medium and maturity 5G adoption cases, due again to the higher tower density of the former with respect to the latter.

In the following step, we compare the exposure in the specific groups (children/teenagers, adults, schools, medical buildings) vs. the exposure collected over the whole set of samples (entire set of buildings and entire set of population). More technically, Fig.~\ref{fig:ecdf_avg_emf_comparison} details the exposure levels $E^{(.)}_b$ and $E^{(.)}_i$ vs. percentage of the following categories: \textit{i}) children/teenagers vs. adults vs. entire population (Fig.~\ref{fig:ecdf_pop_max_emf_sp},\ref{fig:ecdf_pop_max_emf_pp}), \textit{ii}) schools vs. medical buildings vs. all buildings (Fig.~\ref{fig:ecdf_building_max_sp},\ref{fig:ecdf_building_max_pp}). Three considerations hold by analyzing Fig.~\ref{fig:ecdf_avg_emf_comparison}. First, all the categories of people receive similar exposure levels. This outcome derives from the distributions of children/teenagers and adults over the territory, which are rather similar in the scenarios under considerations. Second, schools and medical buildings tend to receive different exposure levels compared to the category including all the buildings. Naturally, this is due to the fact that schools and medical buildings represent a  fraction of the total buildings, and therefore the positioning of each ``sensitive'' building, as well as the relative positioning of the towers in its surroundings, play a key role in determining the actual exposure levels. Third, the composite exposure is overall lower than the {28~[V/m] - 61~[V/m]} \ac{ICNIRP} general public {whole-body} limits \cite{ICNIRPGuidelines:20} over the considered frequency range, thus guaranteeing adherence to international exposure guidelines.

\subsection{Spatial Exposure Levels}
\label{sec:spatial_exposure}

\begin{figure}[t]
\centering
\subfigure[Spinaceto]
{
\includegraphics[width=8cm]{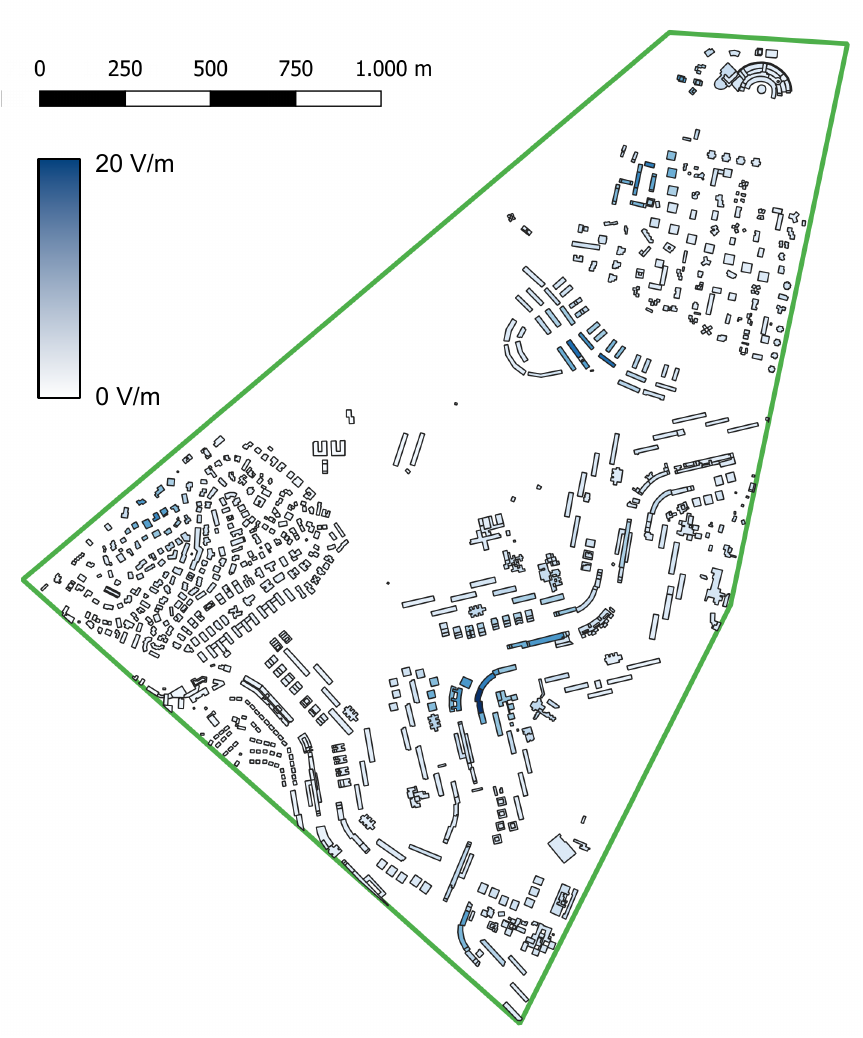}
\label{fig:emf_exposure_buildings_Spinaceto}
}
\subfigure[Ponte-Parione]
{
\includegraphics[width=8cm]{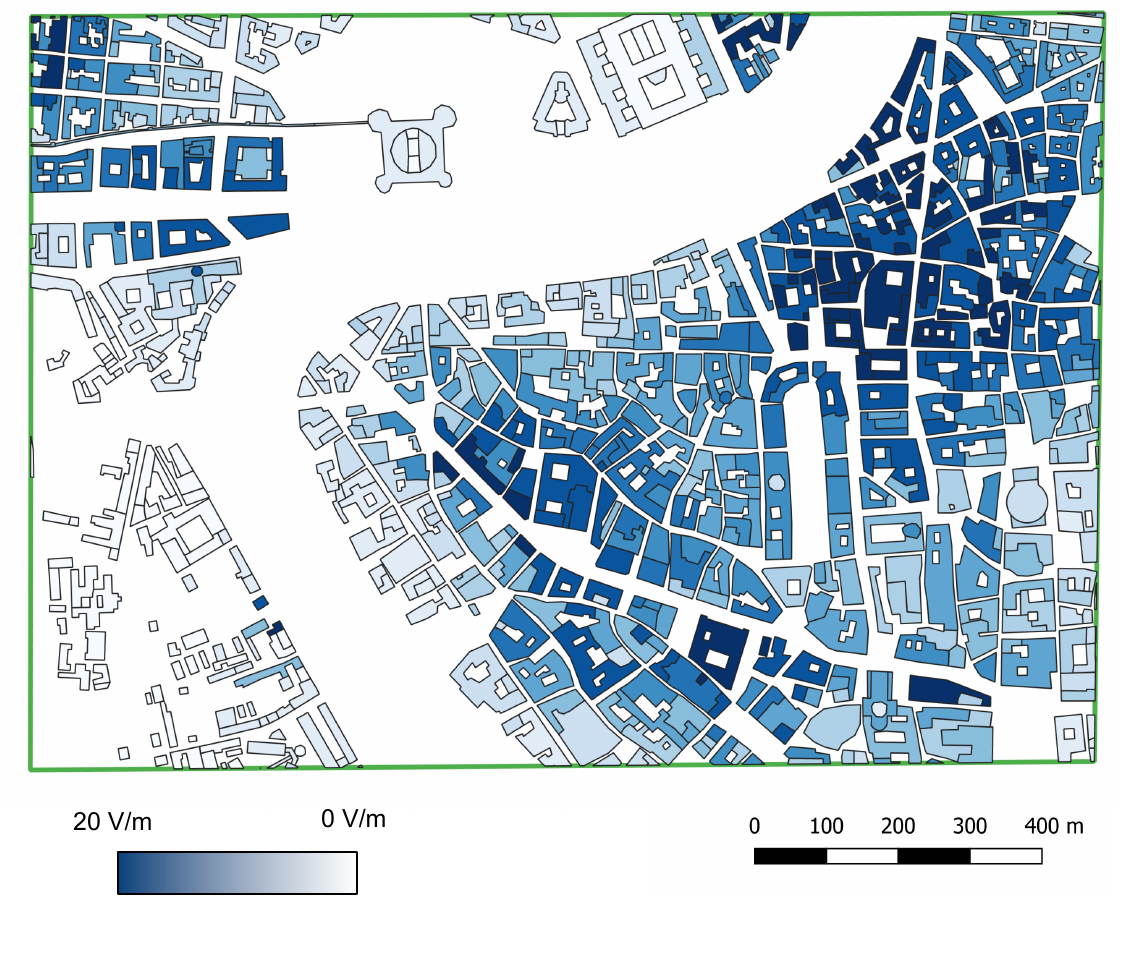}
\label{fig:emf_exposure_buildings_PP}
}
\caption{EMF exposure $E^{\text{TOT}}_b$ over the buildings (5G maturity setting).}
\label{fig:emf_exposure_buildings}
\vspace{-4mm}
\end{figure}

We then provide a visual representation of total exposure $E^{\text{TOT}}_b$ over the buildings, by considering the most conservative settings (i.e., maturity level for 5G).
 Fig.~\ref{fig:emf_exposure_buildings} visually highlights $E^{\text{TOT}}_b$ in Spinaceto and Ponte-Parione, respectively. Interestingly, $E^{\text{TOT}}_b$ strongly varies across the two scenarios. In more detail, Spinaceto is characterized by a relatively low amount of exposure, being the highest value (colored in dark blue color) limited to the few buildings close to the tower installations. On the other hand, the relatively higher tower density in Ponte-Parione compared to Spinaceto results in higher and more uniform exposure levels (as expected). Clearly, the zones in Ponte-Parione that are characterized by a low tower density (like the one show in bottom-left part of Fig.~\ref{fig:emf_exposure_buildings_PP}) are also the ones in which the buildings receive a lower amount of exposure.

\subsection{Impact of Building {and Weather} Attenuation}
\label{sec:building_attenuation}

\begin{figure}[t]
\centering
\subfigure[{Traditional Buildings}]
{
\includegraphics[width=6cm]{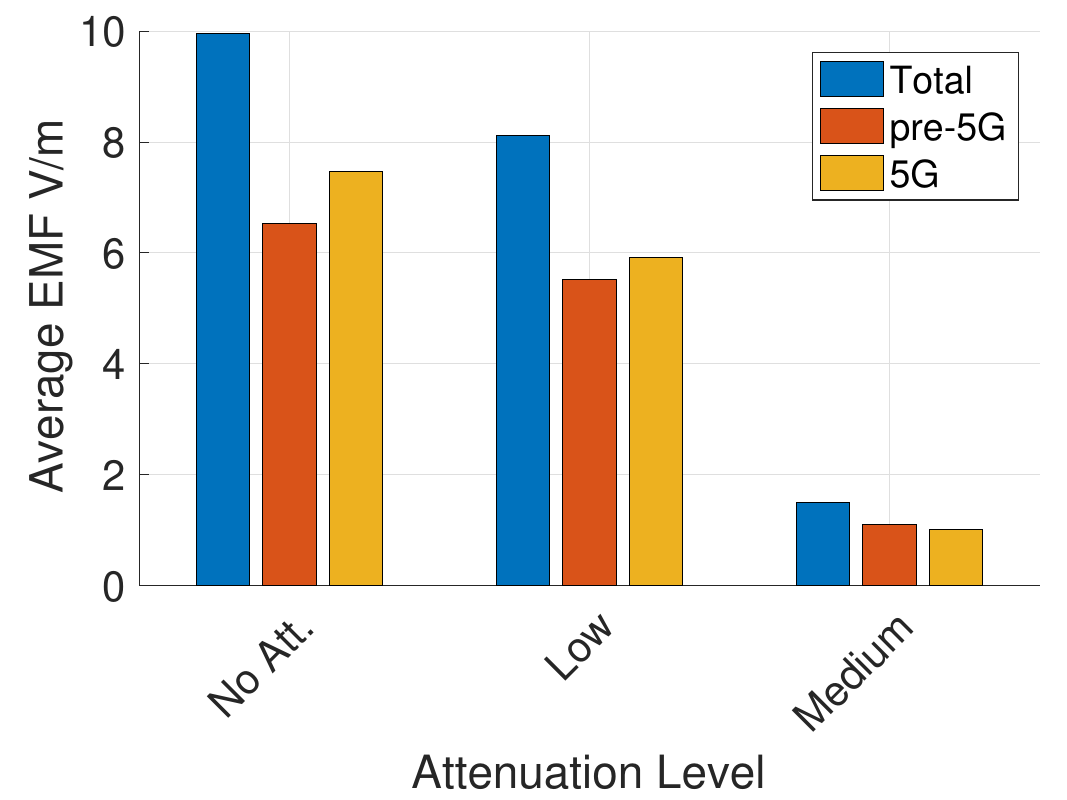}
\label{fig:var_att_chd_tgn_pp_maturity_tr_b}
}
\subfigure[{Thermally-efficient Buildings}]
{
\includegraphics[width=6cm]{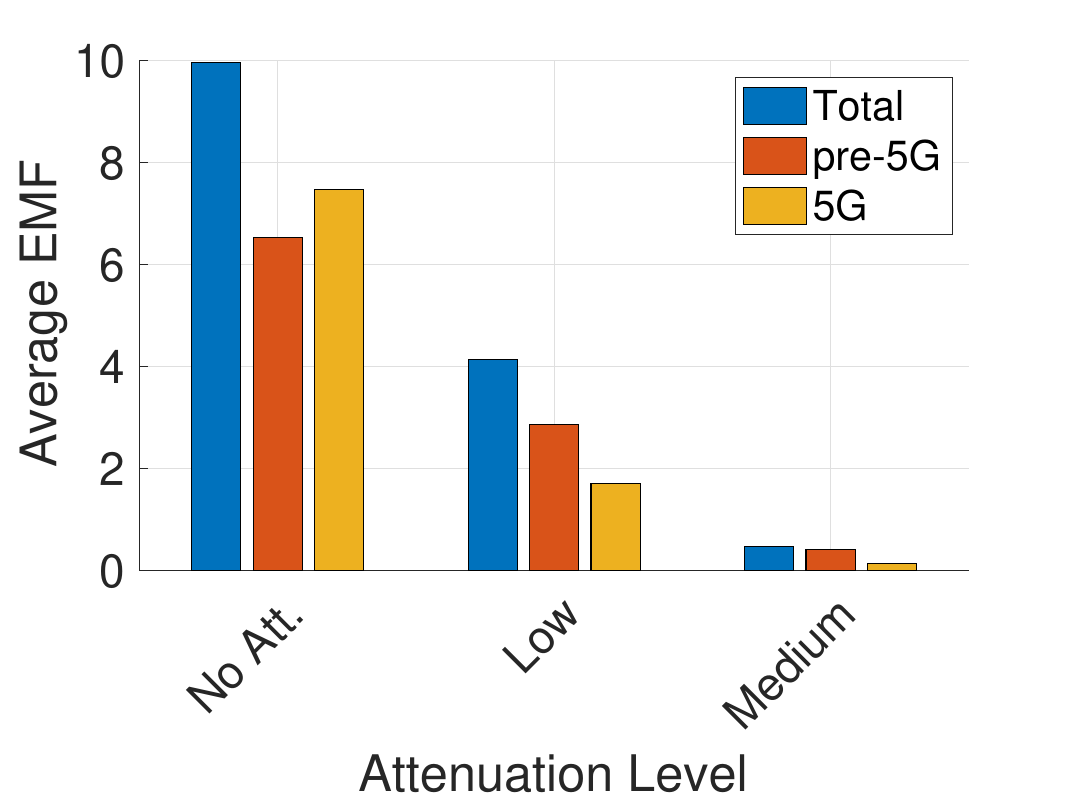}
\label{fig:var_att_chd_tgn_pp_maturity_ee_b}
}
\caption{Impact of introducing the building attenuation on the EMF exposure over children and teenagers $E^{\text{TOT}}_{\text{AVG-CHD-TN}}$, $E^{\text{PRE-5G}}_{\text{AVG-CHD-TN}}$, $E^{\text{5G}}_{\text{AVG-CHD-TN}}$ (Ponte-Parione scenario - 5G maturity setting).}
\label{fig:var_att_chd_tgn_pp_maturity}
\vspace{-4mm}
\end{figure}

In the {following} part of our analysis, we investigate the impact of including different attenuation levels $A_{(c,t,s,o,p,a)}=1$ on the \ac{EMF} assessment over children and teenagers. More in depth, we exploit the attenuation model defined in the \ac{ITU} recommendation P.2109-1 \cite{entryloss} to compute the attenuation values. Obviuously, computing the exact attenuation values for each child/teenager is a very challenging task. In particular, the attenuation level depends on multiple building features (such as building materials, walls thickness, windows/doors type and positioning, glass type, etc.), as well as the positioning of children/teenagers inside the building (e.g., proximity vs. farness with respect to the windows), which are (obviously) not under our control. Therefore, rather than targeting the exact computation of the attenuation for each child/teenager, we consider two representative cases, namely: \textit{i}) low attenuation level (computed with the \ac{ITU} model by setting 1\% of probability) and \textit{ii}) medium attenuation level (computed by averaging the attenuation values retrieved from the \ac{ITU} model between 1\% and 99\% of probability).  {Focusing on the building type, we consider the two categories introduced in the \ac{ITU} model \cite{entryloss}, namely:}
\begin{enumerate}
\item {thermally-efficient buildings, which largely exploit metallized glass and foil-backed panels as materials;}
\item {traditional buildings, not exploiting the aforementioned materials.} 
\end{enumerate}


Fig.~\ref{fig:var_att_chd_tgn_pp_maturity} shows the average \ac{EMF} exposure over children and teenagers $E^{\text{TOT}}_{\text{AVG-CHD-TN}}$, $E^{\text{PRE-5G}}_{\text{AVG-CHD-TN}}$, $E^{\text{5G}}_{\text{AVG-CHD-TN}}$ in Ponte-Parione (with 5G maturity setting),  {for the two building types. Each subfigure also includes} the case without attenuation for comparison.  {Focusing on the evaluation with traditional buildings (Fig.~\ref{fig:var_att_chd_tgn_pp_maturity_tr_b})}, the attenuation level has a large impact on the received exposure over children and teenagers, resulting in a notable exposure decrease for the medium setting compared to the other cases. In addition, the attenuation strongly affects the relative share of 5G exposure with respect to pre-5G one. Without attenuation, the 5G exposure is higher than pre-5G one. When attenuation is introduced, the share of 5G exposure with respect to pre-5G one is reduced, and eventually the exposure from 5G becomes lower than the pre-5G one when a medium attenuation is assumed. This effect is expected, as 5G includes mid-band and and mm-Wave frequencies, which are higher than the ones in use by pre-5G, and are therefore subject to stronger propagation losses.  {Focusing then on the evaluation with thermally-efficient buildings  \ref{fig:var_att_chd_tgn_pp_maturity_ee_b}, this case introduces a strong reduction on the exposure levels, even for the minimum attenuation case. In addition, the average exposure from 5G sources is always lower than the pre-5G in both the minimum and average cases.}   

{Finally,  we have  investigated the additional attenuation that is introduced by the weather conditions (rain,  snow) on the exposure levels.  We refer the interested reader to Appendix~B for the details,  while here we summarize the salient outcomes. In brief, snow events are extremely rare conditions in the city of Rome,  and therefore the impact of snow on the exposure levels can be neglected.  Focusing then on rain,  such events are more frequent in Rome, and therefore their effect on the exposure levels has to be carefully assessed.  However,  the additional results reported in Appendix~B reveal that even a violent rain condition does not dramatically influence the exposure levels. This is due to the following reasons: \textit{i}) the distance between each tower and a pixel is rather short; consequently,   large attenuation factors per km are then translated into low attenuation levels over the (short) distance,  and \textit{ii}) rain condition substantially affects only the exposure from mm-Wave frequencies,  while exposure over lower frequencies is not substantially changed.}

\section{{Discussion}}
\label{sec:discussion}

{In this section, we provide more insights of our work about the following (important) aspects: \textit{i)} influence of path loss model, \textit{ii)} uplink vs. downlink exposure, \textit{iii)} user equipment mobility, \textit{iv)} indoor exposure, \textit{v)} transmit power variation, and \textit{vi)} \ac{QoS} metrics.}  

\subsection{{Influence of Path Loss Model}}

{The exposure computation of Eq.~(\ref{eq:power_density}) is based on the point-source model, a widely accepted solution in the academic literature (see e.g., the survey \cite{9518367}) and in the exposure regulations (see e.g.,  IEC \cite{iecexclusion} and ITU \cite{itutk70} guidelines) to compute the power density from the radiated power of the base stations. More technically, let us focus on a generic pixel and a generic source of radiation. The point-source model can be split into the following sub-tasks:}
\begin{enumerate}
\item {computation of the received power (expressed in [W]) by applying the Friis propagation model \cite{friis1946note}:}
\begin{equation}
\label{eq:pr_prop}
P^R = P^T G^T G^R \left(\frac{\lambda}{4\pi r}\right)^2
\end{equation}
{Where $P^R$ is the power received by the receiver antenna,  $P^T$ is the power radiated by the transmitter antenna,  $G^T$ is the gain of the transmitter antenna,  $G^R$  is the gain of the receiver antenna,  $\lambda$ is the wavelength and $r$ is the source-to-pixel 3D distance.}
\item {computation of the antenna effective area (expressed in [m$^2$]).}
\begin{equation}
\label{eq:ae_area}
 A^E = G^R \left(\frac{\lambda^2}{4\pi}\right)
\end{equation}
\item {computation of the power density (expressed in [W/m$^2$]) by dividing the received power of Eq.~(\ref{eq:pr_prop}) for the antenna effective area of Eq.~(\ref{eq:ae_area}).}
\begin{equation}
\label{eq:pd_prop}
S = \frac{P^R}{A^E}  = \frac{P^T G^T}{4\pi r^2}
\end{equation}
\end{enumerate}
{It is worth noting that the path loss model is adopted in Eq.~(\ref{eq:pr_prop}) and consequently transferred to Eq.~(\ref{eq:pd_prop}.) In more detail,  the Friis propagation model employs a free space path loss.  In our case, a free space propagation correctly captures the path loss conditions, as we remind that our evaluation is done at the eaves level of the buildings, in which \ac{LOS} conditions with most of cellular towers are experienced, thanks also to the mostly flat terrain of the two scenarios and the short tower-to-pixel distance. In addition, it is worth noting that the Friis propagation model is almost equivalent to more complex channel models (like the ones defined by 3GPP \cite{3gppmodel}), under \ac{LOS} and below breakeven distance assumptions - two conditions normally experienced at the eaves level in our scenarios.}

\subsection{{Uplink vs.  Downlink Exposure}}

{The focus of our work is on the downlink exposure from cellular towers, and not on the uplink exposure generated by user terminals. This choice is motivated by the following reasons: \textit{i}) cellular towers represent the major source of concern for the population \cite{9518367}, and \textit{ii}) exposure from user terminals is radically different than the one radiated by cellular towers \cite{9518367} (in terms of metrics, patterns, evaluations, and impact on the population), thus requiring in general a different approach for the exposure assessment compared to the one adopted in this work.}

{To this aim, the main standardization activities for the human exposure from personal devices are reviewed by Hirata \textit{et al.} \cite{hirata2021assessment}. Colombi \textit{et al.} \cite {colombi2018rf} investigate the exposure from mm-Wave devices in near-field conditions. The impact of international regulations on the maximum power and \ac{EIRP} of user equipment are investigated by He \textit{et al.} \cite{he2020implications}. Lundgren \textit{et al.} \cite{lundgren2020near} propose an innovative technique for measuring exposure from user equipment in the near-field region of the radiating antenna. Finally, Castellanos \textit{et al.} \cite{castellanos2022multi} target a multi-objective cellular planning problem that integrates downlink and uplink exposure, concluding that: \textit{i}) uplink exposure is higher than the downlink one, and \textit{ii}) both metrics should be considered during the plannig phase of the network. }

\subsection{{User Equipment Mobility}}

{The moving of a user equipment inside the same coverage of a given sector and/or across multiple sectors may potentially impact the exposure levels that are received by the population. In our work, the statistical reduction factors that are applied to mid-band and mm-Wave 5G antennas are set by imposing the low-mobility scenario described in the \ac{IEC} standards \cite{iecexclusion,iec62669}. Such setting represents the most conservative one for deriving the values of Tab.~\ref{tab:stat_reduction} and it is inline with the national regulation for evaluating exposure from 5G and 4G antennas employing massive \ac{MIMO} and beamforming \cite{deliberasnpa}. On the other hand, a mobility increase of 5G user equipment would result in a decrease of exposure levels, as, intuitively, the temporal and spatial variations of the beams over the territory are increased. Focusing then on the pre-5G and sub-GHz 5G antennas, we point out that the evaluation of such sources is done at the maximum power, and therefore mobility does not affect the presented outcomes.}

\subsection{{Indoor Exposure}}

{Our work includes a coarse evaluation of indoor exposure, by investigating the effect of building type on the exposure levels at the eaves level of the building (as shown by Fig.~\ref{fig:var_att_chd_tgn_pp_maturity}). Obviously, a detailed indoor evaluation (e.g.,  over the whole set of floors for each building,  and over each building room) would dramatically increase the complexity of the evaluation,  requiring a lot of building details (like building planimetry,  thickness of internal walls,  presence of doors,  installation of other sources like \ac{DAS},  etc.) that are not under our control and cannot be easily extracted from our scenarios - which we remind include hundreds to thousands of buildings (as reported in Tab. ~\ref{tab:buildings_pop_metrics}). This specific aspects should be treated in a future work, possibly focusing on smaller areas with a lower building density,  from which the indoor sources can be more easily extracted.}

\subsection{{Transmit Power Variation}}

{In practice,  5G towers or terminals can artificially adjust the \ac{EMF} exposure level while ensuring the quality of communication service \cite{jiang2022hybrid}.  Therefore,  the instantaneous exposure level is dependent on the amount of traffic that is flown on the tower-to-\ac{UE} communication link \cite{liu2021field,chiaraviglio2022emf}.  Differently from these works,  our approach is based on the exposure assessment over longer time scales (e.g.,  dozens of minutes and/or hours). Therefore,  rather than focusing on the instantaneous exposure variation,  our goal is to compute exposure levels that result from an average over a long time scale.  This is also an essential point for the comparison of the exposure levels against the (long-term) whole-body exposure limits. } 

{In addition,  we point out that the temporal variation of \ac{EMF} levels from 5G sources operating in the mid-bands and over mm-Wave frequencies is intrinsically captured by the statistical reduction factors of Tab.~\ref{tab:stat_reduction},  which introduce a scaling to the maximum power in accordance to the temporal and spatial variation of exposure over the territory.  In particular, the power reduction factors that are associated to the early 5G adoption level result in a low transmit power of the antenna on the panel,  while, obviously,  the factors employed in the medium and maturity cases generate an higher transmit power of the antenna. }

{Focusing instead on pre-5G and sub-GHz 5G sources,  our evaluation is done at the maximum level.  This assumption is motivated by the fact that pre-5G sources have already reached a level of maturity,  and therefore a large amount of traffic is carried by pre-5G networks. Moreover, the primary goal of sub-GHz 5G sources is not to sustain high level of traffic,  but rather to provide coverage over the territory.  Consequently, the implementation of the smart antenna features (like \ac{MIMO} and beamforming) is limited for such equipment. \footnote{{The implementation of \ac{MIMO} and beamforming in turn influences the statistical reduction factor of 5G antennas operating on mid-band and mm-Wave.}}  Therefore,  the \ac{EMF} exposure evaluation for such sources is done in our work under conservative settings.}

\subsection{{QoS Metrics}}

{The set of towers and their radio configuration have obviously an impact on the \ac{QoS}/coverage metrics over the territory.  In particular,  such metrics are derived from physical layer parameters of the radio channel (like \ac{RSSI},  \ac{RSRP},  \ac{RSRQ}) that are evaluated over each pixel. Our simulator, obviously, does not reach such level of detail, because our goal is to evaluate the average exposure that is received over the population,  and not to compute complex \ac{QoS} metrics like the RSSI,  which would require a complete simulation of the physical layer in the time-frequency domain for each considered antenna.  We also point out that the methodology for the assessment of human exposure followed through this work is fully compliant with widely adopted international standards (like the IEC 62232:2017 \cite{iecexclusion}), where the aforementioned metrics are not included in the list of parameters for the radio-frequency exposure assessment of base stations. Therefore,  this aspect could be treated by a future research activity, more tailored to the investigation of user-centric metrics from the 5G deployments. }

\begin{table*}[h]
    \caption{{Main Notation.}}
    \label{tab:notation}
    \scriptsize
    \centering
    \begin{tabular}{|m{1.5cm}|m{6cm}|m{1.5cm}|m{6cm}|} 
\hline
\rowcolor{Linen} \textbf{Symbol} & \textbf{Explanation} & \textbf{Symbol} & \textbf{Explanation}\\
\hline

$(.)$ & Possible sets $\{\text{TOT},\text{5G},\text{PRE-5G}\}$& $N^{\text{MED-CENTER}}$ & Number of medical centers\\
$A_{(c,t,s,o,p,a)}$ & Attenuation over pixel $c$ from antenna $a$  on panel $p$ from sector $s$ of operator $o$ on tower $t$& $N^{\text{SCHOOL}}$ & Number of schools \\
$\mathcal{A}$ & Set of antennas& $\mathcal{O}$ & Set of operators\\
$\mathcal{B}$ & Set of buildings &  $P^{\text{MAX}}_{(t,s,o,p,a)}$ & Maximum output power of antenna $a$ on panel $p$ from sector $s$ of operator $o$ on tower $t$\\
$\mathcal{B}_n$ & Subset of buildings belonging to census zone $n$& $\mathcal{P}$ & Set of panels\\
$\alpha^{\text{5G}}_p$ & 5G statistical reduction factor for panel $p$ & $\mathcal{P}^{\text{EAVES}}_b$ & Eaves plane of building $b$ (set of points)\\
$C_b$ & Type of building $b$&  $\mathcal{P}^{\text{SHAPE}}_b$ & 3D shape of building $b$ (set of points) \\
$\mathcal{C}$ & Set of pixels & $r_{(c,t,s,o,p)}$ & 3D distance between pixel $c$ and panel $p$ from sector $s$ of operator $o$ on tower $t$ \\
$\mathcal{C}_b^{\text{EAVES}}$ & Subset of pixels belonging to the eaves plane of building $b$& $R_{(t,s,o,p,a)}$ & Power reduction factor for antenna $a$ of panel $p$ from sector $s$ of operator $o$ on tower $t$\\
$C_p$ & Type of panel $p$& $\rho_{(t,s,o)}$ &  Normalized orientation of sector $s$ of operator $o$ on tower $t$ \\
$C_t$ & Type of tower $t$& $\rho^\text{CLOCK}_{(t,s,o)}$ &  Orientation of sector $s$ of operator $o$ on tower $t$ \\
$D^{\text{H}}_{(t,s,o,p,a)}$ & Horizontal diagram of antenna $a$ on panel $p$ from sector $s$ of operator $o$ on tower $t$& $S_{(c,t,s,o,p,a)}$ & Power density on pixel $c$ received by antenna $a$  on panel $p$ from sector $s$ of operator $o$ on tower $t$\\
$D^{\text{V}}_{(t,s,o,,p,a)}$ & Vertical diagram of antenna $a$  on panel $p$ from sector $s$ of operator $o$ on tower $t$& $S^\text{TOT}_c$ & Total power density on pixel $c$\\
$E^{(\cdot)}_{\text{AVG}}$ & Linear average building exposure over the possible sets $(.)$&  $S^\text{5G}_c$ & Total power density from 5G sources on pixel $c$ \\
$E^{(\cdot)}_{\text{AVG-AD}}$ & Linear average adults exposure over the possible sets $(.)$&  $S^\text{PRE-5G}_c$ & Total power density from pre-5G sources on pixel $c$\\
$E^{(\cdot)}_{\text{AVG-CHD-TN}}$ & Linear average children and teenager exposure over the possible sets $(.)$& $\mathcal{S}$ & Set of sectors\\
$E^{(\cdot)}_{\text{AVG-MED}}$ & Linear average medical center exposure over the possible sets $(.)$& $\sigma^\text{E}_{(t,s,o,p,a)}$ & Electrical tilt of antenna $a$ on panel $p$ from sector $s$ of operator $o$ on tower $t$\\
$E^{(\cdot)}_{\text{AVG-POP}}$ & Linear average population exposure over the possible sets $(.)$& $\sigma^\text{M}_{(t,s,o,p)}$ & Mechanical tilt of panel $p$ from sector $s$ of operator $o$ on tower $t$\\
$E^{(\cdot)}_{\text{AVG-SCHOOLS}}$ & Linear average schools exposure over the possible sets $(.)$& $T_a$ & Technology of antenna $a$\\
$E^{(\cdot)}_b$ & Total electric field on building $b$ over the possible sets $(.)$& $\mathcal{T}$ & Set of towers \\
$E^{(\cdot)}_c$ & Total electric field on pixel $c$ over the possible sets $(.)$& $\theta_{(c,t,s,o,p)}$ & Vertical plane angle  of pixel $c$ w.r.t.  panel $p$ from sector $s$ of operator $o$ on tower $t$\\
$E^{(\cdot)}_i$ & Total electric field on individual $i$ over the possible sets $(.)$&  $V_b$ & Volume of building $b$ \\
$F_a$ & Frequency of antenna $a$& $V^{\text{TOT}}_n$ & Total volume of building in census zone $n$ \\
$\phi_{(c,t,s,o,p)}$  & Horizontal plane angle of pixel $c$ w.r.t.  panel $p$ from sector $s$ of operator $o$ on tower $t$& $z_b$ & global height of building $b$ (eaves level)\\
$G_{(c,t,s,o,p,a)}$ & Received gain on pixel $c$ from antenna $a$ on panel $p$ from sector $s$ of operator $o$ on tower $t$&   $z_{(t,s,o,p)}$ & Global height of panel $p$ from sector $s$ of operator $o$ on tower $t$ \\
$G^{\text{MAX}}_{(t,s,o,p,a)}$ & Maximum gain of antenna $a$ on panel $p$ from sector $s$ of operator $o$ on tower $t$& $z^{\text{CE}}_{(t,o)}$ & Electrical center of operator $o$ on tower $t$ \\
$\Gamma_{(t,s,o,p)}$ & Rotation matrix of panel $p$ from sector $s$ of operator $o$ on tower $t$&  $z^{\text{EAVES}}_b$ & Height above ground of eaves of building $b$ \\
$\mathcal{I}^{\text{CHD-TN}}_b$ & Set of children/teenagers in building $b$&  $z^{\text{LOCAL}}_{(s,p)}$ & Relative positioning of  panel $p$ on sector $s$ w.r.t. the electrical center \\
$\mathcal{I}^{\text{AD}}_b$ & Set of adults in building $b$& $z^{\text{SEA}}_b$ & Altitude above sea level of building $b$ \\
$N^{\text{AD}}_b$ & Number of adults in building $b$  & $x_b$, $y_b$ & Building $b$ positioning in global UTM coordinates \\
$N^{\text{AD}}_n$ & Number of adults in census zone $n$ & $x_c$, $y_c$  & Pixel $c$ positioning in UTM coordinates \\
$N^{\text{CHD-TN}}_b$ & Number of children/teenagers in building $b$ &  $x_t$, $y_t$ & Tower $t$ positioning in UTM coordinates\\
$N^{\text{CHD-TN}}_n$ & Number of children/teenagers in census zone $n$ & $\tilde{x}_{(c,t,s,o,p)}$, $\tilde{y}_{(c,t,s,o,p)}$, $\tilde{z}_{(c,t,s,o,p)}$ & Local coordinates of pixel $c$ w.r.t. panel $p$ from sector $s$ of operator $o$ on tower $t$ \\
$\mathcal{N}$ & Set of census zones & & \\
\hline
\end{tabular}
\end{table*}

\section{Summary and Future Work}
\label{sec:conclusions}


We have assessed the impact of 5G towers on the \ac{EMF} exposure over children/teenagers, school buildings and medical buildings. After introducing a novel methodology for the analysis of exposure over population and buildings, we have applied it in two meaningful scenarios that are subject to different urbanization levels as well as tower deployments. Our results reveal that, although 5G exposure radiated by towers is initially lower than the pre-5G one, 5G will become the dominant source of exposure from cellular towers when a maturity level will be reached. However, the scaling factors applied to the maximum power radiated by mid-band/mm-Wave antennas, the tower distribution and the positioning of the buildings are important aspects heavily influencing the exposure levels over young people and ``sensitive'' places. Eventually, the actual level of exposure over children and teenagers is largely impacted by the building attenuation level, which has a stronger effect over mid-band and mm-Wave 5G frequecies compared to pre-5G ones. Overall, our results indicate that the total exposure levels are always lower than the \ac{EMF} limits reported in international regulations. Moreover, children and teenagers receive similar amount of exposure compared to the whole population. Eventually, the positioning of the ``sensitive'' building has an impact on the exposure level, but, however, the observed exposure trends are similar compared to the ones of the whole set of buildings.

As future work, we plan to extend our assessment to entire municipalities/cities, including zones covered by 5G small cells. In addition, as propagation has a strong effect over the exposure received by children and teenagers, massive campaigns of \ac{EMF} measurements from 5G towers should be performed, especially inside the buildings.  Finally, the investigation of joint uplink and downlink 5G exposure is another avenue of research. 

\section*{{Appendix A: Main Notation}}
\label{app:notation}

{Tab.~\ref{tab:notation} reports the list of symbols used in this work.}

\section*{{Appendix B: Impact of Weather Conditions}}
\label{app:rain_attenuation}

{Snow and rain generally influence the path loss between a user and a cellular tower and consequently the exposure levels on the user. Focusing on the former, snowfall events are extremely rare in the city of Rome, and in any case lasting at most for few hours at most. Since our goal is to evaluate exposure over longer time scales, the impact of snow can be neglected in our analysis. On the contrary, rain conditions are more frequent in the considered scenarios, and therefore the impact of rain must be evaluated. More specifically, we start from the ITU-R P.838-3 recommendation \cite{ITURP8383}, which reports a widely recognized model to express the rain attenuation as a function of the operating frequency (in GHz) and the rain intensity (in mm/h). We then consider the set of frequencies of the panels adopted in this work and the following rain intensities:}
\begin{enumerate}
\item {light rain, corresponding to 1 mm/h;}
\item {moderate rain, corresponding to 6 mm/h;}
\item {{heavy rain, corresponding to 30 mm/h;}
\item violent rain, corresponding to 60 mm/h.}
\end{enumerate}

\begin{figure}[t]
\centering
\includegraphics[width=8cm]{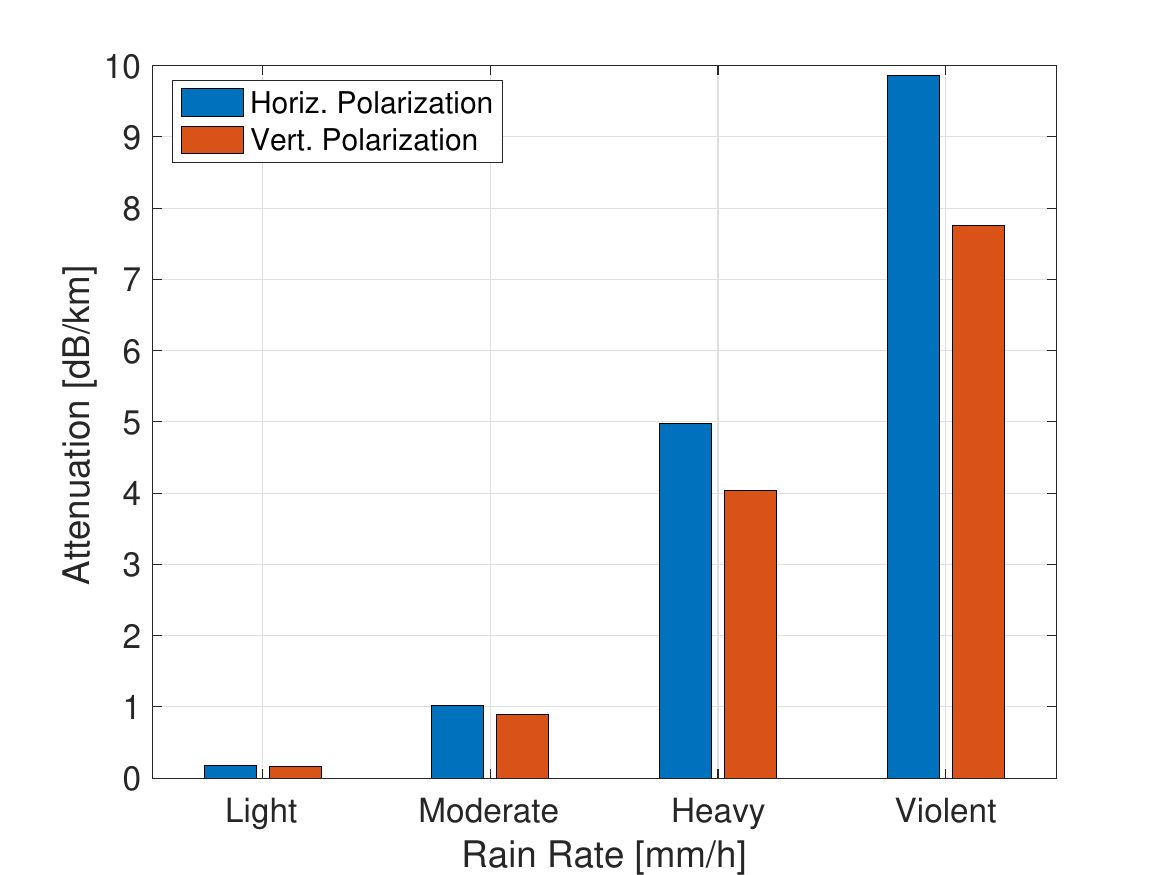}
\caption{{Impact of rain rate on the attenuation per km at 26~[GHz].}}
\label{fig:attenuation_rain}
\end{figure}

{We then apply the ITU model \cite{ITURP8383} to compute the rain attenuation (expressed in dB/km). Interestingly, the computed rain attenuation for the frequencies up 3.7 GHz is always lower than 0.06 dB/km, and hence it is negligible in our analysis. On the contrary, the rain attenuation strongly depends on the rain intensity when considering the 26 GHz frequency, as shown in Fig.~\ref{fig:attenuation_rain} for both horizontal and vertical polarizations. Interestingly, light rain does not substantially affect the path loss, since the rain attenuation in this case is strongly lower than 1~[dB/km]. On the contrary, larger attenuation values are experiences for the moderate, heavy, and violent rain cases, which may (likely) result in a reduction of the experienced exposure over the buildings and over the population.} 

{In order to evaluate the impact of the rain attenuation on the population exposure, we have proceeded as follows:}
\begin{enumerate}
\item {we have assumed three different rain attenuation values, corresponding to the aforementioned moderate, heavy and violent cases and the horizontal polarization case (i.e., the most conservative one);\item for each value of attenuation (expressed in dB/km), each pixel and each source, we have computed the actual rain attenuation loss in dB;}
\item {we have included the rain attenuation loss in the power density computation of Eq.~\ref{eq:power_density};}
\item {we have run our simulator on the most complex scenario (Ponte-Parione), under 5G maturity settings.}
\end{enumerate}

\begin{figure}[t]
\centering
\includegraphics[width=6cm]{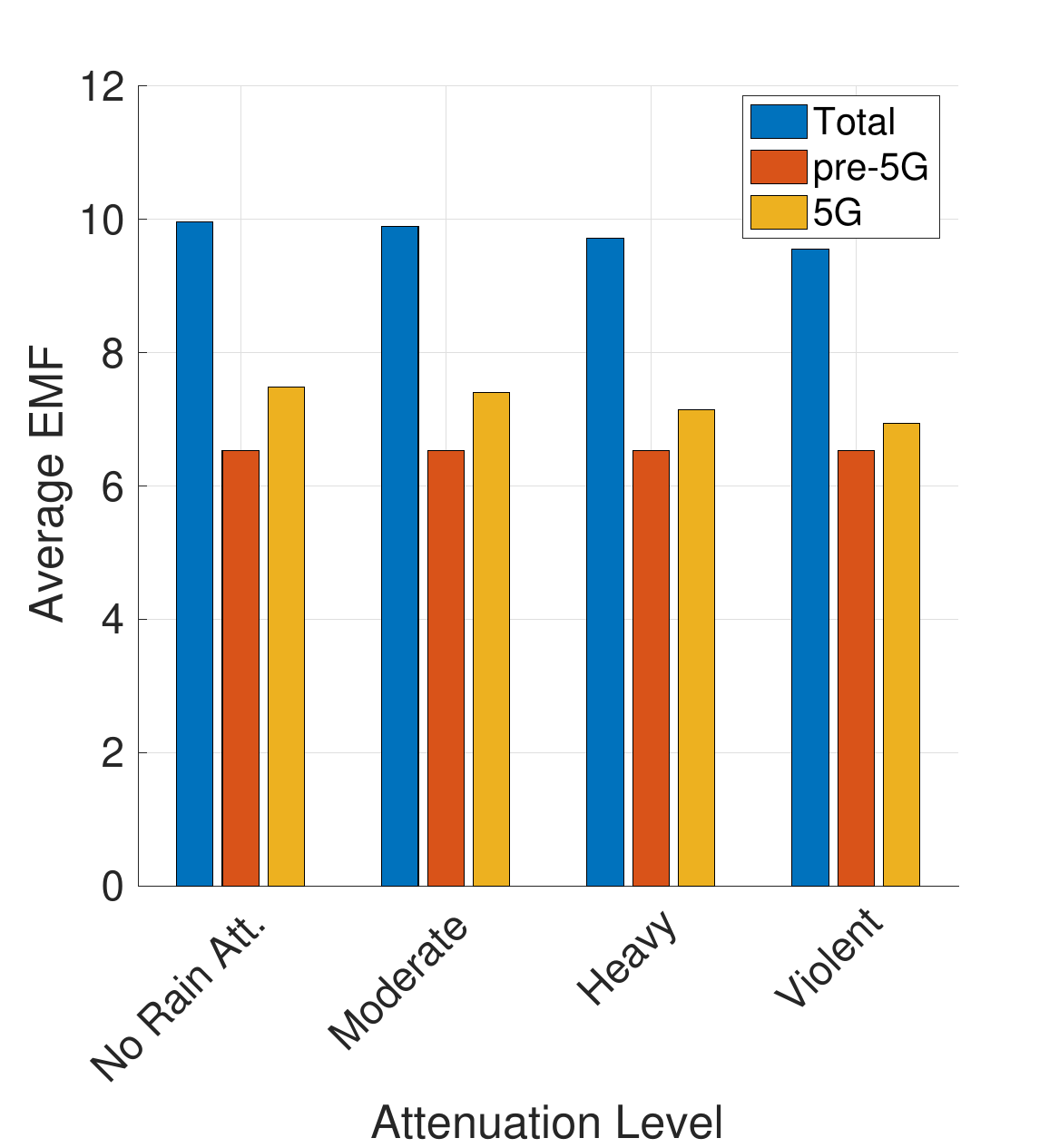}
\caption{{Impact of rain attenuation on the exposure over children and teenagers (Ponte-Parione scenario) - figure best viewed in colors.}}
\label{fig:attenuation_rain_impact}
\end{figure}

{Fig.~\ref{fig:attenuation_rain_impact} reports the average exposure over children and teenagers vs. the different rain attenuation levels. Interestingly, the attenuation introduced by the rain tends to decrease the average exposure (blue bars in the figure). However, the maximum reduction, observed with the violent rain setting, is overall lower than 1~[V/m] on average compared to the case without rain attenuation. This outcome may be explained by the fact that the distance between each pixel and each cellular tower is always rather limited in our scenarios (i.e., few hundred meters at most). Consequently, although the attenuation per km is not marginal (as shown in Fig.~\ref{fig:attenuation_rain}), the actual attenuation that is experienced on the tower-to-pixel path is always rather low. Focusing then on 5G and pre-5G exposure terms (red and yellow bars), the rain does not affect pre-5G exposure (as expected), while a decrease is observed when considering 5G exposure. The set of 5G sources includes in fact frequencies at 26~[GHz], which are subject to a non-negligible attenuation per km when introducing the heavy and violent rain condition. However, the overall exposure reduction of the whole set of 5G sources - which include also sub-6 GHz frequencies - is always rather limited. }

\bibliographystyle{ieeetr}

\begin{IEEEbiographynophoto}{Luca Chiaraviglio} (M'09-SM'16) 
received the Ph.D. degree in Telecommunication and Electronics Engineering from
the Politecnico di Torino, Italy. He is currently an Associate Professor
with the University of Rome ``Tor Vergata'', Italy. He has co-authored more than
150 articles published in international journals, books, and conferences. His current research topics cover
5G networks, optimization applied to telecommunication networks,
electromagnetic fields, and health risks assessment of 5G communications.
Dr. Chiaraviglio received the Best Paper Award at the IEEE Vehicular Technology Conference (VTC)-Spring 2020, the IEEE VTC-Spring
2016, and the Conference on Innovation in Clouds, Internet and
Networks (ICIN) 2018, all of them appearing as the first author.
Some of his papers are listed as the Best Readings on Green
Communications by the IEEE. Moreover, he has been recognized as
an Author in the Top 1\% Most Highly Cited Papers in the Information
and Communication Technology (ICT) field worldwide and top 2\% world scientists according to the 2021 
update of the science-wide author databases of standardized citation indicators.
\end{IEEEbiographynophoto}

\begin{IEEEbiographynophoto}{Chiara Lodovisi}
 is a researcher at CNIT (Italy) and at the University of Rome Tor Vergata (Italy). She holds a Ph.D. in Engineering Electronics, obtained at the University of Rome Tor Vergata (Italy). She worked for five years as an RF Engineer consultant for H3G mobile operator. For five years she worked on optical communications, study and implementation of submarine and satellite optical links and radio over fiber.  Her research topics concern 5G networks, health risk assessment of 5G communications, interoperability over fiber between TETRA/LTE systems and 5G networks. 
\end{IEEEbiographynophoto}

\begin{IEEEbiographynophoto}{Daniele Franci} 
received the M.Sc. degree (cum laude) and the Ph.D. degree in nuclear and subnuclear physics from the Sapienza University of Rome, Rome, Italy, in 2007 and 2011, respectively. From 2009 to 2011, he was a Technology Analyst with Nucleco S.p.A, involved in the radio-logical characterization of radioactive wastes from the decommissioning of former Italian nuclear power plants. He joined Agenzia per la Protezione Ambientale del Lazio (ARPA Lazio), in 2011, being involved in RF-EMF human exposure assessment. Since 2017, he has been involved in the activities with the Italian Electro Technical Committee (CEI) for the definition of technical procedures for EMF measurement from 4G/5G mMIMO sources.
\end{IEEEbiographynophoto}

\begin{IEEEbiographynophoto}{Enrico Grillo}  received the M.Sc. degree in electronics engineering from the Seconda Universit{\'a} di Napoli, Aversa, Italy, in 1999.
In 2000, he works on RF to grow up the first 3G telecommunication radio network. Since 2005, he has been involved as a Research Technician in the prevention and monitoring of electromagnetic
pollution with Agenzia per la Protezione Ambientale del Lazio (ARPA Lazio), the local environmental agency of the Lazio Region.
\end{IEEEbiographynophoto}

\begin{IEEEbiographynophoto}{Settimio Pavoncello} 
was born in Rome, Italy, in 1973. He received the M.Sc. degree in Telecommunications Engineering from the Sapienza University of Rome, Rome, in 2001. Since 2002, he has been working with the EMF Department, Agenzia per la Protezione Ambientale del Lazio (ARPA Lazio), Rome. He is specialized in electromagnetic field measurements and EMF projects evaluation related to radios, TVs, and mobile communications systems maturing huge experience in the use of broadband and selective instruments. In past years, he has deepened in the issues related to measurements on LTE and NB-IoT signals. Since 2018, he has been actively involved in the working group of the Italian Electrotechnical Committee aimed at defining measurement procedures for mobile communications signals and is currently engaged in various projects concerning measurement on 5G signals.
\end{IEEEbiographynophoto}

\begin{IEEEbiographynophoto}{Tommaso Aureli} 
received the M.Sc. degree in biological science from the Sapienza University of Rome, Rome, Italy, in 1985. He joined Agenzia per la Protezione Ambientale del Lazio (ARPA Lazio), in 2002. From 2004 to 2018, he was the Director of the EMF Division, being involved in both measurement and provisional evaluation EMF from civil sources. He is currently the Director of the Department of Rome of ARPA Lazio.
\end{IEEEbiographynophoto}

\begin{IEEEbiographynophoto}{Nicola Blefari-Melazzi} 
is currently a Full
Professor of telecommunications with the University
of Rome ``Tor Vergata'', Italy. He is currently
the Director of CNIT, a consortium of 37 Italian
Universities. He has participated in over 30 international
projects, and has been the principal investigator
of several EU funded projects. He has been
an Evaluator for many research proposals and a
Reviewer for numerous EU projects. He is the
author/coauthor of about 200 articles, in international
journals and conference proceedings. His research interests include
the performance evaluation, design and control of broadband integrated
networks, wireless LANs, satellite networks, and of the Internet.
\end{IEEEbiographynophoto}

\begin{IEEEbiographynophoto}{Mohamed-Slim Alouini}  (S'94-M'98-SM'03-F'09) was born in Tunis, Tunisia. He received the Ph.D.degree in Electrical Engineering from the California Institute of Technology (Caltech), Pasadena, CA, USA, in 1998. He served as a faculty member in the University of Minnesota, Minneapolis, MN, USA, then in the Texas A\&M University at Qatar,Education City, Doha, Qatar before joining King Abdullah University of Science and Technology (KAUST), Thuwal, Makkah Province, Saudi Arabia as a Professor of Electrical Engineering in 2009. His current research interests include modeling, design, and performance analysis of wireless communication systems.
\end{IEEEbiographynophoto}

\end{document}